%% file: tamuthesis.tex
\documentclass[12pt]{report}
\usepackage[letterpaper]{geometry}
 \usepackage[doublespacing]{setspace}
 \usepackage{tocloft}
 \usepackage[rm, tiny,center, compact]{titlesec}
 \usepackage{indentfirst}
 \usepackage{etoolbox}
\usepackage{tocvsec2}
 \usepackage[titletoc]{appendix}
 \usepackage{appendix}
 \usepackage{tamuconfig}
\usepackage{rotating}
\usepackage[utf8]{inputenc}
\usepackage{amsmath}
\usepackage{siunitx}
\usepackage{float}
\usepackage{multirow}
\usepackage[square,comma,numbers,sort]{natbib}
\usepackage{algorithm}
\usepackage{algpseudocode}
\usepackage{varwidth}
\DeclareRobustCommand\nocite[1]{{\def\cite##1{\ignorespaces}#1}}
\newcommand\nocitecaption[1]{\caption[\nocite{#1}]{#1}}

\begin{document}

\renewcommand{\tamumanuscripttitle}{Inclusive jet longitudinal double-spin asymmetry $A_{LL}$ Measurements in 510 G\lowercase{e}V Polarized $\lowercase{pp}$ Collisions at STAR}
\renewcommand{\tamupapertype}{Dissertation}
\renewcommand{\tamufullname}{Zilong Chang}
\renewcommand{\tamudegree}{Doctor of Philosophy}
\renewcommand{\tamuchairone}{Carl A. Gagliardi}
\renewcommand{\tamumemberone}{Rainer J. Fries}
\newcommand{\tamumembertwo}{Saskia Mioduszewski}
\newcommand{\tamumemberthree}{Joseph B. Natowitz}
\newcommand{\tamumemberfour}{Robert E. Tribble}
\renewcommand{\tamudepthead}{Peter M. McIntyre}
\renewcommand{\tamugradmonth}{December}
\renewcommand{\tamugradyear}{2016}
\renewcommand{\tamudepartment}{Physics}

\include{titlepage} 
\include{abstract}
\include{dedication}
\include{acknowledgements}

\include{lists}  

\include{chapter1}
\include{chapter2}
\include{chapter3}
\include{chapter4}
\include{chapter5}
\include{chapter6}
\include{chapter7}

\let\oldbibitem\bibitem
\renewcommand{\bibitem}{\setlength{\itemsep}{0pt}\oldbibitem}
\include{bibliography}

\end{document}

%% file: titlepage.tex
%
%
%


\providecommand{\tabularnewline}{\\}

\begin{titlepage}
\begin{center}
\MakeUppercase{\tamumanuscripttitle}
\vspace{4em}

A \tamupapertype

by

\MakeUppercase{\tamufullname}

\vspace{4em}

\begin{singlespace}

Submitted to the Office of Graduate and Professional Studies of \\
Texas A\&M University \\

in partial fulfillment of the requirements for the degree of \\
\end{singlespace}

\MakeUppercase{\tamudegree}
\par\end{center}
\vspace{2em}
\begin{singlespace}
\begin{tabular}{ll}
 & \tabularnewline
& \cr
Chair of Committee, & \tamuchairone\tabularnewline
Committee Members, & \tamumemberone\tabularnewline
 & \tamumembertwo\tabularnewline
 & \tamumemberthree\tabularnewline
 & \tamumemberfour\tabularnewline
Head of Department, & \tamudepthead\tabularnewline

\end{tabular}
\end{singlespace}
\vspace{2em}

\begin{center}
\tamugradmonth \hspace{2pt} \tamugradyear

\vspace{3em}

Major Subject: \tamudepartment \par
\vspace{3em}
Copyright \tamugradyear \hspace{.5em}\tamufullname 
\par\end{center}
\end{titlepage}
\pagebreak{}

%% file: abstract.tex
%
%
%

\chapter*{ABSTRACT}
\addcontentsline{toc}{chapter}{ABSTRACT} 

\pagestyle{plain} 
\pagenumbering{roman} 
\setcounter{page}{2}

\indent With the early \(e-p\) deep inelastic scattering (DIS) measurements and the development of Quantum Chromo-dynamics (QCD), the proton is revealed to be not just composed of the three quarks, \(u\), \(u\), and \(d\), that give its quantum numbers, but also thousands of quark and anti-quark pairs and gluons that mediate the strong forces among quarks. The densities of partons - quarks, anti-quarks, and gluons - inside the proton are given by the parton distribution functions (PDFs).  The PDFs are functions of the momentum fraction, \(x\), carried by the partons within the proton and the scale, \(Q^2\), at which the densities are probed.

In the longitudinally polarized lepton-nucleon and nucleon-nucleon scatterings, the polarized PDFs also depend on the spin orientation of the parton relative to the proton, like and unlike the proton spin, in addition to \(x\) and \(Q^2\). The early polarized DIS experiments carried out by the EMC collaboration and later experiments at HERMES and COMPASS showed that the quarks inside the proton only contribute approximately 30\% of the total proton spin. As proposed by the Jaffe-Manohar sum rule, the proton spin receives contributions not only from the quarks, but also from the gluons and from the orbital angular momentum of the quarks and gluons. This leaves an open question to further explore the gluon and orbital momentum contributions.

The Relativistic Heavy Ion Collider (RHIC) at Brookhaven National Laboratory is a facility that collides protons polarized in both longitudinal and transverse directions at energies up to \(\sqrt{s} = 510\) GeV. The STAR and PHENIX detectors, located at two separate locations on the RHIC ring, can both provide useful constraints on the gluon distributions for \(x\) as low as 0.02. In particular, over the last decade STAR has constrained the gluon polarization with measurements of the longitudinal double-spin asymmetry, \(A_{LL}\), for inclusive jet production in \(\sqrt{s} = 200\) GeV \(pp\) collisions.  The results provide the first evidence, at the level of \(\sim 3 \sigma\), that the gluons in the proton with \(x>0.05\) are polarized.

In this analysis, I perform the first ever measurement of \(A_{LL}\) for inclusive jet production in \(pp\) collisions at the higher beam energy of \(\sqrt{s}\) = 510 GeV, based on data that STAR recorded during 2012.  The higher beam energy extends the sensitivity to gluon polarization down to \(x \sim 0.02\).  The high statistics of the data set and the small size of the physics asymmetries, compared to the previous measurements at 200 GeV, required the development of several new or improved analysis procedures in order to minimize the systematic uncertainties.  These include:  the first implementation by STAR of an underlying event subtraction during jet reconstruction, a much improved technique to estimate the trigger and reconstruction bias effects, a detailed optimization of the PYTHIA tune that provides a much better match between the experimental data and simulated Monte Carlo events, and a new procedure to estimate the uncertainties associated with the PYTHIA tune parameters.

The results for inclusive jet \(A_{LL}\) \textit{vs}.\@ jet \(p_T\) in 510 GeV \(pp\) collisions are presented.  They are found to be consistent with predictions from recent global analyses of the polarized PDFs that included prior RHIC data in the fit.  They are also consistent with the previous STAR inclusive jet \(A_{LL}\) measurements at \(\sqrt{s}\) = 200 GeV in the region where the kinematics for the two beam energies overlap.  These results will provide important new constraints on the gluon polarization in the proton in the \(x\) region below that sampled in 200 GeV \(pp\) collisions.

\pagebreak{}

%% file: dedication.tex
%
%
%

\chapter*{DEDICATION}
\addcontentsline{toc}{chapter}{DEDICATION}  

\indent This dissertation is dedicated to my parents, Huirong Li and Cunbao Chang, and my younger brother, Ziqing Chang, for their unconditional love and unwavering support.

\pagebreak{}

%% file: acknowledgements.tex
%
%
%

\chapter*{ACKNOWLEDGEMENTS}
\addcontentsline{toc}{chapter}{ACKNOWLEDGEMENTS}  

\indent I would like to thank my advisor, Dr. Carl Gagliardi, for his continuous guidance and support. I have made steady progress over the past six years under his advice, and now I am well equipped with knowledge and skills. Thank you for the support from all my committee members, Dr. Rainer Fries, Dr. Saskia Mioduszewski, Dr. Joseph Natowitz and Dr. Robert Tribble.

I am grateful for the help from former group members, Dr. Pibero Djawotho, especially for his help at the early stage of my analysis and establishing the data analysis framework, Dr. James Drachenberg, Dr. Mriganka Mondal, and Liaoyuan Huo. I would also like to thank members of the STAR Spin Physics Working Group, Dr. Renee Fatemi and her group at University of Kentucky, including Kevin Adkins and Suvarna Ramachandran, Dr. Elke Aschenauer, Dr. Bernd Surrow, Dr. Scott Wissink,  Dr. Stephen Trentalange, Dr. Ernst Sichtermann, Dr. Anselm Vossen, Dr. Brian Page, Dr. Grant Webb, Dr. Yuxi Pan, Dr. Jinlong Zhang, Chris Dilks, Danny Olvitt, the members of STAR Software Computing Group headed by Dr. Jerome Lauret, and Dr. Eleanor Judd from the STAR Trigger Group.

In addition, I would like to thank faculty, staff and fellow graduate students at the Cyclotron Institute, Dr. Che-Ming Ko, Dr. Yiu-Wing Lui, Dr. Feng Li, and Ms. Paula Barton. I am thankful for all the friendship that I have made over the past six years. I'm lucky enough to be able to finish my physics graduate study at Texas A\text{\&}M University and spend four college years at University of Science and Technology of China. 

This material  is based upon work supported in part by the United States Department of Energy, under grant number DE-FG02-93ER40765.

\pagebreak{}

%% file: lists.tex
%
%
%

\phantomsection
\addcontentsline{toc}{chapter}{TABLE OF CONTENTS}  

\begin{singlespace}
\renewcommand\contentsname{\normalfont} {\centerline{TABLE OF CONTENTS}}


\setlength{\cftaftertoctitleskip}{1em}
\renewcommand{\cftaftertoctitle}{%
\hfill{\normalfont {Page}\par}}

\tableofcontents

\end{singlespace}

\pagebreak{}


\phantomsection
\addcontentsline{toc}{chapter}{LIST OF FIGURES}  

\renewcommand{\cftloftitlefont}{\center\normalfont\MakeUppercase}

\setlength{\cftbeforeloftitleskip}{-12pt} 
\renewcommand{\cftafterloftitleskip}{12pt}

\renewcommand{\cftafterloftitle}{%
\\[4em]\mbox{}\hspace{2pt}FIGURE\hfill{\normalfont Page}\vskip\baselineskip}

\begingroup

\begin{center}
\begin{singlespace}
\setlength{\cftbeforechapskip}{0.4cm}
\setlength{\cftbeforesecskip}{0.30cm}
\setlength{\cftbeforesubsecskip}{0.30cm}
\setlength{\cftbeforefigskip}{0.4cm}
\setlength{\cftbeforetabskip}{0.4cm} 

\listoffigures

\end{singlespace}
\end{center}

\pagebreak{}

%
\phantomsection
\addcontentsline{toc}{chapter}{LIST OF TABLES}  

\renewcommand{\cftlottitlefont}{\center\normalfont\MakeUppercase}

\setlength{\cftbeforelottitleskip}{-12pt} 

\renewcommand{\cftafterlottitleskip}{12pt}

\renewcommand{\cftafterlottitle}{%
\\[4em]\mbox{}\hspace{4pt}TABLE\hfill{\normalfont Page}\vskip\baselineskip}

\begin{center}
\begin{singlespace}

\setlength{\cftbeforechapskip}{0.4cm}
\setlength{\cftbeforesecskip}{0.30cm}
\setlength{\cftbeforesubsecskip}{0.30cm}
\setlength{\cftbeforefigskip}{0.4cm}
\setlength{\cftbeforetabskip}{0.4cm}

\listoftables 

\end{singlespace}
\end{center}
\endgroup
\pagebreak{}  

%% file: chapter1.tex
%
%
%


\pagestyle{plain} 
\pagenumbering{arabic} 
\setcounter{page}{1}

\chapter{\uppercase {Introduction: The Proton Internal Structure}}

\section{Parton Model of the Proton}

The picture of a proton which is composed by two up quarks (\(u\)) with charge \(\frac{2}{3}\)e and spin \(\frac{1}{2}\) and one down quark (\(d\)) with charge \(-\frac{1}{3}\)e and spin \(\frac{1}{2}\), the so called quark model, seems to depict the proton quantum numbers perfectly \cite{ZweigOne,ZweigTwo,GMann1964}. The charge sum matches with the proton charge +e. From the Pauli principle, the three quarks exactly make up the total proton spin \(\frac{1}{2}\). However, a series of experiments in the past three decades has shown the proton internal structure is far more abundant and intriguing than the simple three quark model.

Deep inelastic scattering (DIS) experiments in the late 1960s at Stanford Linear Accelerator Center (SLAC) confirmed that quarks are the  constituents of the proton as suggested by the proton quark model \cite{SLACiep930,SLACiep935,Feynman1969,Bjorken1969}. The development of Quantum-chromo-dynamics (QCD) in the 1970s demonstrated that quarks are confined inside the proton and the interactions among quarks are intermediated by gluons \cite{DGross1973, HPolitzer1973}. Gluons can also split into quark and anti-quark or gluon-gluon pairs. Therefore inside the proton, beside the three quarks mentioned above that contribute proton's quantum numbers, known as valence quarks, there are gluons and quark and anti-quark pairs, the so called sea quarks. This picture is generally accepted as the parton model of the proton.

The partons - valence quarks, sea quarks and gluons - are distributed by certain forms of functions called parton distribution functions (PDFs). Since partons cannot break free from each other, the parton distribution function is probed at a certain energy scale, for example the momentum transfer \(k\) between two partons. In general the PDFs are expressed as a function of the momentum fraction carried by the parton, often denoted as Bjorken \(x\), at the energy scale \(Q^{2}\) = \(k^{2}\). At a fixed energy scale \(Q^{2}\), the \(u\) and \(d\) quarks obey the following equations:
\begin{align}
\int[u(x)-\bar{u}(x)]dx= 2, \int [d(x)-\bar{d}(x)]dx=1,
\end{align}
where \(u(x)\) and \(d(x)\) are the \(u\) and \(d\) quark distributions and \( \bar{u}(x)\) and \(\bar{d}(x)\) are the anti-\(u\) and anti-\(d\) quark distributions at fixed \(Q^2\). Also the momentum sum rule needs to be satisfied.
\begin{align}
\int x[\sum_{q}((q(x)+\bar{q}(x)) +g(x)]dx=1,
\end{align}
where \(q\) represents the possible flavors of quarks and \(g(x)\) is the gluon distribution at fixed \(Q^{2}\).

At high energy, perturbative quantum-chromo-dynamics (pQCD) is able to calculate the two-parton cross-sections at certain precision and the lepton-hadron and hadron-hadron scattering cross-sections can be approximately expressed as the convolution of hadron PDFs and partonic cross-sections. The proton PDFs have been explored through various experiments, for example DIS experiments and hadron collider experiments, by measuring scattered products in particle detectors. One common method is to compare the cross-section of measured scattered products with theoretical calculations to un-convolute the PDFs. One common technique is to assume certain function forms with several undetermined variables for PDFs at initial momentum transfer, \(Q_0^{2}\), then use DGLAP evolution equations\cite{Gribov1972,Dokshitzer1977,Altarelli1977} to evolve the PDFs to the \(Q^{2}\) of the experiment data, convolute the proper PDFs with the pQCD partonic cross-sections to get the theoretical cross-sections, and then fit the data with the theoretical cross-sections to determine the free parameters in order to obtain the PDFs.

\section{Notable Experiments}
Several recent DIS experiments at the Hadron Electron Ring Accelerator (HERA) during the past two decades provided precise measurements on the proton PDFs covering a wide \(x-Q^{2}\) range where \(0.045<Q^{2}<30000\) \(\textrm{GeV}^{2}\) and \(6 \times 10^{-5} < x < 0.65 \). HERA had the capability to collide electrons or positrons up to 30 GeV with high energy protons up to 920 GeV. The neutral current cross-section, \(ep \rightarrow eX\) via a photon or \(Z\) boson exchange, charge current cross-section, \(ep\rightarrow \nu X\) via a \(W^{\pm}\) boson exchange, inclusive jet production and open charm production were studied by the H1 and ZEUS collaborations to determine proton PDFs \cite{h1zeus2010}.

The Tevatron, a hadronic collider, also gives extra constraints on the proton PDFs. Protons and anti-protons with center of mass energy 1.96 TeV collided with each other at Tevatron.  Inclusive jet measurements have been done by the CDF and D0 collaborations, which are noteworthy to provide constraints on the high-\(x\) gluon distribution inside the proton \cite{cdfjets2008,d0jets2012}. In addition the lepton charge asymmetry from \(W\) decay and \(Z\) boson rapidity distribution are sensitive to the quark distributions inside the proton \cite{cdfw2005,d0w2008,cdfz2010,d0z2007}. The Drell-Yan dimuon production from E866/NuSea at Fermilab is another measurement to access the anti-quark distribution in the proton.  The experiment measured the ratio of muon pairs from an 800 GeV proton beam incident on liquid hydrogen and deuterium targets.  The ratio directly unfolds the ratio of \(\bar{d}\) to \(\bar{u}\) distributions inside the proton, and showed \(\bar{d}> \bar{u}\) at \(0.015 < x < 0.35\) \cite{e866nuesea2001}.

\section{Global QCD Analysis}
A Global analysis is a theoretical framework to predict the PDF from the global experimental data. The global analysis assumes certain functional-form dependences on \(x\) at its initial \(Q_0^2\) for the quarks and anti-quarks with flavor \(u\), \(d\) and \(s\) and the gluons.  The parameters of those functions are fitted to the experimental data by using the PDF evolution techniques and the leading order (LO), the next-to-leading (NLO) or the next-to-next-to-leading order (NNLO) theoretical cross-sections. The results of PDFs are often called the LO PDF, the NLO PDF or the NNLO PDF based on the choice of the LO, NLO, or NNLO theoretical calculations. Nowadays most global analyses provide NNLO PDFs for unpolarized protons. But in this document, only NLO PDF will be discussed because that is the state of the art for polarized protons. The results of a global analysis give the PDF as a function of \(x\) and \(Q^2\) and its uncertainties for the quarks and anti-quarks with flavor \(u\), \(d\), \(s\) and the gluons.

\begin{figure}[H]
\centering
\includegraphics[scale=0.5]{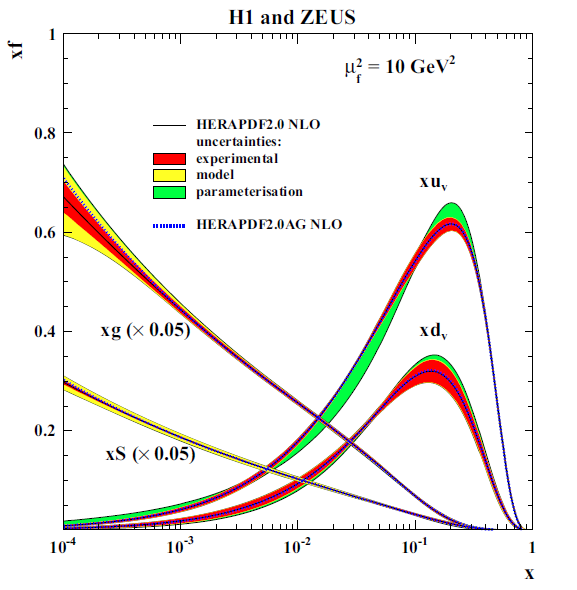}
\nocitecaption{The NLO HERAPDF 2.0 fits where $xS = 2x(\bar{u} + \bar{d} + \bar{s})$ at $Q^2 = 10$ $\textrm{GeV}^2$ \cite{HERAPDF20}.}
\label{fig:herapdfnlo}
\end{figure}

One NLO analysis is HERAPDF, which uses the datasets from the H1 and ZEUS collaborations at HERA. The newest NLO HERAPDF2.0 analysis uses various combined datasets from the two collaborations with minimal \(Q^2 \) of 3.5 GeV, which includes charged and neutral current cross-sections and inclusive jet production \cite{HERAPDF20}. The charged and neutral current cross-sections sufficiently extract the valence and sea quark distributions and the gluon distribution from scaling violation. Though the gluon distribution extracted from scaling violation correlates strongly with the coupling constant \(\alpha_s\), the jet cross-section data provide an independent measurement of the gluon distribution. The NLO fit gives a \(\chi^2\) per degree of freedom of 1.2 and agrees well with the measured HERA data. Figure \ref{fig:herapdfnlo} shows the valence quark, sea quark and gluon distribution from the recent HERAPDF2.0 at $Q^2 = 10$ $\textrm{GeV}^2$.

The NLO CT14 fit is another global QCD analysis from the CTEQ collaboration, based upon its several previous versions such as CT10, CTEQ6, and so on \cite{cteq6l2002, CT10, CT14}. Not only does it have the DIS data from HERA, the lepton  asymmetry from W boson and inclusive jet data from the Tevatron, Drell-Yan measurements from E866, but it also includes data from the LHC. It gives the best fit by minimizing the global \(\chi^{2}\) among experiment data.  The non-perturbative effects such as higher-twist effects or nuclear corrections are reduced by putting certain kinematic cuts on the experimental data. The new PDF gives a good description for the inclusive jet cross-sections at the LHC, which helps to constrain the gluon distribution.

\begin{figure}[H]
\centering
\includegraphics[scale=0.6]{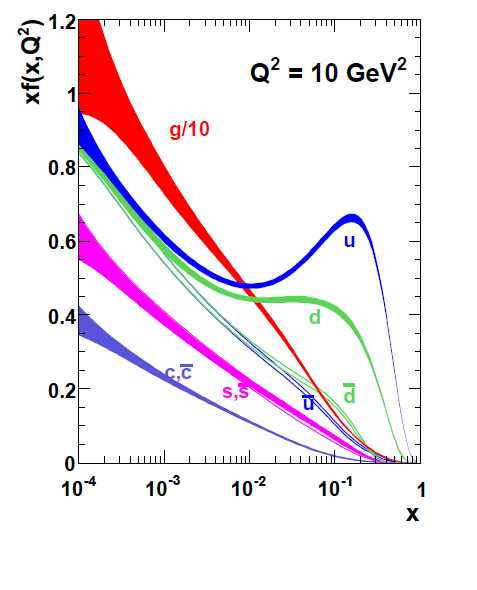}
\nocitecaption{The NLO MSTW fits at $Q^2 = 10$ $\textrm{GeV}^2$ \cite{MSTW}.}
\label{fig:mstw2008nlo}
\end{figure}

The NLO MSTW global analysis provides another useful set of PDFs \cite{MSTW}. A variety of data were selected but with a cut to reduce higher-twist effects. The dimuon production from neutrino-nucleon scattering experiments at NuTeV provide constraints on the strange quark and anti-quark distributions. The lepton charge asymmetry from \(W\) boson decay and \(Z\) boson rapidity distribution measurements at Tevatron constrain the \(d\) quark distribution. The inclusive jet data at the Tevatron prefer a small gluon distribution at high \(x\).  Figure \ref{fig:mstw2008nlo} shows the quark and anti-quark and the gluon distribution from the MSTW NLO predictions.

The neural network technique is also applied to determine the PDFs, such as the NNPDF model \cite{NNPDFunp}. The NNPDF group has produced its most recent version NNPDF3.0 with LHC data. However in this document, the older version NNPDF2.3 is chosen as the reference for the un-polarized proton PDF. The NNPDF2.3 uses DIS data from HERA and fixed-target experiments, Drell-Yan data from FermiLab, and boson and inclusive jet production at Tevatron. The techniques can be summed up into two stages, first produce a replicate set of data by Monte Carlo (MC) sampling of the probability distribution of the input dataset; second construct PDFs for every replica by using a neural network fit. The parameterization is made flexible and unbiased with 37 free parameters per flavor. The group also developed a new algorithm to speed the DGLAP evolution. A thousand replicas are produced and the PDFs are determined from the best fit of each replica. The final result PDF is the average of the best fits from 1000 replicas and the standard deviation is taken as the uncertainty on the PDF. The model gives compatible results with the MSTW and CTEQ6 models except larger uncertainties on quark distributions, larger gluon uncertainties at small-\(x\), and smaller uncertainties on the difference between \(\bar{u}+ u\) and \(\bar{d} + d\).

\section{Exploring Polarized Proton PDFs}
Though it is interesting enough to understand how partons are distributed inside the proton, it is not yet complete without understanding how partons make up the proton spin \(\frac{1}{2}\). The simple static quark model with two \(u\) quarks and one \(d\) quark explains the \(\frac{1}{2}\) spin quantum number well. However based on the parton model, it's straightforward to imagine inside the proton some partons have their spin directions along the proton spin, the others have their spin directions opposite to the proton spin and all the partons revolving around. The net effect of the parton spin and their orbital motions is the 1/2 proton spin, the so-called proton sum rule \cite{JaffeManohar1990},
\begin{align}
\frac{1}{2}= \frac{1}{2} \Delta\Sigma + \Delta G + L_{q,\bar{q},g},
\end{align}
where \(\Delta\Sigma = \int \sum_{q}[(q(x)^{+} -  q(x)^{-})+ (\bar{q}(x)^{+} - \bar{q}(x)^{-}]dx\), \(\Delta G= \int [ g(x)^{+} -g(x)^{-}]dx\), and \(L_{q,\bar{q},g}\) is the orbital angular momentum of quarks, anti-quarks, and gluons. The \(q(x)^{+(-)}\), \(\bar{q}(x)^{+(-)}\), and \(g(x)^{+(-)}\) are the spin-dependent or polarized PDFs of quarks, anti-quarks and gluons, where \(+\) means the parton spin direction is along the proton spin direction and \(-\) means the parton spin direction is opposite to the proton spin direction. Several important experiments show the proton spin is far more complex than the simple three quark model could explain.

At European Muon Collaboration (EMC), high energy longitudinally polarized muon beams impinging on longitudinally polarized proton targets were used to study the spin-dependent parton distributions. The muons beams ranging from 50 GeV to 300 GeV were produced from pion decay and the polarization could be up to 82\% at 200 GeV. Irradiated ammonia was used as the target material because of abundant free protons and high resistance to radiation damage. The target polarization was about 75\% to 80\% \cite{emcexp1981}. The spin-asymmetry\( A=\frac{(\sigma^{\uparrow\uparrow}-\sigma^{\uparrow\downarrow})}{(\sigma^{\uparrow\uparrow}+\sigma^{\uparrow\downarrow})}\), was measured where \(\sigma^{\uparrow\uparrow(\uparrow\downarrow)}\) is the cross-section when the polarization of muons and protons are along(opposite) with each other.

The spin-asymmetry is related to the virtual photon-nucleon spin asymmetry \(A_{1}= \frac{\sigma^{1/2}-\sigma^{3/2}}{\sigma^{1/2}+\sigma^{3/2}}\), where \(\sigma^{1/2(3/2)}\) is the photo-absorption cross-section when the projection of the total angular momentum of the virtual photon-nucleon system along the virtual photon direction is \(\frac{1}{2} (\frac{3}{2})\). \(A_{1}\) is directly related to the spin-dependent structure function \(g_{1}^{p}(x) = \frac{1}{2} \sum e_{q}^2 [q(x)^{+} - q(x)^{-}]\) in the scaling limit. \(A_{1}\) was measured at different incident muon beam energies, at 100, 120 and 200 GeV, which covers the \(x\) range from 0.01 and 0.7 and the \(Q^{2}\) range from 1.5 to 70 \(\textrm{GeV}^{2}\). The integral of \(g_{1}^{p} (x)\)  over \(x\) from 0.01 to 0.7 is calculated as \(0.114 \pm 0.012\) (stat.) \(\pm 0.026 \) (syst.). Also the integral of neutron \(g_{1}^{n}(x)\) can be calculated assuming the validity of Bjorken sum rule \cite{bjokensum1970}. Based on the integral of \(g_{1}^{p} (x) \) and \(g_{1}^{n} (x)\) and ignoring the strange sea quark contribution, the total contribution from \(u\) and \(d\) quarks is \(14 \pm 9 \) (stat.) \(\pm 21 \) (syst.) percent of the proton spin. The calculated integral of \(g_{1}^{p}(x)\) is smaller than the Ellis-Jaffe sum rule prediction \cite{ellisjaffe1974}. Assuming the difference is contributed by strange sea quark polarization, then the total contribution from \(u\), \(d\) and \(s\) is \(1 \pm 12\) (stat.) \(\pm 24\) (syst.) percent of the proton spin. In summary quarks and anti-quarks in the proton carry a small fraction of the total proton spin, and the other larger part should be carried by gluons and the orbital angular moment \cite{emc1988}.

The COMPASS experiment at CERN is also using longitudinally polarized high energy muon beams and longitudinally polarized fixed targets to study spin-dependent proton structure. The incident muon beam energy varies between 140 and 180 GeV with
polarization about 80\%. The target is a solid state target. The irradiated ammonia (\(NH_{3}\)) provides the polarized protons with polarization about 85\%. The \(^{6}LiD\) is used to provide polarized deuterons with polarization about 50\%, because \(^{6}Li\) is regarded as a system of a deuteron and a helium-4 (\(^{4}He\)) and has essentially the same magnetic moment as the deuteron. The targets are placed in two or three separate cells around the beam line and the polarization (\(\rightarrow\) or \(\leftarrow\) ) in the cells can be different from each other, so the beam can hit the targets with both polarizations simultaneously. The polarization of the targets can be flipped from longitudinal to transverse. The scattered muons and hadrons are captured in its detector system \cite{compassexp2007}.

The inclusive measurements of the spin asymmetry \(A_{1}^{p}\) and \(A_{1}^{n}\) by using proton and deuteron targets respectively have been performed in the kinematic region \(0.004 < x < 0.7\) and \(Q^{2} > 1\) \(\textrm{GeV}^{2}\) \cite{compassg1p2010,compassg1n2007}. In order to allow flavor separation in exploring quark distribution functions, the semi-inclusive measurements of \(A_{1}^{p}\) and \(A_{1}^{n}\) with charged pions (\(\pi^{+,-}\)) and kaons (\(K^{+,-}\)) have also been performed in the same kinematic region, except the \(x\) of \(A_{1}^{n}\) extend from 0.004 to 0.3 \cite{compasssidisg1p2010,compasssidisg1n2009}. The spin-dependent structure function \(g_{1}^{p(n)} (x)\) is then calculated from \(A_{1}^{p(n)}\), which is used to extract the spin dependent PDFs in the further analysis, for example the NLO global analysis. The polarized gluon distribution \(\Delta g\) can be accessed from the \(Q^{2}\) dependent \(g_{1}^{p(n)} (x)\) in the above measurements, however it only covers a small range of \(Q^{2}\) which limits the ability to constrain \(\Delta g\). The virtual photon-gluon fusion, \(\gamma^{*} g\rightarrow q\bar{q}\), makes it possible to access \(\Delta g\) \cite{compassphoton2006}. The charm production, which is reconstructed from decays to charged pions (\(\pi^{+,-}\)) and kaons (\(K^{+,-}\)) \cite{compassdcharm2009,compasscharm2013} and the high-\(p_T\) hadron pairs \cite{compasshadronpairs2013} due to the process, are measured.

The HERMES experiment at DESY is another DIS experiment to study the spin-dependent proton structure. It uses an innovative technique for its targets, the gaseous targets of polarized atoms of hydrogen and deuteron. The direction of polarization can be flipped within milliseconds. It can achieve about 85\% polarization for longitudinally polarized targets and about 75\% for transversely polarized targets \cite{hermestargets2005}. The electron and positron beams are operating at the energy of 27.5 GeV. The un-polarized electron or position beams become spontaneously transversely polarized by the emission of synchrotron radiation. The polarization can go up to 60\% as the beam develops. A spin rotator can be applied to make the beam longitudinally polarized. The scattered lepton and hadron are detected by its detector system with good particle identification capability \cite{hermesexp1998}.

HERMES measures the spin dependent \(A_{1}^{p(n)}\) to extract \(g_{1}^{p(n)}\) at \(0.0041 < x < 0.9\) and \(0.18 \textrm{ GeV}^{2} < Q^{2} < 20 \textrm{ GeV}^{2}\) with the polarized positron beams and hydrogen and deuteron targets \cite{hermesdis2007}. The semi-inclusive measurements of \(\pi^{+}\) and \(\pi^{-}\) with hydrogen targets and \(\pi^{+}\), \(\pi^{-}\), \(K^{+}\) and \(K^{-}\) with deuteron targets allow to access the flavor separated spin-dependent quark distribution functions at \(0.023 < x < 0.6\) and \(Q^2 > 1\) \(\textrm{GeV}^{2}\) \cite{hermes2005}. The asymmetry of virtual photon-production of the charged high \(p_T\) hadron pairs (\(h_{1} h_{2}\)) with \(p_{T}^{h_{1}}>1.5 \)GeV and \(p_{T}^{h_{2}} >1.0 \) GeV is also measured to access the spin-dependent gluon distributions at the averaged \(x\), \(<x> = 0.17\) and the averaged \(Q^{2}\), \(<Q^2> = 0.06\) \(\textrm{GeV}^{2}\) \cite{hermeshighpt2010}.

Relativistic Heavy Ion Collider (RHIC) is capable to colliding high energy polarized proton beams to access the polarized proton structure. The STAR and PHENIX experiments at RHIC are equipped to serve this purpose. The W boson asymmetry of both experiments allows to measure flavor-separated spin-dependent distribution of \(u\) and \(d\) quarks, especially the \(u\) and \(d\) sea quark distributions. The inclusive jet measurements at STAR which will be discussed intensively in the following sections, and the \(\pi^{0}\) measurements at PHENIX are designed to extract the \(\Delta g\) inside the proton.

\section{The NLO Global Polarized PDF Analysis}

\begin{figure}[H]
\centering
\includegraphics[scale=1.0]{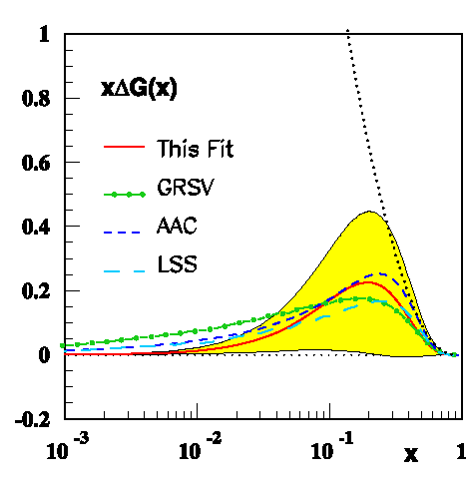}
\nocitecaption{$x\Delta G(x)$ predicted by BB model with other global models at $Q^2 = 4$ $\textrm{GeV}^2$ \cite{BB10}.}
\label{fig:bb2010deltaG}
\end{figure}

One NLO analysis developed by Bl\"{u}mlein and B\"{o}ttcher, the BB model, used inclusive DIS experimental data to study the polarized PDF in the proton \cite{BB10}. The analysis is based on the spin-dependent structure functions \(g_{1}^{(p,n)}(x)\) which are extracted from the longitudinal spin asymmetry in the DIS experiments, for example the EMC proton data, the proton and deuteron data from HERMES, and the proton and deuteron data from COMPASS. The PDFs are parameterized with a common certain functional form for \(\Delta u_{v}\), \(\Delta d_{v}\), \(\Delta \bar{S}\) and \(\Delta g\) at the initial \(Q_0^{2} = 4\) \(\textrm{GeV}^{2}\) with seven free parameters. The free parameters and the QCD scale constant \(\Lambda_{QCD}\) are determined from a fit to the experimental data. The statistical uncertainties from the data are propagated to the calculated PDFs. The systematic uncertainties due to data and theory are evaluated. Its results show that at \(0.005 < x < 0.75\) and \(Q_0^{2} = 4\) \(\textrm{GeV}^{2}\), the quark and anti-quark contribution and gluon contribution to proton spin are \(\Delta \Sigma = 0.193 \pm 0.075\) and \(\Delta G = 0.462 \pm 0.430\), which indicates the inclusive DIS data constrain the quark and anti-quark contribution well but provides loose constraints on gluon contribution. Figure \ref{fig:bb2010deltaG} shows the \(x\Delta G(x)\) compared with other global fits at \(Q^2 = 4\) \(\textrm{GeV}^2\).

\begin{figure}[H]
\centering
\includegraphics[scale=1.0]{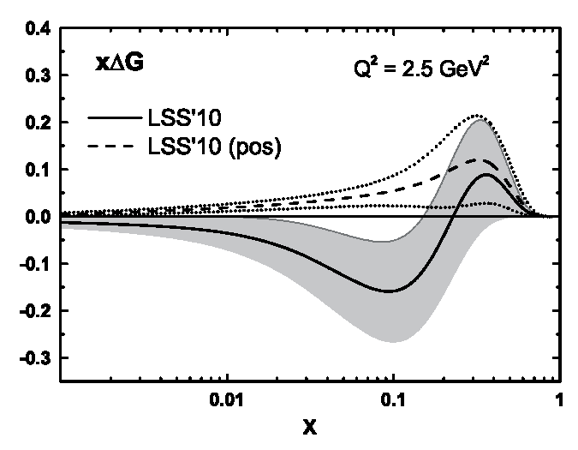}
\nocitecaption{$x\Delta G(x)$ predicted by LSS model with at $Q^2 = 2.5$ $\textrm{GeV}^2$ \cite{LSS10}.}
\label{fig:lss2010deltaG}
\end{figure}

The LSS model named after its authors, Leader, Sidorov and Stamenov, is another global analysis by including polarized DIS experimental data to extract the polarized PDF inside the proton \cite{LSS10}. It takes inclusive DIS and semi-inclusive data as its input, for example data from EMC, HERMES and COMPASS, as well as lower-\(Q^2\) data from Jeffereson Lab. It considers the target mass correction and higher twist effects for the inclusive DIS data. Certain functional forms are assumed for \(\Delta u + \Delta \bar{u}\), \(\Delta d + \Delta \bar{d}\), \(\Delta \bar{u}\), \(\Delta \bar{d}\),\( \Delta \bar{s}\) and \(\Delta g\) at the initial \(Q_0^{2} = 1\) \(\textrm{GeV}^{2}\) to fit its data. In this analysis, the semi-inclusive data play a role in determining the sea quark distribution without addition assumption. Two different shapes are considered for \(\Delta g\), one with a sign-changing node and one that is positive definite. Both fits find \(\Delta \bar{d} < 0\), \(\Delta \bar{u}\)  is positive below \(x \sim 0.2\) and negative above \(x \sim 0.2\) and \(\Delta \bar{s}\)  changing signs over the measured \(x\). The inclusive and semi-inclusive data poorly constrain the gluon distribution \(\Delta g\). The fits to sign-changing and positive \(\Delta g\) give comparable \(\chi^{2}\). At \(Q_0^{2} = 4\) \(\textrm{GeV}^{2}\), it gives \(\Delta \Sigma = 0.254 \pm 0.042\) and \(\Delta G = -0.34 \pm 0.46\) for sign-changing \(\Delta g\) and \(\Delta \Sigma = 0.207 \pm 0.034\) and \(\Delta G = 0.32 \pm 0.19\) for positive \(\Delta g\).  Figure \ref{fig:lss2010deltaG} shows the \(x\Delta G(x)\) at \(Q^2 = 2.5\) \(\textrm{GeV}^2\).

\begin{figure}[H]
\centering
\includegraphics[scale=0.8]{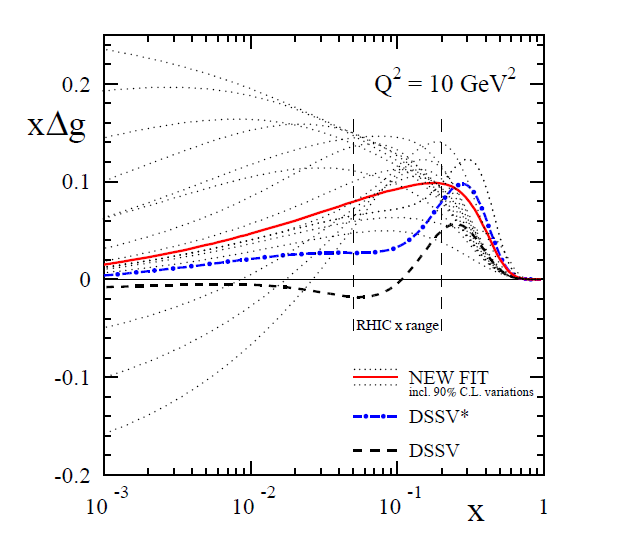}
\nocitecaption{$x\Delta G(x)$ predicted by DSSV with (red) and without (blue) 2009 RHIC data and an earlier version of its fit (dashed) at $Q^2 = 10$ $\textrm{GeV}^2$ \cite{DSSV14}.}
\label{fig:dssvdeltaG}
\end{figure}

The global analysis developed by de Florian, Sassot, Stratmann and Vogelsang, known as the DSSV models, uses not only the inclusive DIS and semi-inclusive DIS data but also hadronic collider data from RHIC to extract the polarized PDFs \cite{DSSV08,DSSV09,DSSV14}. Data from EMC, HERMES and COMPASS are included in their analysis as well as the inclusive \(\pi^{0}\) data from PHENIX and inclusive jet data from STAR at RHIC. The parameterization functions are chosen for \(\Delta u + \Delta \bar{u}\), \(\Delta d + \Delta \bar{d}\), \(\Delta \bar{u}\), \(\Delta \bar{d}\), \(\Delta \bar{s}\) and \(\Delta g\) with 19 free parameters to be determined by the fitting procedure at the initial \(Q_0^{2} = 1\) \(\textrm{GeV}^{2}\). The higher twist effects are ignored in calculating the spin dependent structure function \(g_{1}^{p,n}\), but target mass correction is considered. The uncertainties are calculated by the standard Hessian method and Lagrange multiplier method, and both methods give consistent results. The results show that the light sea quark polarization \(\Delta \bar{u} > 0\), \(\Delta \bar{d} < 0\) and \(\Delta \bar{u} - \Delta \bar{d} > 0\). The strange sea quark distribution changes signs from positive to negative as \(x\) approach below 0.02, which implies a large negative strange sea quark contribution to the proton. The \(\Delta \Sigma\) is about 0.37 by allowing the SU(3) flavor asymmetry to be broken at \(Q^{2} = 10\) \(\textrm{GeV}^{2}\). The earlier analysis study without including the recent 2009 STAR 200 GeV inclusive jet data show very small gluon polarization in the accessed \(x\) range at the same \(Q^{2}\). However the newest release of DSSV model with the 2009 data included showed the truncated \(\Delta G\) from 0.05 to 1 to be about three \(\sigma\) above zero.  Figure \ref{fig:dssvdeltaG} shows the \(x\Delta G(x)\) for the current DSSV model at \(Q^2 = 10\) \(\textrm{GeV}^2\) with and without RHIC data compared with an earlier version of DSSV .The \(\Delta G\) below 0.05 is loosely constrained by the current data, however. The higher center mass energy, 510 GeV, data from RHIC will provide constraints at smaller \(x\).

\begin{figure}[H]
\centering
\includegraphics[scale=1.0]{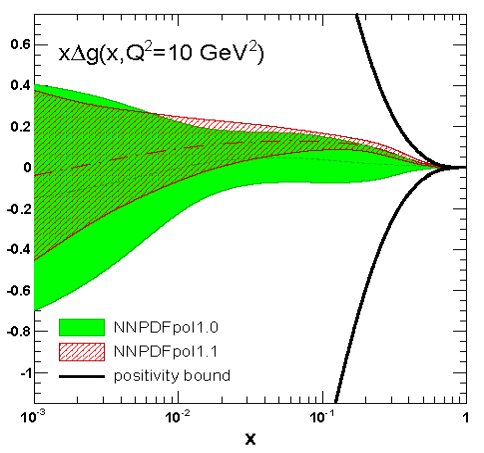}
\nocitecaption{$x\Delta G(x)$ predicted by NNPDF1.1 and NNPDF1.0 at $Q^2 = 10$ $\textrm{GeV}^2$ \cite{NNPDFpol}.}
\label{fig:nnpdfdeltaG}
\end{figure}

Like the un-polarized PDFs, the NNPDF group also provides their polarized PDFs by using the same techniques \cite{NNPDFpol}. In their earlier version, NNPDF1.0 the inclusive DIS data were only included so it could not separate the parton distributions between quarks and anti-quarks \cite{NNPDFpol10}. The \(u\) and \(d\) quark and anti-quark distributions they obtained, \(\Delta u + \Delta \bar{u}\), and \(\Delta d + \Delta \bar{d}\), agree well with DSSV and BB model, but the strange quark and anti-quark distributions \(\Delta s + \Delta \bar{s}\) have larger uncertainties. The gluon distributions have larger uncertainties at small \(x\) compared to the other models. Their latest version, NNPDF1.1 however includes the W boson asymmetry data from RHIC, which allows the quark and anti-quark separation, the inclusive jet measurements from RHIC and the open-charm data from COMPASS, both of which help to constrain gluon distributions. The extracted quark and anti-quark distributions between the two versions are rather similar, at \(Q^{2} = 10\) \(\textrm{GeV}^{2}\) and \(0.001 < x < 1\), the \(\Delta \Sigma = 0.23 \pm 0.15\) and \(\Delta \Sigma = 0.25 \pm 0.10\) for the earlier and later version respectively. The major highlight of the latest version is the constraints placed on \(\Delta G\) when including the inclusive jet data from RHIC, the truncated \(\Delta G\) where \(0.05 < x < 0.2\) at \(Q^{2} = 10\) \(\textrm{GeV}^{2}\) improved from \(0.05 \pm 0.15\) to \(0.17 \pm 0.06\). This also suggests the positive gluon polarization at the accessed \(x\) range, which is consistent with what the recent DSSV model finds.  Figure \ref{fig:nnpdfdeltaG} shows the \(x\Delta G(x)\) for the current NNPDF1.1 and the old NNPDF1.0 at \(Q^2 = 10\) \(\textrm{GeV}^2\).

Another comprehensive global QCD analysis of spin-dependent parton distributions is developed by the Jefferson Lab Angular Momentum Collaboration (JAM) \cite{JAM}. The analysis uses the latest high-precision DIS data collected from Jefferson Lab (JLab) and others data from EMC, HERMES, COMPASS etc. A generic parametrization for the \(\Delta u + \Delta \bar{u}\), \(\Delta d + \Delta \bar{d}\), \(\Delta s + \Delta \bar{s}\), \(\Delta g\), flavor-separated twist-3 distributions, and \(d_2\) moment of the nucleon is assumed at the initial input scale \(Q_0^2\). An iterative Monte Carlo fitting technique is applied to extract the fitting parameters. The JAM PDF describes the global inclusive DIS data very well overall. It also constrains the quark and anti-quark distributions well, which yields \(\Delta \Sigma = 0.28 \pm 0.04\) at \(Q^2 = 1\) \(\text{GeV}^2\) over the extrapolated full \(x\) range. Like other DIS fits, the JAM PDF found it difficult to constrain gluon polarizations, however it suggests a positive \(\Delta g\) with a small spread over \(x \approx\) 0.1 to 0.5, as supported by the JLab data. Though JLab data plays an important role in reducing the uncertainty band for the polarized quark and anti-quark distributions and higher twist contributions, a call for polarized \(pp\) data from RHIC to constrain the gluon polarization is pointed out.

%% file: chapter2.tex
%
%
%


\chapter{\uppercase {Inclusive Jet Measurements at Hadronic Collider}}

\section{Inclusive Jet and Its Asymmetry}

\begin{figure}[H]
\centering
\includegraphics[scale=0.6]{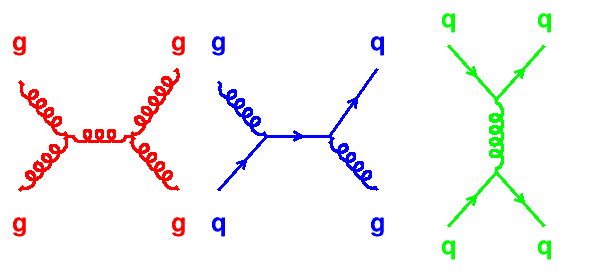}
\caption{Feynman diagram for the $gg$ (red), $qg$ (blue) and $qq$ (green) sub processes.}
\label{fig:subfeynman}
\end{figure}

%
In addition to the inclusive jet study in DIS experiments via the quark-gluon fusion process, inclusive jet measurements in the hadronic collider are another effective way to study the internal structure of the proton, especially at wider kinematic range. In the proton-proton (\(pp\)) or proton-anti-proton (\(p\bar{p}\)) collisions, the inclusive production process can be denoted as \(pp(\bar{p}) \rightarrow jet+X\), where the \(X\) can be any hadronic product. The jets are contributed by the \(2 \rightarrow 2\) hard scattering, such as quark-quark (anti-quark), \(qq(\bar{q}\)), quark-gluon, \(qg\), and gluon-gluon, \(gg\) scattering shown in Figure \ref{fig:subfeynman}.

For un-polarized collisions, the inclusive jet cross-section is measured to extract un-polarized PDFs in the proton. The cross-section of inclusive jet can be expressed as,
\begin{align}
\frac{d\sigma}{dp_{T}} = \sum_{a,b} \int dx_{a} dx_{b} \times f_{a}(x_{a})f_{b}(x_{b}) \frac{d\sigma^{a+b \rightarrow jet+X}}{dp_{T}} \nonumber
\end{align}
where \(f_{a(b)}(x_{a(b)})\) is the PDF of parton \(a\)(\(b\)) and \(\frac{d\sigma^{a+b\rightarrow jet+X}}{dp_{T}}\) is the partonic scattering cross-section for partonic process \(a+b \rightarrow jet+X\). Recent inclusive jet cross-section measurements from the CDF and D0 experiments at the Tevatron play an important role in determining the gluon distribution function in the proton by the NLO global analysis \cite{cdfjets2008, d0jets2012}.

For polarized collisions, the inclusive jet longitudinal double-spin asymmetry \(A_{LL}\) is measured to access the polarized PDFs in the proton. \(A_{LL}\) is defined by
\begin{align}
A_{LL} = \frac{\sigma^{++}-\sigma^{+-}}{\sigma^{++}+\sigma^{+-}}, \label{eq:aLLtheory}
\end{align}
where \(\sigma^{++(-)}\) is the inclusive jet cross-section when two beams have the same (opposite) helicity. The numerator can be written as the integral of the differential cross-section by jet \(p_{T}\), which is \(d\Delta \sigma/(dp_T )=\frac{1}{2}(\frac{d\sigma^{++}}{dp_{T}} - \frac{d\sigma^{+-}}{dp_{T}})\). The \(\frac{d\sigma^{++}}{dp_{T}}\) can be written as,

\begin{align}
\frac{d\sigma^{++}}{dp_{T}} &= \sum_{ab} \int dx_{a} dx_{b} \times \{ [ f_{a}^{+}(x_{a})f_{b}^{+}(x_{b})+f_{a}^{-}(x_{a}) f_{b}^{-}(x_{b}) ] \times \frac{d\hat{\sigma}_{ab \rightarrow jet+X}^{++}}{dp_{T}} \nonumber \\
&\qquad  + [f_{a}^{+}(x_{a})f_{b}^{-} (x_{b})+f_{a}^{-}(x_{a}) f_{b}^{+}(x_{b})] \times \frac{d\hat{\sigma}_{ab \rightarrow jet+X}^{+-}}{dp_{T}} \}, 
\end{align}

\noindent where \(d\hat{\sigma}_{ab \rightarrow jet+X}^{++(-)}{dp_{T}}\) is the two parton scattering cross-section with the scattering partons having parallel (anti-parallel) spin directions. Likewise for \(\frac{d\sigma^{+-}}{dp_{T}}\), there is
 
\begin{align}
\frac{d\sigma^{+-}}{dp_{T}} &=\sum_{ab} \int dx_{a}dx_{b} \times \{ [f_{a}^{+}(x_{a})f_{b}^{-}(x_{b})+f_{a}^{-}(x_{a})f_{b}^{+}(x_{b})] \times \frac{d\hat{\sigma}_{ab \rightarrow jet+X}^{++}}{dp_{T}} \nonumber \\
&\qquad + [f_{a}^{+}(x_{a})f_{b}^{+}(x_{b})+f_{a}^{-}(x_{a})f_{b}^{-}(x_{b})] \times \frac{d\hat{\sigma}_{ab \rightarrow jet+X}^{+-}}{dp_{T}} \}.
\end{align}

Then \(d\frac{\Delta \sigma}{dp_{T}} \) can be expressed as the following

\begin{align}
\frac{d \Delta \sigma}{dp_{T}} &= \frac{1}{2} \times \sum_{ab} \int dx_{a}dx_{b} \times \frac{1}{2} \times \sum_{ab} \int dx_{a} dx_{b} \times \{ \nonumber \\
&\qquad [ f_{a}^{+}(x_{a}) - f_{a}^{-}(x_{a}) ] [f_{b}^{+}(x_{b}) - f_{b}^{-}(x_{b}) ] \times \frac{d\hat{\sigma}_{ab\rightarrow jet+X}^{++}}{dp_{T}} \nonumber \\
&\qquad - [f_{a}^{+}(x_{a})-f_{a}^{-}(x_{a})] [f_{b}^{+}(x_{b}) - f_{b}^{-} (x_{b})] \times \frac{d\hat{\sigma}_{ab \rightarrow jet+X}^{+-}}{dp_{T }} \} \nonumber \\
&=\frac{1}{2} \times \sum_{ab} \int dx_{a}dx_{b} \times \{ \nonumber \\
&\qquad [ f_{a}^{+}(x_{a}) - f_{a}^{-} (x_{a})] [f_{b}^{+}(x_{b}) - f_{b}^{-} (x_{b})] \nonumber \\
&\qquad \times [ \frac{d\hat{\sigma}_{ab \rightarrow jet+X}^{++}}{dp_{T}} - \frac{d\hat{\sigma} _{ab \rightarrow jet+X}^{+-}}{dp_{T}} ] \} \nonumber \\
&= \frac{1}{2} \times \sum_{ab} \int dx_{a} dx_{b} \times \Delta f_{a}(x_{a})\Delta f_{b}(x_{b}) \frac{d \Delta \hat{\sigma}_{ ab \rightarrow jet+X}}{dp_{T}},
\end{align}

\noindent where \(\Delta f_{a(b)}(x)= f_{a(b)}^{+}(x) - f_{a(b)}^{-}(x)\) is the polarized parton distribution function for parton \(a\) (\(b\)) and \( \frac{d \Delta \hat{\sigma}^{ab \rightarrow jet+X}}{dp_{T}} = \frac{d\hat{\sigma}_{ab \rightarrow jet+X}^{++}}{dp_{T} } - \frac{d\hat{\sigma}_{ab \rightarrow jet+X}^{+-}}{dp_{T}} \) is the spin dependent two parton scattering cross-section. Therefore \(A_{LL}\) is sensitive to the polarized PDFs in the proton.

\section{Jet Finding Algorithm}
Jets are clusters of final particles after hadron collisions. Jets can be defined at the parton level as well if the combinations are made on the partons produced after the scattering. The jet cross-section calculations depend on the algorithm used to find jets. The algorithm needs to be chosen carefully to avoid divergence in the cross-section calculations. The jet finding algorithm is also necessary to find jets from the detector response collected during experiments. Several jet finding algorithms have been developed during the last two decades, and they can be categorized into two types, cone algorithm and \(k_{T}\) algorithm.
The cone algorithm is based on finding stable cones that encapsulate particles within certain area around their centroid. The centroid of a cone which has N particles is defined by,
\begin{equation}
\eta^{c} = \frac{\sum_{i}E_{T}^{i} \eta^{i}}{E_{T}^{c}},
\phi^{c}= \frac{\sum_{i}E_{T}^{i} \phi^{i}}{E_{T}^{c}},
E_{T}^{c} = \sum_{i}E_{T}^{i},
\end{equation}
where \(\eta^{i}\), \(\phi^{i}\), and \(E_{T}^{i}\) are the pseudo-rapidity, azimuthal angle and transverse energy of the \(i\)-th particle of the N particles. There are several versions of cone algorithms trying to find stable cone centroids \cite{conesalgo}. One variant of these algorithm is addition of midpoints in the starting seed list. The initial seed list is constructed from individual measured particles such as calorimeter towers with a minimal energy cut. Then the list is expanded by adding mid-points from all the possible combinations of each initial seed for example \(p_{i} + p_{j}\), \(p_{i} + p_{k}\), \(p_{j} + p_{k}\), \(p_{i} + p_{j} + p_{k}\), etc. where \(p_{i,j,k}\) is the momentum of particles deposited in tower \(i\),\(j\), and \(k\) converted by its \(E_T\). The algorithm starts with the points in the list as the centroid one by one and tries to compare the particles falling inside the cone radius \(R\) and the particles that construct the point. If they agree, then a candidate jet is found. If they don't, the point is discarded from the list and the algorithm continues to the next point. The process is iterated until the list exhausts.

A splitting and merging procedure is applied to the candidate jets found in the above steps. The candidate jets are sorted from the highest to the lowest by their \(E_{T}\). A nominal fraction of the shared \(E_{T}\) by the neighboring jet \(f\) is assumed. From the highest \(E_{T}\) jet candidate, if the fraction of the shared energy with another candidate is greater than the nominal value, the two jets will be merged, otherwise the sharing towers will be split to the two candidates based on their distance to each candidate. If a candidate shares energy with more than one neighboring candidates, choose the highest \(E_{T}\) neighbor. The merged or split jets re-enter the candidate list and the list is sorted by \(E_{T}\) again. The above procedure is repeated until no jet shares energy with the others in the list.

The \(k_{T}\) algorithm tries to find jets on a list of pre-clusters which could be particles or partons \cite{kt1993, ktellis1993}. For each pre-cluster in the list, the energy \(E\) and momentum \(\overrightarrow p\) are known. First define the distance,
\begin{align}
d_{i} = p_{T,i}^{2}
\end{align}
and
\begin{align}
d_{ij} &= min(p_{T,i}^{2}, p_{T,j}^{2}) \times \frac{\Delta R_{ij}^{2}}{R^{2}} \nonumber \\
&= min(p_{T,i}^{2},p_{T,j}^{2}) \times \frac{(y_{i}-y_{j})^{2}+ (\phi_{i}-\phi_{j})^{2}}{R^{2}},
\end{align}
where \(p_{T, i(j)}\),\(y_{i(j)}\), and \(\phi_{i(j)}\) are transverse momentum, rapidity and azimuthal angle of the \(i\)-th and \(j\)-th pre-cluster and \(R\) is the jet parameter. Then the algorithm calculates all the \(d_{i}\) and \(d_{ij}\), then finds the minimum \(d_{min}\) of all the \(d_{i}\) and \(d_{ij}\). If \(d_{min}\) is one of the \(d_{ij}\), combine the \(i\)-th and \(j\)-th pre-cluster together by \(E_{ij}= E_{i}+E_{j}\) and \(\overrightarrow{p_{ij}} = \overrightarrow{p_{i}} + \overrightarrow{p_{j}}\), replace them with the combined pre-cluster with \(E_{ij}\) and \(\overrightarrow{p_{ij}}\) and re-calculate the \(d_{i}\) and \(d_{ij}\) for the new list. Otherwise, if \(d_{min}\) is one of the \(d_{i}\), remove the \(i\)-th pre-cluster from the list as a jet found. The process continues until the pre-cluster list is empty.

There are two additional variants of the \(k_{T}\) algorithm depending on the definition of the \(d_{i}\) and \(d_{ij}\). The Cambridge/Aachen algorithm defines \(d_{i} = 1\) and \(d_{ij} = \frac{\Delta R_{ij}^{2}}{R^{2}}\) \cite{cajet1997, cajet1999}. The anti-\(k_{T}\) algorithm defines \(d_{i}=\frac{1}{p_{T,i}^{2}}\) and \(d_{ij} = min (\frac{1}{p_{T,i}^{2}},\frac{1}{p_{T,j}^{2}}) \times \frac{\Delta R_{ij}^{2}}{R^{2}}\). When the jets are clustered by some hard particles coming from the hard scattering and some soft particles not coming from the hard scattering, the anti-\( k_{T}\) algorithm is less susceptible to the diffusion of soft radiation and underlying events because those events tend have smaller \(p_{T}\). All the three \(k_{T}\) type algorithms yield the same inclusive jet cross-sections in NLO pQCD calculations.

\section{Inclusive Jet Measurements at STAR}
\begin{figure}[H]
\centering
\includegraphics[scale=0.6]{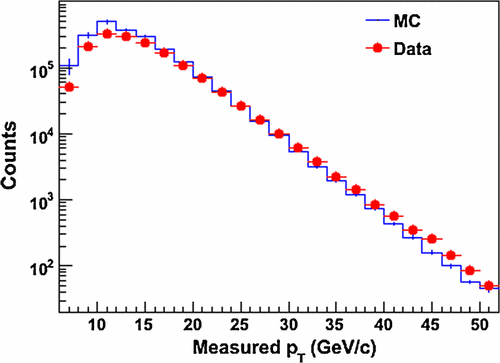}
\nocitecaption{Comparison of jet yields vs. jet $p_T$ from $pp$ collisions at $\sqrt{s} = 200$ GeV \cite{run6aLL2012}.}
\label{fig:run6jetyields}
\end{figure}

\begin{figure}[H]
\centering
\includegraphics[scale=0.6]{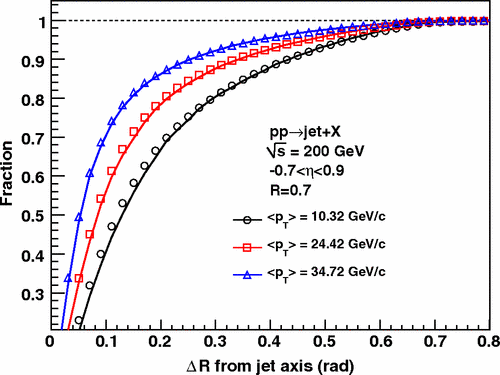}
\nocitecaption{Comparison of jet transverse energy fraction within a cone radius of $\Delta R$ centered on the reconstructed jet thrust axis from $pp$ collisions at $\sqrt{s} = 200$ GeV \cite{run6aLL2012}.}
\label{fig:run6jetprofile}
\end{figure}

At RHIC, with the capabilities of its detectors STAR has measured inclusive jet production from the longitudinally polarized \(pp\) collisions at the center of mass energy \(\sqrt{s} =\) 200 GeV and 500 GeV since the 2003 RHIC run. Previous inclusive jet studies demonstrate the jet reconstruction is well understood at RHIC kinematics \cite{run3run4res2006, run6aLL2008, run6aLL2012}. For example the comparison between data and simulation agree well for the jet yields vs. jet \(p_T\) as in Figure \ref{fig:run6jetyields} and the transverse energy fraction within a cone radius of \(\Delta R\) centered on the reconstructed jet thrust axis as in Figure \ref{fig:run6jetprofile}.

\begin{figure}[H]
\centering
\includegraphics[scale=1.0]{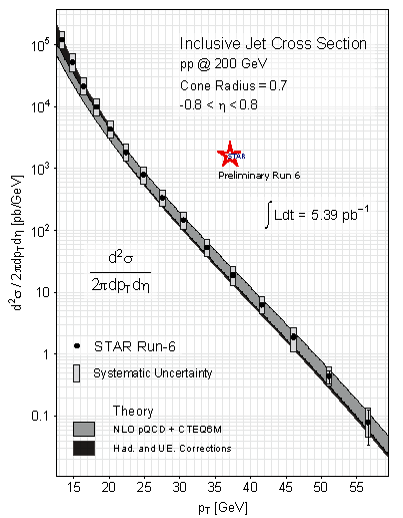}
\nocitecaption{STAR 2006 inclusive jet cross-section from $pp$ collisions at $\sqrt{s} = 200$ GeV \cite{sakumarun6}.}
\label{fig:run6crs}
\end{figure}

\begin{figure}[H]
\centering
\includegraphics[scale=1.0]{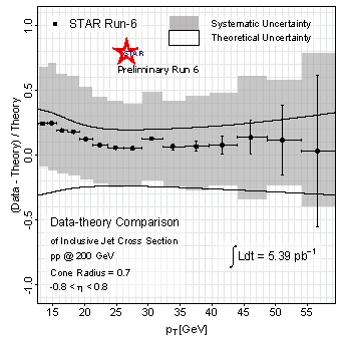}
\nocitecaption{STAR 2006 inclusive jet cross-section from $pp$ collisions at $\sqrt{s} = 200 GeV$ compared with theory \cite{sakumarun6}.}
\label{fig:run6crsthe}
\end{figure}

The recent inclusive jet cross-section measurement at mid-pseudo-rapidity, \(|\eta| < 1\), from the 2006 STAR \(pp\) collisions at \( \sqrt{s} =\) 200 GeV is shown in Figure \ref{fig:run6crs} \cite{sakumarun6}. The difference between the measured cross-section and the theoretical prediction is well within the systematic uncertainty, as seen in Figure \ref{fig:run6crsthe}. The analysis uses the CDF mid-point cone algorithm with seed energy 0.5 GeV and merge/split fraction 0.5. The inclusive jet cross-sections agree well with the NLO theoretical calculations after hadronization and underlying event corrections.

\begin{figure}[H]
\centering
\includegraphics[scale=0.6]{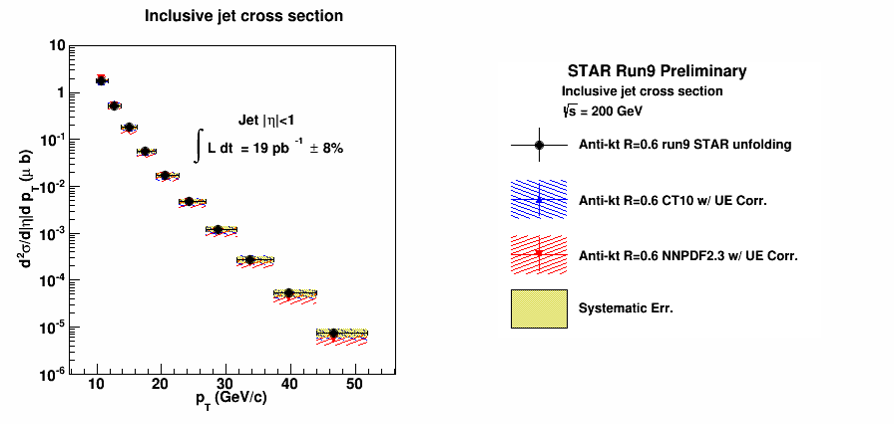}
\nocitecaption{STAR 2009 inclusive jet cross-section from $pp$ collisions at $\sqrt{s} = 200$ GeV \cite{run9crsdis}.}
\label{fig:run9crs}
\end{figure}

The inclusive jet cross-section at the mid pseudo-rapidity \(|\eta| < 1\) is also measured from the 2009 STAR \(pp\) collisions at \(\sqrt{s} = 200\) GeV with a larger dataset than the year 2006. The anti-\(k_T\) algorithm with jet parameter \(R\) = 0.6 is used for the jet reconstruction. The results agree well with the NLO theoretical calculations after hadronization and underlying event corrections as shown in Figure \ref{fig:run9crs} \cite{run9crsdis}. The inclusive jet cross-section is also divided into two sub pseudo-rapidity ranges \(|\eta| < 0.5\) and \(0.5 < |\eta| < 1.0\) to provide reference for the double spin asymmetry analysis with the same dataset.

Di-jet analysis is also performed at STAR to determine the gluon density inside the proton. Di-jet cross-sections at the mid pseudo-rapidity \(|\eta| < 0.8\) from the STAR 2009 data at \(\sqrt{s} = 200\) and \(500\) GeV are measured by using the anti-\(k_T\) algorithm with jet parameter \(R = 0.6\) \cite{run9dijet500}. The di-jet cross-sections show the excellent agreement with the NLO calculations after hadronization and underlying event corrections.

\begin{figure}[H]
\centering
\includegraphics[scale=1.0]{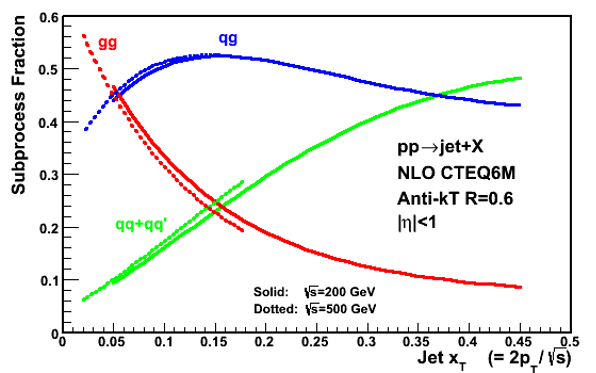}
\nocitecaption{Inclusive jet cross-section fractions due to subprocesses $gg$, $qg$, and $qq$ from NLO calculations at $\sqrt{s}$ = 200 and 500 GeV \cite{nlojet2012}.}
\label{fig:subpro}
\end{figure}

The inclusive jet \(A_{LL}\) measurements are one of the highlights of the STAR spin physics program. As discussed in Section 2.1, the inclusive jet production is contributed by three partonic scattering processes, \(qq\), \(qg\), and \(gg\). Figure \ref{fig:subpro} shows the fraction of inclusive jet production at \(\sqrt{s} =\) 200 GeV and 500 GeV due to individual processes over the jet \(x_{T} = \frac{2p_T}{\sqrt{s}}\) between 0.02 and 0.5 for jets in the jet mid-pseudo-rapidity range \(|\eta| < 1\). The jet cross-sections are calculated at the NLO by using the anti-\(k_{T}\) algorithm with jet parameter \(R =\) 0.6, using the code from \cite{nlojet2012}. At the low \(x_{T}\) region, the \(qg\) and \(gg\) dominate the jet production. The \(gg\) contribution drops down significantly at \(x_{T}\) around 0.15, in contrast the \(qq\) contribution grows steadily as \(x_{T}\) increases. However at the point where \(x_{T}\) near 0.3 the contributions from \(qg\) and \(qq\) are equal, the total jet cross-section has dropped four orders of magnitude relative to that at low \(x_{T}\) around 0.1. The partonic longitudinal double spin asymmetry \(\hat{a}_{LL}\) is relatively large for \(gg\) and \(qg\) processes over the corresponding kinematics. Therefore the inclusive jet \(A_{LL}\) is sensitive to the polarized gluon distributions \(\Delta g\) in the proton.

\begin{figure}[H]
\centering
\includegraphics[scale=0.6]{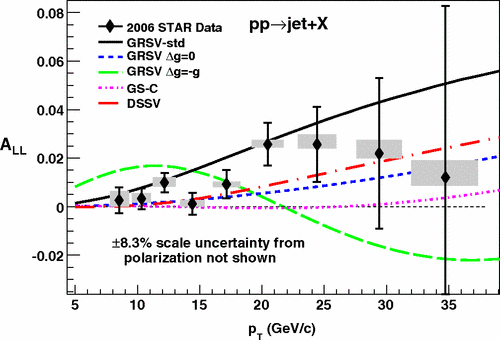}
\nocitecaption{STAR 2006 inclusive jet $A_{LL}$ from longitudinally polarized $pp$ collisions at $\sqrt{s} = 200$ GeV \cite{run6aLL2012}.}
\label{fig:run6aLL}
\end{figure}

The STAR inclusive jet \(A_{LL}\) from the 2006 RHIC longitudinally polarized \(pp\) collisions at \(\sqrt{s} = 200\) GeV is shown in Figure \ref{fig:run6aLL} \cite{run6aLL2012}. The same jet algorithm is used for the \(A_{LL}\) analysis as the cross-section study. The sampled gluon \(x_{gluon}\) by the \(A_{LL}\) is down to as low as 0.05 in this analysis. The early DSSV model shows relatively small gluon polarization in the covered \(x\) range \(0.05 < x < 0.2\) with the 2006 data included but with a large uncertainty. The results exclude several theoretical models that predict large gluon polarization.

In the year 2009, STAR collected a large data sample of 200 GeV longitudinally polarized \(pp\) data during the RHIC run. The event statistics used in the inclusive jet \(A_{LL}\) analysis was about 20 times larger than the 2006 analysis. This arose from increases in the trigger rates enabled by improvement to the data acquisition system, combined with increases in the trigger acceptance and efficiency. The trigger improvements also led to reduced trigger bias. The analysis uses the anti-\(k_{T}\) algorithm with jet parameter 0.6, instead of the CDF mid-point cone algorithm with cone radius 0.7. A change in the way the jet reconstruction corrected for hadronic energy deposits in the electro-magnetic calorimeter improved the jet momentum resolution from 23\% to 18\%. 

\begin{figure}[H]
\centering
\includegraphics[scale=0.6]{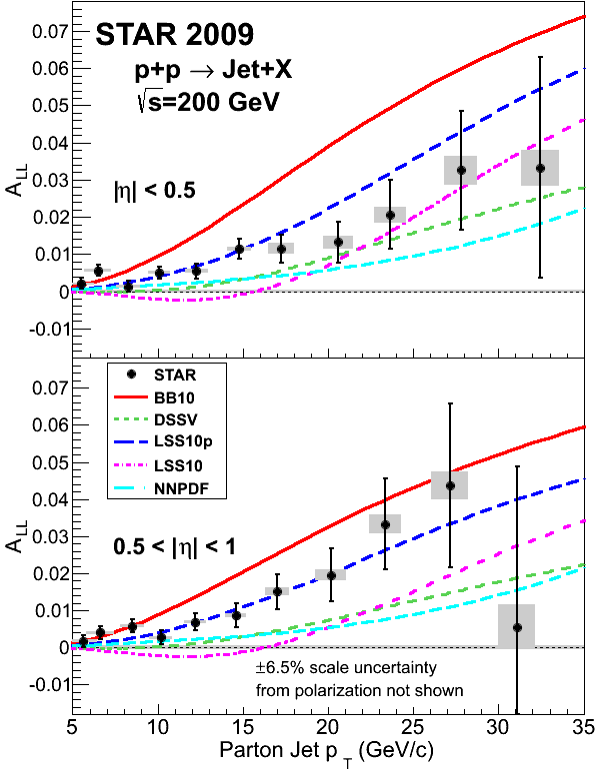}
\nocitecaption{STAR 2009 inclusive jet longitudinal $A_{LL}$ at $\sqrt{s} = 200$ GeV \cite{run9aLL2015}.}
\label{fig:run9aLL}
\end{figure}

The results of the STAR 2009 inclusive jet \(A_{LL}\) is shown in Figure \ref{fig:run9aLL} \cite{run9aLL2015}. The \(A_{LL}\) is divided into two sub-\(\eta\) ranges,\( |\eta| < 0.5\) and \(0.5 < |\eta| < 1.0\), since the theory predicts about 20\% difference in \(A_{LL}\) at the same jet \(p_{T}\) in those two ranges. The 2009 results are a factor of at least four more precise than the 2006 results at low jet \(p_{T}\) and a factor of three at high jet \(p_{T}\). The measured \(A_{LL}\) fall among the recent model predictions \cite{BB10, LSS10, DSSV08, DSSV09, NNPDFpol10}. Noticeably the \(A_{LL}\) is sitting above the DSSV prediction, which includes the STAR 2006 inclusive jet \(A_{LL}\) data, but well within its quoted uncertainty. It is easy to image that the more precise STAR 2009 results will push the DSSV prediction up. Fortunately the newly released DSSV model includes them in their new fit and gives \(\Delta G = 0.19_{-0.05}^{+0.06}\) for \(x > 0.05\) at 90\% confidence limit \cite{DSSV14}. The NNPDF group also find \(\Delta G = 0.23 \pm 0.07\) for \(0.05 < x < 0.5\) and the uncertainty band on \(x \Delta g(x)\) shrinks when including the STAR 2009 results in their analysis \cite{NNPDFpol}.

STAR was scheduled to take longitudinally polarized \(pp\) collision at \(\sqrt{s} =\) 510 GeV during the 2012 RHIC run and has fulfilled its expectation. The inclusive jet \(A_{LL}\) measurements will allow to access the polarized gluon distribution at lower sampled \(x\) gluon. The details of this analysis will be discussed in the following sections.

%% file: chapter3.tex
%
%
%

\chapter{\uppercase{RHIC AND STAR DETECTORS}}


\section{RHIC Facility}
Relativistic Heavy Ion Collider (RHIC) is a world leading facility that has the capability to collide a wide range of ions for example, uranium (\(U\)), gold (\(Au\)), copper (\(Cu\)), helium (\(^3He\)), aluminum (\(Al\)), deuteron (\(d\)) and proton (\(p\)) at high center of mass energy \(\sqrt{s}\). More impressively, it is the only facility that can collide polarized proton beams up to \(\sqrt{s} = \)510 GeV at the present time. It was built inside a 2.4 mile circumference underground tunnel on the site of Brookhaven National Laboratory (BNL). Two beams circulate in opposite directions and are brought to collide at certain intersection points. Detectors are built at each intersection point to detect particles produced in the collisions. The following Figure \ref{fig:rhic} shows the layout of the RHIC facility \cite{rhicproj,rhicdesign}.

\begin{figure}[H]
\centering
\includegraphics[scale=0.36]{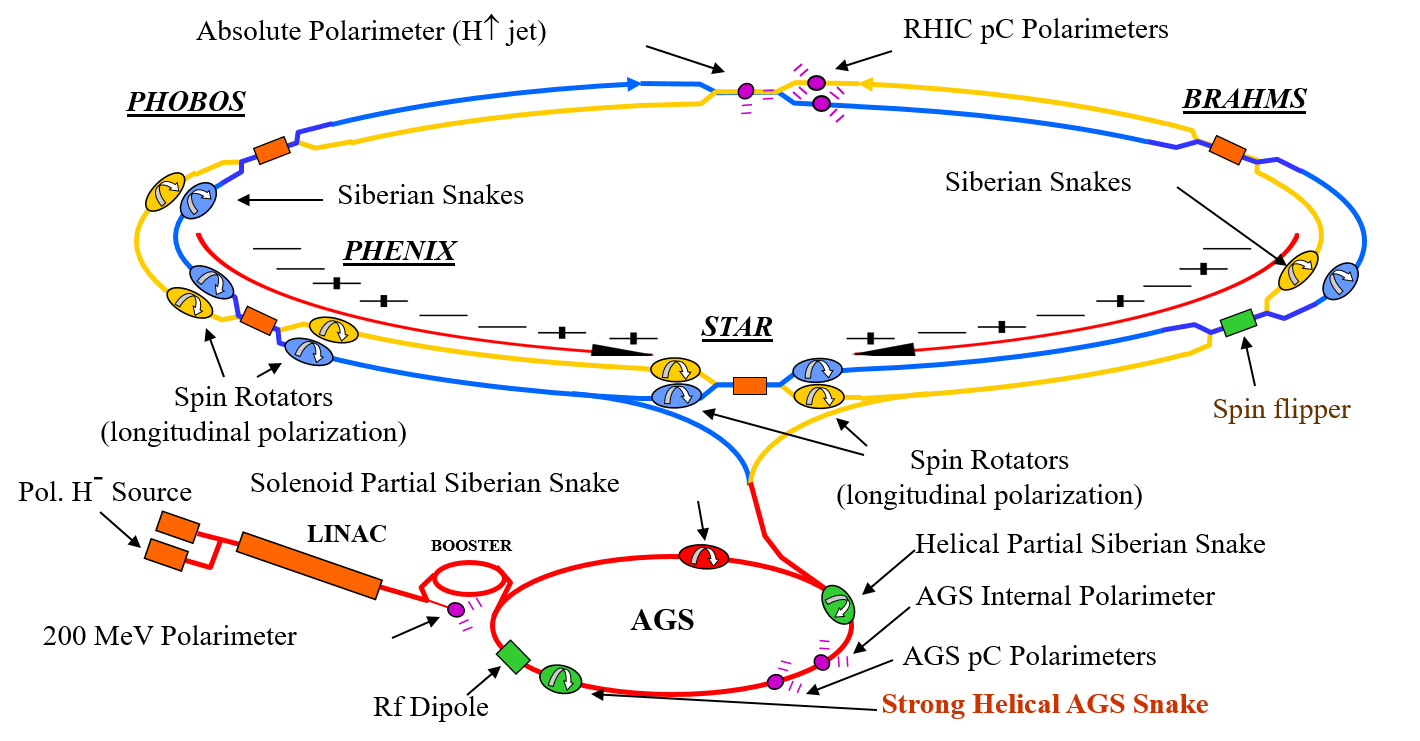}
\nocitecaption{The layout of RHIC facility for plarized proton operation \cite{rhicproj, rhicdesign}.}
\label{fig:rhic}
\end{figure}

The optically pumped polarized ion source (OPPIS), produces a 500 \(\mu\)A \(H^-\) ion current in a single 300 \(\mu\)s pulse with 80\% polarization \cite{rhicpolarize}. The polarized \(H^{-}\) ion pulse is accelerated to 200 MeV with the LINAC, and then strip-injected and captured in a single bunch in the Alternating Gradient Synchrotron (AGS) Booster. The bunch in the Booster contains about \(4 \times 10^{11}\) polarized protons with a normalized 95\% beam emittance about \(10\pi\) \(\mu m\). The bunch of polarized protons is accelerated to 1.5 GeV in the Booster and then transferred to the AGS.

The polarized proton bunch in the AGS is accelerated to 25 GeV. A partial Siberian Snake and RF dipole are used to keep the proton bunch from depolarizing. When the proton bunch reaches the desired energy, it is injected to RHIC through the AGS-to-RHIC transfer line with better than 50\% overall efficiency of the acceleration and beam transfer. There are two rings in RHIC allowing proton beams circulating in the opposite directions, clock-wise and counter-clock-wise, known as the blue and yellow beams respectively. 120 bunches of each ring are repeatedly filled. Since each bunch is accelerated independently, the polarization direction of each bunch can be optional. Both rings are then accelerated to the full energy requested by the physics goal. It takes about 10 minutes together to fill both rings.

Two major detectors are built at the intersection points at 6 o'clock and 8 o'clock, named STAR and PHENIX experiments. A pair of Siberian Snakes located near the 3 and 9 o'clock of each ring keep the beams from depolarizing. Pairs of spin rotators are installed at both ends of the two experiments for each ring. One rotator rotates the proton spin direction from the vertical to the horizontal, and the other rotates it back to the vertical. The spin rotators grant flexibility to both experiments to collide polarized proton beams with transverse or longitudinal polarization at their choice.

\begin{figure}[H]
\centering
\includegraphics[scale=0.8]{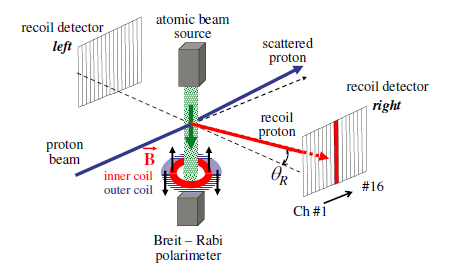}
\nocitecaption{$H$-jet polarimeter layout \cite{hjetpol}.}
\label{fig:hjet}
\end{figure}

The beam polarization is measured by the RHIC \(pC\) and Hydrogen jet (\(H\)-jet) polarimeters located at the 12 o’clock intersection point for both rings. The \(H\)-jet polarimeter \cite{hjetpol} is composed of a polarized atomic beam source, two recoil detectors parallel to the beam and a Breit-Rabi polarimeter as shown in Figure \ref{fig:hjet}. The recoil detector is an array of silicon detectors. The \(p-C\) detector \cite{pCpol} consists of an ultra-thin carbon ribbon target and six silicon detectors located at \SI{90}{\degree} to the beam direction. The \(pC\) polarimeter is cheap to maintain and can provide fast measurements at full luminosity to allow bunch by bunch measurements. When the transversely polarized proton beam hits polarimeter targets, both polarimeters measure recoiled targets through elastic scattering. The elastic scattering is dominated by the Coulomb-Nuclear Interference (CNI) between the polarized beam and the target at this RHIC kinematics.

\section{STAR Detectors}

\begin{figure}[H]
\centering
\includegraphics[scale=0.6]{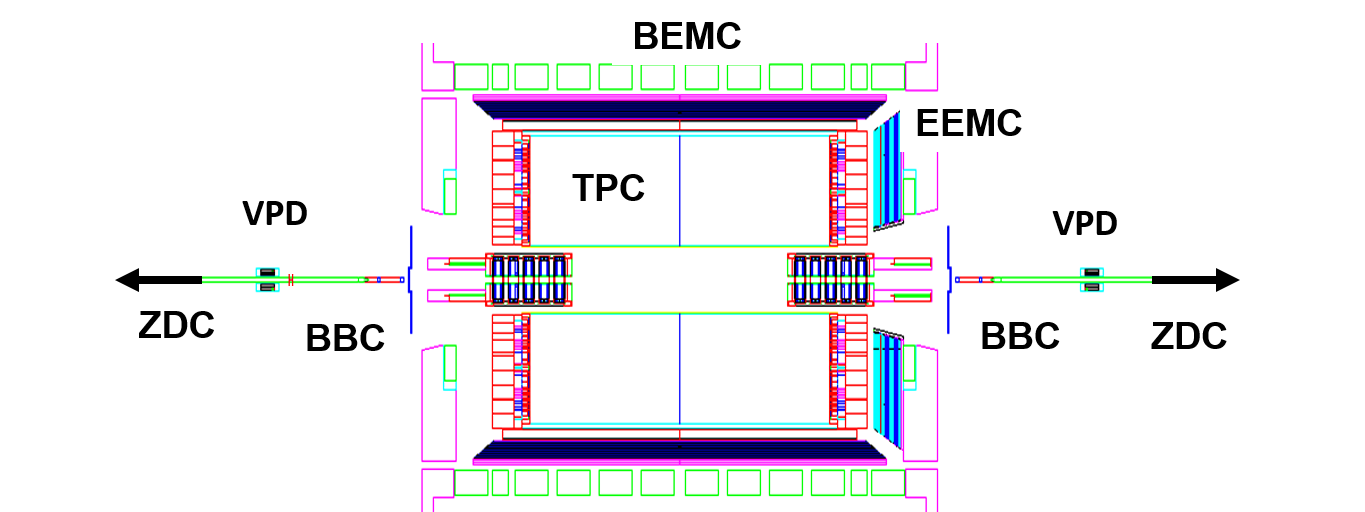}
\caption{Cross sectional view of STAR detectors.}
\label{fig:star}
\end{figure}

The Solenoidal Tracker at RHIC (STAR) is a large detector system built at the 6 o'clock intersection point of the two rings \cite{rhicstar}. The detectors at STAR are designed to better understand the fundamental structure of hadronic interactions. Figure \ref{fig:star} shows the STAR detector system. The STAR magnet can be operated at full field of 0.5 T and half field of 0.25 T to provide tracking ability for charged particles. The Time Projection Chamber (TPC) is the main part of the system to measure charged particle tracks after collisions. The Barrel and Endcap Electro-magnetic Calorimeter (BEMC and EEMC) allow to measure hadronic and photonic energy deposition in the calorimeter towers. The Beam-Beam Counter (BBC), Vertex Position Detector (VPD) and Zero-degree Calorimeter (ZDC) are used to monitor collision luminosity and beam polarimetry. These detectors will be introduced in the following sections.

\subsection{TPC}
The TPC is the central part of the STAR detector system \cite{startpc}. It is a cylindrical detector with 4 m in diameter and 4.2 m in length built around the beam-line. Thousands of particles can be produced after high center of mass energy heavy-ion collisions. The charged particles of them are deflected by the STAR magnet in a helical motion. The TPC is able to record those tracks, measure their momenta and identify particles by their ionization energy loss (\(dE/dx\)). Its acceptance covers \(2\pi\) in azimuthal angle \(\phi\) and from approximately \(-\)1.3 and +1.3 in pseudo-rapidity \(\eta\). It is capable to measure particle momentum from 0.1 GeV to 30 GeV and provide particle identification over a wide momentum range.

\begin{figure}[H]
\centering
\includegraphics[scale=1.0]{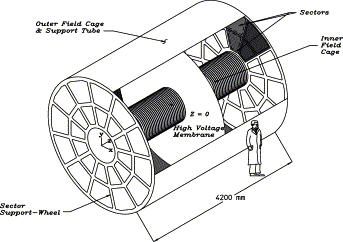}
\nocitecaption{The layout of the STAR TPC \cite{startpc}.}
\label{fig:tpc}
\end{figure}

Figure \ref{fig:tpc} shows the layout of the STAR TPC. It consists of a central membrane, an outer and inner field cage and two end-cap planes. The empty space between the central membrane and two end-caps is filled with gas. When charged particles pass through the TPC gas, the ionized secondary electrons drift toward the two end-caps in the uniform electric field provided by the central membrane and the end-caps. The drifted electrons are collected at the end-caps. The uniform electric field is maintained by the central membrane serving as a cathode, which is operated at 28 kV and the end-caps at ground. The inner and outer field cage confine the TPC gas and define the boundary of the electric field. The TPC gas is P10 gas (10\% methane, 90\% argon) regulated at constant pressure. It makes the drift velocity stable and insensitive to pressure and temperature changes by operating at the peak drift velocity. The value of the central membrane voltage is optimized for this purpose.

\begin{figure}[H]
\centering
\includegraphics[scale=1.0]{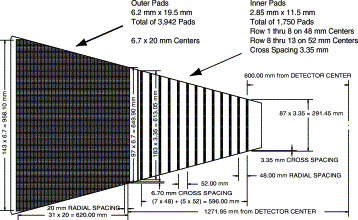}
\nocitecaption{The design of a TPC readout sector at both endcaps \cite{startpc}.}
\label{fig:tpcsec}
\end{figure}

The readout endcaps are based on Multi-Wire Proportional Chambers (MWPC) with readout pads. The drift electrons avalanche in the high fields due to the 20 \(\mu\)m anode wire, then the created positive ions in the avalanche induce image charges on the pads, and the image charges are read out by the digital system. There are 12 read out sectors arranged on a clock on each side of the endcaps with 3 mm small space between them. For each sector, there are two sub-sectors, inner-sectors and outer-sections, due to higher track densities in the inner section and lower track density in the outer section. Figure \ref{fig:tpcsec} shows the geometry and design of one TPC readout sector. There are in total 45 pad rows, 13 in the inner sector and 32 in outer sector. Each pad has granular size to determine the (\(x\), \(y\)) position of the drifting electrons. The arrival time of the drifting electrons is measured at the endcap. Together with the starting time of the collision, the \(z\) position of the drift origin can be determined. By having the (\(x\), \(y\), \(z\)) coordinates of the drifting electrons, one can reconstruct tracks produced by collisions and determine the track momentum from the track curvature.

\subsection{BEMC}
The BEMC is a major upgrade to the STAR baseline detector\cite{starbemc}. It allows to trigger on and study high \(p_{T}\) events like the jet events studied here, leading hadrons, isolated photons (\(\gamma\)), heavy quark production and \(W/Z\) boson decay. Its acceptance is \(|\eta| < 1\)  in pseudo-rapidity and \(2\pi\) in azimuthal angle \(\phi\). The front face of BEMC is at the radius of about 220 cm from the beam-line outside of the STAR TPC and inside the STAR magnet. The detector is based on the alternating lead and plastic scintillator layers with 20 times radiation length (\(X_{0}\)) at \(\eta =\) 0.  The BEMC includes a shower maximum detector (SMD). The shower maximum detector gives precise spatial information to reconstruct \(\pi^{0}\) and \(\eta\) mesons, isolated photons and single electrons and electron pairs in intense hadron backgrounds.

\begin{figure}[H]
\centering
\includegraphics[scale=0.8]{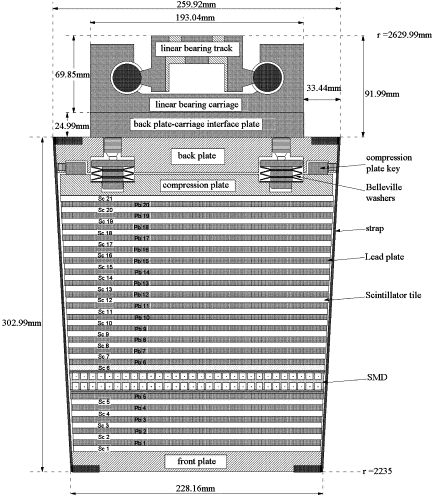}
\nocitecaption{The side view of the BEMC module \cite{starbemc}.}
\label{fig:bemcmod}
\end{figure}

The design of the BEMC includes 120 calorimeter modules each extending 0.1 in \(\phi\) and 1.0 in \(\eta\), which is about 26 cm wide and 293 cm long. The total depth of a module is about 30 cm. There are 120 modules, 60 in \(\phi\) by 2 in \(\eta\) to comprise the whole detector. Each module is segmented into 40 towers, 2 in \(\phi\) by 20 in \(\eta\), covering 0.05 in \(\phi\) and 0.05 in \(\eta\) for each tower. Figure \ref{fig:bemcmod} shows the side view of a BEMC module.

The BEMC consist of lead-scintillator stack with 20 layers of 5 mm thick lead and 21 layers of scintillators. The first 2 layers are 6 mm thick and the last 19 layers are 5 mm thick. The lead-scintillator stacks are held together between the front and back plates. The SMD is located between the fifth lead layer and the sixth scintillator layer. It is a gas amplification proportional wire counter with strip readout.

\begin{figure}[H]
\centering
\includegraphics[scale=0.8]{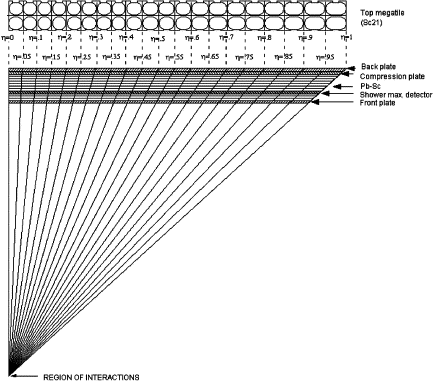}
\nocitecaption{The side view of the BEMC module and layout of the 21st scintillator layer mega-tile \cite{starbemc}.}
\label{fig:bemctow}
\end{figure}

The material of the scintillator is Kuraray SCSN81. The scintillator is machined in the form of mega-tile sheets with 40 optically isolated tiles in each layer corresponding to the individual towers in the module. The signal from each tile is read out with a wave-length shifting fiber. The signal in the wave-length shifting fiber is then transported from the detector through the STAR magnet to decoder boxes outside the magnet by a multi-fiber optical cable. In the decoder boxes, the signal from 21 scintillator layers composing a single tower are merged onto a single photomultiplier tube (PMT) which is also outside of the magnet. Figure \ref{fig:bemctow} shows the layout the 21st mega-tile layer.

The readout of BEMC is used as a part of STAR trigger system to trigger on high-\(p_{T}\) events, for example the jet triggers, because there is no dead-time for the detector at RHIC bunch crossing. For each tower, the BEMC uses a 12-bit flash ADC. The STAR trigger system doesn't use the full BEMC data. Instead it groups BEMC towers into 300 trigger patches covering a region 0.2 in \(\phi\) by 0.2 in \(\eta\) and uses two sets of trigger primitives from those patches. The first set is 300 tower sums digitized to 6 bits each, and the second set is 300 high tower values of 6 bits from the single largest tower within each trigger patch.

\subsection{EEMC}
Similar to BEMC, the EEMC extends the pseudo-rapidity coverage for high-\(p_{T}\) events to \(1 < \eta < 2\) with \(2\pi\) in azimuthal angle \(\phi\) \cite{stareemc}. It is built at the west side of the STAR detector with a toroidal shape around the beam-line and 270 cm from its front face to the collision point. It also includes a shower maximum detector (SMD), together with pre-shower and post-shower detectors to discriminate \(\pi^{0}/\gamma\) and electrons/hadrons.

\begin{figure}[H]
\centering
\includegraphics[scale=1.0]{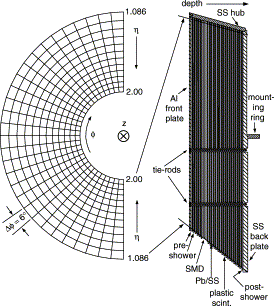}
\nocitecaption{STAR EEMC with the schematic tower structure on the left and the cut view of the EEMC at constant $\phi$ on the right \cite{stareemc}.}
\label{fig:eemc}
\end{figure}

Figure \ref{fig:eemc} shows the one half of the STAR EEMC with the schematic tower structure and the cut view of the EEMC at a fixed \(\phi\). The EEMC is built in fact at \(\eta\) from 1.086 to 2.000, allowing a small gap between BEMC and EEMC needed for services to exit the solenoid. The detector uses the alternating lead/stainless steel and plastic scintillator layers with 24 4 mm thick scintillator layers and 23 5 mm thick lead and stainless steel laminate layers. The total thickness is roughly equivalent to 21.4 radiation length. The scintillator material is the same as used in the BEMC. The whole detector is divided into 12 \SI{30}{\degree} modules and each module has 60 towers with each tower spanning 0.1 (\SI{6}{\degree}) in \(\phi\) and varying size from 0.057 to 0.099 in \(\eta\). Each module is constructed in the mega-tile form with two \SI{12}{\degree} mega-tiles at the ends and one \SI{6}{\degree} mega-tile in the middle. The \SI{12}{\degree} mega-tile has 24 trapezoidal tiles and the \SI{6}{\degree} mega-tile has 12, corresponding to each tower. The wavelength fiber is attached to each tile to readout the scintillation light. The wavelength fiber is connected to a clear fiber which bundles the signals from the 24 scintillator layers, then the clear fiber runs outside of the STAR magnet and the signal from the 24 layers is combined in an optical mixer and fed into a photo-multiplier tube (PMT).

The SMD is located after the fifth lead/stainless steel layer about 5 radiation lengths deep. It uses the scintillator strips, instead of proportional wire counter with strip readout in the BEMC. The pre-shower detector is the first two scintillator layers behind the front plate and the post-shower is the last scintillator layer. The signals from each of those three layers are read out by two independent wavelength fibers. One is for constructing the total tower signal and the other as the output signal of the pre-shower and post-shower detector.

The readout of the 12-bit EEMC tower signal is also sent to STAR trigger system at the level 0. The towers are grouped into trigger patches each of which spanning 0.2 in \(\phi\) and 0.2 or 0.4 in \(\eta\). The summed tower ADC and highest tower ADC within a trigger patch are calculated as inputs to the trigger system. Jet patches can not only be formed inside the EEMC but also can be combined with the BEMC to form a overlap jet patch to define jet patch triggers. 

\subsection{BBC}
The BBC is a fast detector to provide signals to the STAR trigger system at the level 0 \cite{starbbc}. It serves the purposes for triggering on minimal bias events, monitoring overall luminosity, measuring relative luminosities due to different spin patterns in bunch crossings and measuring local polarimetry. It is mounted around the beam line outside of the STAR magnet at the east and west side of the collision center about 374 cm from the center.

\begin{figure}[H]
\centering
\includegraphics[scale=0.5]{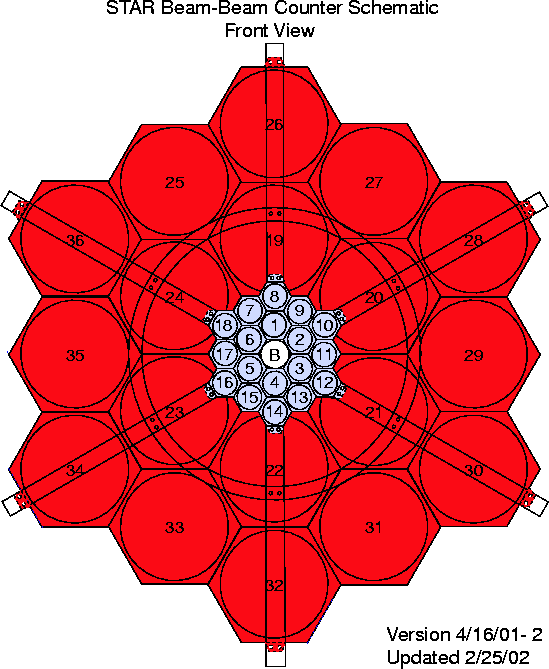}
\nocitecaption{Front view of STAR BBC annuli \cite{starbbc}.}
\label{fig:bbc}
\end{figure}

Figure \ref{fig:bbc} shows the structure of the STAR BBC. There are two annuli of scintillators with each annulus having 18 hexagonal tiles, 6 in the inner ring and 12 in the outer ring. The tiles in the outer annulus are called large tiles and the tiles in the inner annulus are called small tiles. The signals from the large tiles are not used in the following analysis. The coverage of small tiles is \(3.4 < |\eta| < 5.0\) in pseudo-rapidity and \(0 < \phi < 2\pi\) in azimuth. The signals from the small tiles are fed into 16 photo-multiplier tubes (PMT). The outputs of those PMTs are transferred to the STAR trigger system.

\subsection{ZDC}
The ZDC is intended to detect evaporation neutrons from heavy-ion collisions at small angles close to the beam-line, \(\theta <\) 4 mrad \cite{starzdc}. ZDCs are located at the east and west sides of the collision center. Each ZDC has three modules with each 10 cm in width and 13.6 cm in length. The ZDC module has multiple alternating quartz and tungsten layers. The tungsten plate is 0.5 cm thick, corresponding to 2 nuclear interaction length and 50 radiation length for each complete ZDC module. The Cherenkov light produced by charged particles in showers while neutrons hitting the detector are transported by wavelength fiber to a single photo-multiplier tube (PMT). The signals from the east and west side ZDC also flow to the STAR trigger system. These signals are used to trigger on minimum bias events, monitoring overall luminosity, and measuring relative luminosity due to different spin patterns in bunch crossings.

\subsection{VPD}
\begin{figure}[H]
\centering
\includegraphics[scale=0.5]{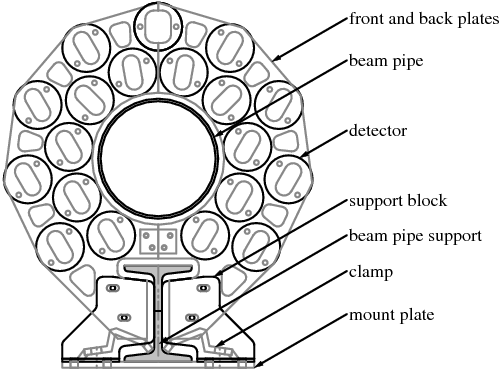}
\nocitecaption{The STAR VPD detector \cite{starvpd}.}
\label{fig:vpd}
\end{figure}

The VPD are also used by the STAR trigger system to serve similar purposes as the BBC and ZDC such as for triggering on minimum bias events and measuring relative luminosity \cite{starvpd}. There are two VPD, one on each side of the collision center about 5.7 m from the center, covering \(4.24 < |\eta| < 5.1\) and \(0 < \phi < 2\pi\). On each side, the VPD has 19 individual detectors. Figure \ref{fig:vpd} shows the individual detectors in one of the VPDs. Each individual detector has an aluminum cylinder with front and back caps. There are a 6.4 mm thick lead absorber, a 10 mm thick scintillator right next to the lead absorber, and a photo-multiplier tube attached to the scintillator by optically transparent silicone adhesive. The lead absorber is about 1.13 radiation length thick. Two sets of signals from the VPD are sent out, one to the STAR trigger system and one to the STAR data acquisition system.

%% file: chapter4.tex
\chapter{\uppercase{Inclusive Jet Longitudinal Double Spin Asymmetry Analysis}}

As shown in Equation \ref{eq:aLLtheory}, the inclusive jet double spin asymmetry \(A_{LL}\) is the fractional difference of jet cross-sections between the like and unlike helicity of the two longitudinally polarized beams.
\begin{align}
A_{LL} = \frac{1}{P_A \times P_B}\frac{(\sigma^{++}+\sigma^{--})-(\sigma^{+-}+\sigma^{-+})}{(\sigma^{++}+\sigma^{--})+(\sigma^{+-}+\sigma^{-+})}.
\end{align}
Instead of directly measuring total jet cross-sections for the four spin states, the number of jets for the four spin states and relative luminosities are used since \(N = L\times \sigma\). In addition, the experimentally observed asymmetry needs to be scaled up to account for the incomplete polarization of the two beams. In equation \ref{eq:aLLexp}, the numerator and denominator are the sums from all the runs and scaled both by the beam polarizations,
\begin{align}
A_{LL} &= \frac{\sum_{run} P_AP_B((N^{++}+N^{--}) - R(N^{+-}+N^{-+}))}{\sum_{run} P_A^2P_B^2((N^{++}+N^{--})+ R (N^{+-}+N^{-+}))}, \label{eq:aLLexp}
\end{align}
where \(R = \frac{L^{++} + L^{--}}{L^{+-}+L^{-+}}\).

\section{The Experimental Data Sample}
In the year of 2012 RHIC run, STAR has taken longitudinally polarized pp collision data at the center of mass energy \(\sqrt{s} =\) 510 GeV. The data taking period extended about six weeks, from March 15th, 2012 to April 18th, 2012. The event triggers were set up for physics goals with the trigger configuration, named "\(pp510\_production\_2012\)''. There were 744 runs recorded with major detectors, TPC, BEMC and EEMC active and in good running status. A run is a data taking period ranging from a few minutes to an hour.

The relevant events for the inclusive jet measurement are triggered by jet patch triggers with three different thresholds. There are about 177 million, 163 million, and 42 million events collected for the jet patch triggers with the three thresholds from the smallest value to the largest value.

\section{Spin Patterns for the 2012 RHIC Longitudinally Polarized $pp$ Run}
For each RHIC ring, there are 120 bunches carrying the proton beam. Only 111 of them are filled with nine left empty known as the abort gap. Bunches in each ring are numbered from 0 to 119, referred as the bunch ID. The following definitions are made:  a) the beam circulating clockwise is color coded as the blue beam and the beam circulating counter clockwise is color coded as the yellow beam; b) at each intersection point a fixed pair of bunches from the two beams collide; c) at the eight o’clock intersection point the \(n\)-th bunch in the blue beam collides with the \(n\)-th bunch in the yellow beam and d) at STAR, six o'clock intersection point, the blue beam IDs are used as the bunch crossing number. From the above definitions, the bunch ID for the yellow beam at STAR can be deduced from the bunch crossing number.

During the 2012 RHIC run, two additional bunches from each ring were left empty, that is bunch ID 38 and 39 in the blue beam and bunch ID 78 and 79 in the yellow beam. At STAR those bunches meet with the abort gap in the other beam. RHIC beams are injected into the RHIC rings bunch by bunch, usually taking about 10 minutes. The duration from when the beams are fully injected into the rings, to when the beams are dumped is called a RHIC fill. One fill usually lasts about eight hours. Bunches can have different spin orientations. However for a fill, a specific spin orientation is fixed when the bunches are filled, which constitutes a spin pattern. A certain spin pattern is carefully chosen before the fill starts.

There are four intended spin patterns for the two beams. The spin pattern repeats every eight bunches. The four patterns are: \(P1\), \(+ - + - - + - +\); \(P2\), \(- + - + + - + -\); \(P3\) \(+ + - - + + - -\); and \(P4\) \(- - + + - - + +\). \(P2\) is the mirror image of \(P1\) and so does \(P3\) of \(P4\). Beams with one of first two patterns collide with one of the last two patterns, therefore there are eight combinations of colliding spin patterns. This provides all the possible collision spin patterns at every bunch crossing which helps to reduce the systematic uncertainty caused by bunch crossing conditions.

At STAR, the spin configurations are number-coded with the rules shown in table \ref{tab:spinbit}. The coded number is also known as the spin bit that implies the spin orientation of the two colliding bunches at the 12 o'clock intersection piont. At STAR, due to  the Siberan snake on the ring, the spin orientation rotates \SI{180 }{\degree} therefore the positive (\(+\)) helicity becomes the negative helicity (\(-\)) and visa versa.

\begin{table}[H]
\centering
\begin{tabular}{|c|c|c|}
\hline
Spin configuration & Blue beam helicity & Yellow beam helicity \\
\hline
5 & + & + \\
\hline
6 & $+$ & $-$ \\
\hline
9 & $-$ & $+$\\
\hline
10 & $-$ & $-$\\
\hline
1 & empty & $+$ \\
\hline
2 & empty & $-$ \\
\hline
4 & $+$ & empty \\
\hline
8 & $-$ & empty \\
\hline
0 & empty & empty\\
\hline
\end{tabular}
\caption{Spin bit maps to the spin configurations at the 12 o'clock intersection point.}
\label{tab:spinbit}
\end{table}

\section{Beam Polarizations}
\begin{figure}[H]
\centering
\includegraphics[scale=0.8]{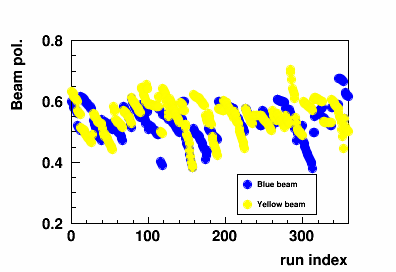}
\nocitecaption{Beam polarization vs. run number where runs in this plot are those runs selected in this analysis \cite{run12polresult}. Patterns follow time dependence within a fill.}
\label{fig:run12pol}
\end{figure}

The proton-Carbon (\(pC\)) \cite{pCpol} polarimeter and \(H\)-jet polarimeter \cite{hjetpol} are used to measure the beam polarization. Both polarimeters are located in the vicinity of the 12 o'clock intersection point. They measure recoiled target nuclei produced by very small-angle elastic scattering of the transversely polarized proton beam. The elastic scattering process in this region is dominated by the Coulomb-Nuclear Interference (CNI) which generates asymmetries \(A_N\) in the yields of recoiled nuclei relative to the polarization orientation. The measured asymmetry \(\epsilon \) is the product of beam polarization \(P\) and \(A_N\), \(\epsilon = P \times A_N\). In the \(H\)-jet polarimeter, the atomic hydrogen beam target can be polarized, and its polarization can be precisely measured by its Breit-Rabi polarimeter. The \(A_N\) with respect to polarized targets can be measured by averaging the polarization of polarized beam. The same way can be done to measure \(A_N\) with respect to the polarized beam. Therefore the \(H\)-jet is able to provide an absolute beam polarization measurement. The \(pC\) polarimeter measures a series of intensity averaged polarizations over a period of time. The measured polarizations are fitted to the form \(P(t)=P_{0}-P'\times t\), where \(P(t)\) is the polarization measured at time \(t\), \(P_{0}\) is the polarization at the fill starting time \(t_{0}\) and \(P'\) is the absolute polarization loss rate \cite{run12polnote}. The fitted parameters \(P_{0}\) and \(P'\) are given fill by fill as well as the starting time \(t_{0}\). The final beam polarization is taken from the \(pC\) measurement scaled by the \(H\)-jet polarization \cite{run12polresult}. For a specific run taken at a certain time, it is easy to calculate the polarization for that run based on the form \(P(t) = P_{0} - P' \times t\). Figure \ref{fig:run12pol} shows the polarizations of the blue and yellow beams for the final selected runs in this analysis.

%
\input{chapter4_rellum}

\section{Jet Patch Trigger Setup}
\begin{figure}[H]
\centering
\includegraphics[scale=0.6]{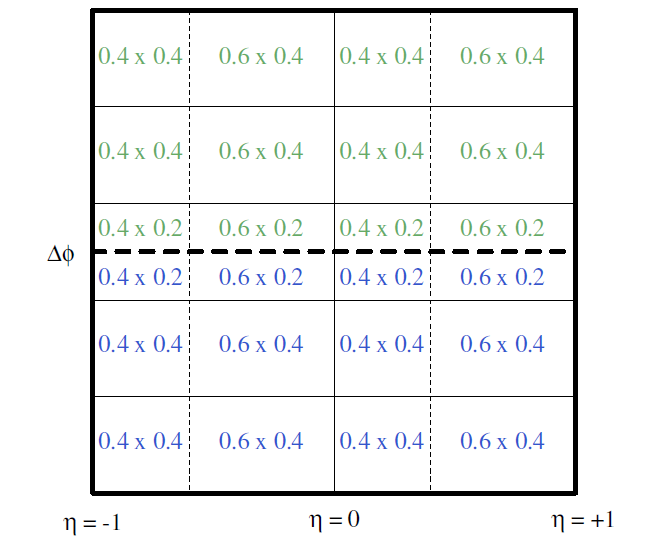}
\caption{BEMC DSM $\eta - \phi$ scheme.}
\label{fig:bemcdsm}
\end{figure}

The STAR BEMC and EEMC serve as the trigger detectors for high \(p_{T}\) and jet event studies. A BEMC tower covers \(0.05 \times 0.05\) in \(\eta\) and \(\phi\). A trigger patch in the BEMC consists of \(4 \times 4\) BEMC towers covering \(0.2 \times 0.2\) in \(\eta\) and \(\phi\). A jet patch is a defined \(1.0 \times1.0\) \(\eta\)-\(\phi\) region, which is contributed by \(5 \times 5\) trigger patches. In the trigger system, a level 0 Data Storage and Manipulation (DSM) board holds 10 channels, each of which comes from a single trigger patch. Each channel receives a six bit patch sum ADC and a six bit high tower ADC from the trigger patch. Figure \ref{fig:bemcdsm} shows the BEMC trigger scheme in a \(2 \times 2\) \(\eta\)-\(\phi\) subset of the full \(2 \times 6\) \(\eta\)-\(\phi\) space. There are 15 level 0 DSM boards on each side of detector, the east side corresponding to negative \(\eta\) and the west side corresponding to positive \(\eta\). The patch sum from each channel is summed up into two groups, lower \(\eta\) sum and higher \(\eta\) sum. The sums are passed to six level 1 DSM boards, each of which has six input channels from level 0 DSM boards. The input channels are formed in a way such to cover 2 units in \(\eta\) and 1 unit in \(\phi\). The combinations of lower \(\eta\) and higher \(\eta\) sums form three jet patches covering respectively \(-1.0 < \eta < 0\), \(-0.6 < \eta < 0.4\) and \(0 < \eta < 1.0\). Those sums are compared with three thresholds to form the threshold bits. The threshold bit is set to one if the sum is greater than the threshold and zero otherwise. These threshold bits from each level 1 BEMC DSM boards are passed to a level 2 DSM board for the further manipulation. A partial sum from \(0.4 < \eta < 1.0\), is also passed to the level 2 to combine with a partial sum from the EEMC.

\begin{figure}[H]
\centering
\includegraphics[scale=0.6]{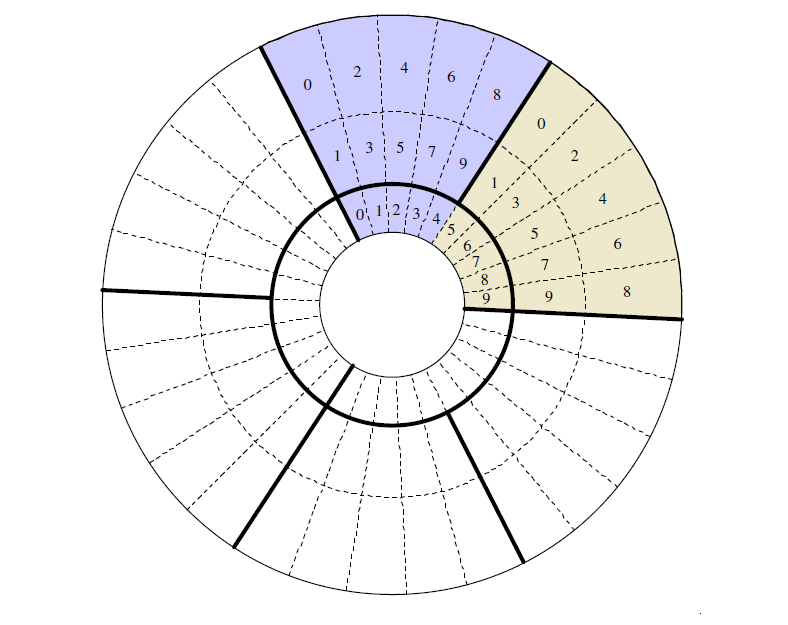}
\caption{EEMC DSM $\eta - \phi$ scheme, seen from the West looking towards the center of STAR.}
\label{fig:eemcdsm}
\end{figure}

The jet patch formed in the EEMC is similar to what is formed in the BEMC. The segments of jet patches in a particular \(\phi\) direction are matched with those in the BEMC to be able to form boundary jet patches between the BEMC and EEMC. Figure \ref{fig:eemcdsm} shows the EEMC trigger scheme in the full EEMC region. A level 0 EEMC DSM board receives inputs from 10 trigger patches. There are six single output DSM boards in the outer ring and three double output DSM boards in the inner ring. The six single output DSM boards calculate a lower \(\eta\) sum and a higher \(\eta\) sum and send them to a level 1 DSM board. The lower \(\eta\) is the EEMC partial jet patch sum covering \(1.0 < \eta < 1.4\) to be combined with the BEMC partial jet patch sum. There are two level 1 DSM boards, each covering one half of the EEMC. One level 1 DSM board receives three inputs from single output level 0 DSM boards and three inputs from double output level 0 DSM boards. Three EEMC jet patch sums are calculated in each level 1 DSM board and are compared with three thresholds 0 to 2 from the lowest to the highest. The threshold bits are passed to the level 2 DSM boards. The largest partial jet patch sum is also passed to the level 2 together with its ID used to identify the \(\phi\) position of the partial jet patch.

At the level 2, the DSM board receives six input channels from the BEMC and two channels from the EEMC. Each of the six BEMC input channels has threshold bits for three BEMC jet patches and one jet patch sum. Each of the two EEMC input channels has threshold bits for three jet patches and one largest partial jet patch sum. The largest partial jet patch sums from the EEMC are added to the corresponding partial jet patch sums from the BEMC in order to form the BEMC-EEMC boundary jet patch sums. These sums are also compared with three ordered thresholds 0 to 2 to form threshold bits as shown in Table \ref{tab:jpth}. In total three threshold bits for 18 BEMC jet patches, six EEMC jet patches and two BEMC-EEMC boundary jet patch are available. The jet patch 0 (JP0) trigger is defined such that at least one of these jet patches passed jet patch threshold 0. The same definitions apply for jet patch 1 (JP1) trigger and jet patch 2 (JP2) trigger.

\begin{table}[H]
\centering
\begin{tabular}{|c|c|c|}
\hline
Trigger & Threshold & Nominal $E_T$ in GeV\\
\hline
JP0 &  28 & 5.4 \\
\hline
JP1 & 36 & 7.3\\
\hline
JP2 & 66 & 14.4\\
\hline
\end{tabular}
\caption{Jet patch trigger thresholds.}
\label{tab:jpth}
\end{table}

\section{Jet Reconstruction at STAR}
The STAR software has the capability to run either a jet cone algorithm or a \(k_{T}\) type algorithm to find jets. The STAR 2006 inclusive jet \(A_{LL}\) analysis used the CDF cone algorithm with cone radius 0.7, seed energy 0.5 GeV and split and merging fraction 0.5. For the STAR 2009 inclusive jet \(A_{LL}\) analysis, the FastJet anti-\(k_{T}\) algorithm \cite{fastjet} with jet parameter 0.6 was used because the algorithm proves to be less susceptible to diffuse soft background from pile-up and underlying events. In this analysis, the FastJet anti-\(k_{T}\) algorithm has been chosen and the jet parameter is reduced to 0.5. The reasons for this will be discussed in the following section.

Both charged particle tracks measured from the TPC and tower deposition energy measured from the BEMC and EEMC are fed into the STAR jet finder. The jet finder is run on an event by event basis. For each event the primary tracks are selected, which are emitted from the reconstructed primary vertex. The primary vertex is the vertex where the two protons collide. The pile-up proof vertex (PPV) finder was used in this analysis. It does not rule out finding several primary vertices in a single event. However the best possible vertex, or in other words the highest ranked vertex is selected in the analysis. The PPV finder assigns vertices that are most likely to belong to pile-up events a negative ranking, therefore the highest ranked vertex chosen is required to have a positive ranking. In addition the highest ranked reconstructed vertex should be within the range of \(\pm\) 90 cm. The table \ref{tab:trackcuts} lists the track cuts used in this analysis to select good tracks for the jet finding process.

\begin{table}[H]
\centering
\begin{tabular}{|c|}
\hline
track flag $>$ 0 (good) \\
\hline
number of hit points  $\geq$ 12 \\
\hline
$\frac{N_{fit}}{N_{poss}} >$ 0.51 \\
\hline
$Dca \leq$ 3 cm \\
\hline
0.2 $<p_T<$ 200\\
\hline
$-$2.5 $ < \eta <$2.5 \\
\hline
Last point distance $>$ 125 cm \\
\hline
\end{tabular}
\caption{Track cuts used for jet reconstruction.}
\label{tab:trackcuts}
\end{table}

\begin{figure}[H]
\centering
\includegraphics[scale=0.6]{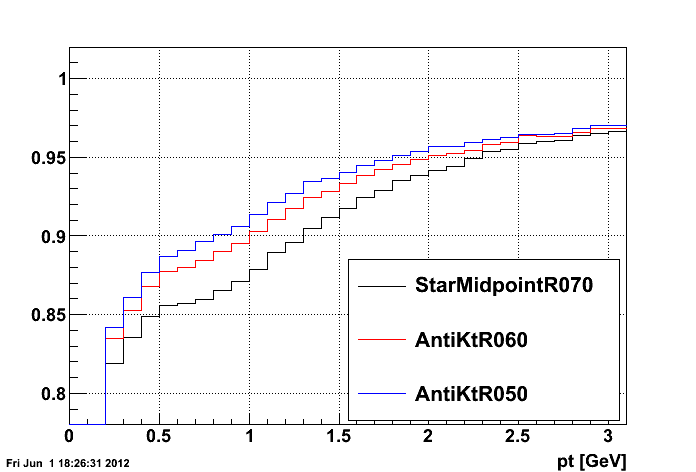}
\caption{The ratio of number of tracks with the $p_T$ dependent three-dimensiona cut over these with the $p_T$ dependent two-dimensional cut for a low luminosity run at $\sqrt{s} = $ 500 GeV. }
\label{fig:dcacutlowlum}
\end{figure}

\begin{figure}[H]
\centering
\includegraphics[scale=0.6]{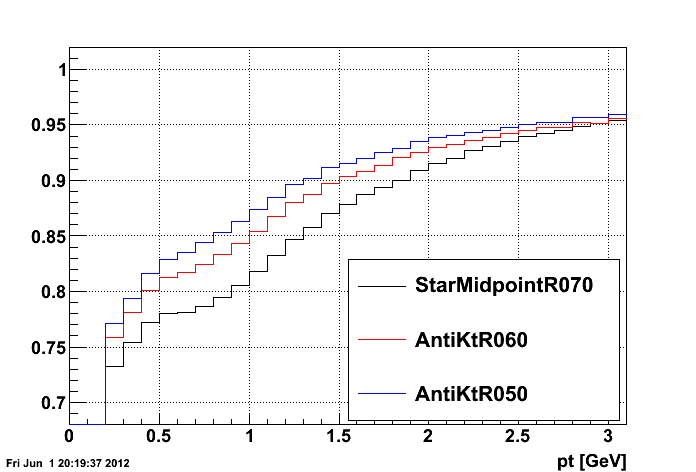}
\caption{The ratio of number of tracks with the $p_T$ dependent three-dimensional cut over these with the $p_T$ dependent two-dimensional cut for a high luminosity run at $\sqrt{s} = $ 500 GeV. }
\label{fig:dcacuthighlum}
\end{figure}

A \(p_T\) dependent distance closest to approach (\(Dca\)) cut as shown  in Equation \eqref{eq:dcacut} helps reduce the pile-up effects. 
\begin{align}
Dca_T <
	\begin{cases}
	2 \text{ cm} & \quad \text{if } p_T < 0.5 \text{ GeV} \\
	-1.0 \text{ cm/GeV} \times p_T + 2.5 \text{ cm} & \quad \text{if } 0.5 < p_T < 1.5 \text{ GeV} \\
	1 \text{ cm} & \quad \text{if } p_T > 1.5 \text{ GeV} \\
	\end{cases}
\label{eq:dcacut}
\end{align}
The full three dimensional Dca is used here, \(Dca_{T} = \sqrt{Dca_{x}^{2} + Dca_{y}^{2} + Dca_{z}^{2}}\), where \(Dca_{x}\) \(Dca_{y}\), and \(Dca_{z}\) are the \(x\), \(y\) and \(z\) components of the track \(Dca\). In previous analysis at \(\sqrt{s}\) = 200 GeV, a similar \(p_{T}\) dependent two dimensional \(Dca\) , \(Dca_{D} = \sqrt{Dca_{x}^{2} + Dca_{y}^{2}}\) was applied. Figures \ref{fig:dcacutlowlum} and \ref{fig:dcacuthighlum} show the ratio of the number of jet tracks found after the \(p_T\) dependent three-dimensional cut to that after the \(p_T\) dependent two-dimensional cut vs. track \(p_{T}\) by different jet finding algorithms for two runs with high and low luminosities, where the ZDC coincidence rates are 200 and 93 kHz respectively. The higher luminosity is more prone to the effects of pile-up tracks. The three-dimensional cut removes a bigger fraction of low \(p_{T}\) tracks in the higher luminosity runs where the pile-up effect is more dominant. Both cuts make only a small difference, less than 5\%, when track \(p_{T}\) above 3 GeV between the higher luminosity run and the lower luminosity run. Also the anti-\(k_{T}\) algorithm with a smaller \(R\) parameter is less affected by the pile-up effects than the mid-point code algorithm with a larger \(R\) parameter.

%

%
Table \ref{tab:towercuts}  lists the tower cuts for the BEMC and EEMC towers. The tracks are also matched to the BEMC and EEMC towers and the full track momentum is subtracted from the matched tower \cite{run9aLL2015}. Tracks with momentum are converted to Lorentz four-vectors by using the pion default mass 0.1396 GeV, and towers with their deposition energy are converted to Lorentz four-vectors by using the massless photon and assuming the momentum is pointing from the primary vertex to the deposited tower. Those two sets of Lorentz four-vectors are merged and taken as input to the jet finder where the jet algorithm is applied. At the end, only those jets that have transverse momentum \(p_{T}\) between 5 GeV and 200 GeV are passed down to the analysis stream.

\begin{table}[H]
\centering
\begin{tabular}{|c|}
\hline
tower status $=$ 1 (good) \\
\hline
tower $E_T >$ 0.2 \\
\hline
tower ADC $-$ ped $>$ 4 \& ADC $-$ ped $>$ RMS \\
\hline
\end{tabular}
\caption{BEMC and EEMC tower cuts used for jet reconstruction.}
\label{tab:towercuts}
\end{table}

The reconstructed jets are also required to pass certain kinematic cuts to ensure the jets fall well within the detector acceptance and the interested kinematic range and those jets are not apparently coming from non-collision backgrounds. The following table \ref{tab:jetcuts} lists the jet cuts used in this analysis. The detector \(\eta\) is defined by projecting the thrust of jet axis from the collision vertex to the BEMC detector and taking the \(\eta\) component of the projected vector as expressed in the STAR detector coordinate.  The minimum sum track \(p_T\) and the neutral energy fraction \(R_t\) is to keep jet candidates from the neutral jets that are constituted by neutral particles. At the measured energy scale, the neutral jets are likely coming from non-collision backgrounds such as cosmic rays. In addition, since the track reconstruction by TPC is not reliable at track \(p_T >\) 30 GeV, jets with a track that has \(p_T >\) 30 GeV are skipped. A upper jet \(p_T\) cuts are also applied for JP0 and JP1 triggered jets, which are 33.6 GeV and 39.3 GeV. For jets above those \(p_T\) limits, the JP2 triggered jets dominate the statistics. For the underlying event corrections, jets are dropped if the underlying event correction \(dp_T\) makes the jet \(p_T\) shift down more than two jet \(p_T\) bins. The jet \(p_T\) bin boundaries in GeV are chosen as the bin width is 17\% of the lower edge, such as 6.0, 7.0, 8.2, 9.6, 11.2, 13.1, 15.3, 17.9, 20.9, 24.5, 28.7, 33.6, 39.3, 46.0 and 53.8.

\begin{table}[H]
\centering
\begin{tabular}{|c|}
\hline
$-$0.9 $<$ jet $\eta$ $<$ 0.9 \\
\hline
$-$0.7 $<$ jet detector $\eta$ $<$ 0.9 \\
\hline
sum track $p_T$ $>$ 0.5\\
\hline
jet neutral energy fraction $R_t$ $<$ 0.94\\
\hline
Individual track $p_T$ $<$ 30 GeV \\
\hline
JP0 jet $p_T < 33.6$ GeV \\
\hline
JP1 jet $p_T < 39.3$ GeV \\
\hline
Underlying event correction $dp_T$ shifts no more than two bins \\
\hline
\end{tabular}
\caption{Jet candidate cuts used in this analysis.}
\label{tab:jetcuts}
\end{table}

\section{Run Quality Assurance}
The run quality assurance (QA) is to select the good runs from all the runs taken during the 2012 RHIC 510 GeV longitudinally polarized \(pp\) running period. The run quality is to make sure all the data in the final data analysis have physics merit to achieve the inclusive jet \(A_{LL}\) measurements. Therefore the jet quantities such as the jet transverse momentum \(p_{T}\), the jet pseudo-rapidity \(\eta\), the jet azimuthal angle \(\phi\), the jet neutral fraction \(R_{t}\), the number of tracks per jet and the number of towers per jet, are inspected to flag the bad runs that will be discarded in the final analysis.

Several general requirements are applied before the QA process starts. Runs that were running with the TPC, BEMC and EEMC active are selected. The duration of runs should be longer than 180s. Runs that have relative luminosity information are kept. A simple hot tower check in the BEMC and EEMC towers is performed to eliminate a few runs that have abnormally large ADC values in BEMC and EEMC towers. Due to the efficiency of the PPV vertex finder dropping significantly as the run luminosity increases, the runs with accidental and multiple corrected BBC coincidence rate greater than 5 MHz are dropped.

The jet quantities are inspected on a run-by-run basis. Traditionally the averaged jet quantities are plotted against run index. The run index usually increments from zero and the order of the run index is based on the time when the runs were taken from the earliest to the latest. In other words, the jet quantities are inspected in a time series. At the beginning of a fill the collision rate is high, and at the end of a fill the collision rate is low. The jet quantities vary with the collision rates. So it is easy to spot the fill pattern in the plot of jet quantity against run index. However in 2012 the collision rates averaged a factor of five higher than during the 2009 RHIC run, and the collision rate coverage was much wider in the 2012 RHIC run. Given that jet quantities change dramatically when the collision rates differ largely, it was necessary to inspect jet quantities against the collision rate. This led led to a new QA selecting process.

\begin{figure}[H]
\centering
\includegraphics[scale=1.0]{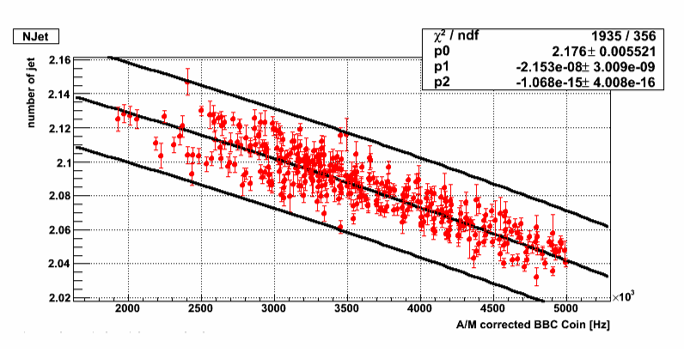}
\caption{The averaged number of jets per event vs. accidental and multiple corrected BBC rate for JP2 events. }
\label{fig:jetqanjets}
\end{figure}

\begin{figure}[H]
\centering
\includegraphics[scale=1.0]{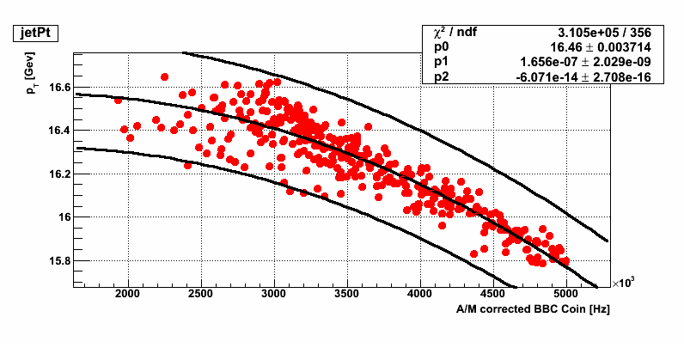}
\caption{Jet $p_T$ vs. accidental and multiple corrected BBC rate for JP2 events. }
\label{fig:jetqapt}
\end{figure}

\begin{figure}[H]
\centering
\includegraphics[scale=1.0]{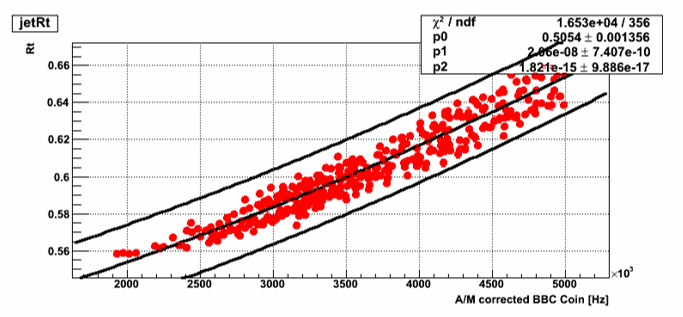}
\caption{Jet neutral fraction $R_t$ vs. accidental and multiple corrected BBC rate for JP2 events. }
\label{fig:jetqart}
\end{figure}

The collision rate is taken from the calculated BBC coincidence rate after the accidental and multiple correction from the scaler system. A number of jet quantities were plotted against the corrected BBC coincidence rate. Figures \ref{fig:jetqanjets}, \ref{fig:jetqapt} and \ref{fig:jetqart} show jet quantities, the averaged number of jets, the jet \(p_T\) and the jet neutral fraction \(R_{t}\) vs. the corrected BBC coincidence rate for the JP2 events. The jet quantities dependence on the collision rate can be seen clearly. For example the jet \(p_{T}\) decreases as the collision rate increases, due to reduced tracking efficiency in the higher collision rate cases. In contrast the jet \(R_{t}\) shows the opposite trends.

The final selection criteria are determined by fitting the jet quantities as a function of collision rates with a second or third order polynominal form. The RMSs of the jet quantities relative to the fitted value are calculated. The runs that have at least one of the jet quantities inspected fall outside the fitted value plus or minus three times RMS region are discarded. Different jet quantities may reflect different issues in the run taken period. For example jet \(\eta\) and \(\phi\) may reflect ineffective coverage of the TPC, BEMC and EEMC, jet track \(p_{T}\) may be distorted by pile-up events and so on. Therefore it is useful to remove questionable runs as many ways as possible by tagging on all sorts of jet quantities and requiring the jet quantity following the general trend vs the collision rate. But the cut should not be strict enough to lose event statistics unnecessarily. A total of 21 jet quantities are considered in this process. The total integrated luminosity for the JP0, JP and JP2 triggers is respectively 1.2 \(pb^{-1}\), 5.5 \(pb^{-1}\), and 46 \(pb^{-1}\) accounting for the trigger pre-scaling.


\section{Underlying Event Correction and Its Contribution}

The underlying event contribution, as one of the background contributions to jet signals measured in this analysis, is also considered. From the picture of the proton parton model, it is easy to imagine the two incident protons as two clusters of partons colliding with each other. The scattering process of most interest is the hardest two partons scattering. The other soft scatterings could also be mixed into the hardest scattering and contribute to the signals that are finally measured. The background generated due to these multiple soft scatterings is classified as underlying events. This background is different from detector pile-up, because these collisions are coming from the same proton-proton collision as the hardest scattering.

There are several methods to measure the underlying event effects in the jet analysis and then make corrections for the jet physical quantities, for example jet cross-section, at LHC. One of these methods developed by the ALICE experiment, called perpendicular cones method \cite{alicecone} is adapted in this analysis. In this analysis it's called the off-axis cone method.

\begin{figure}[H]
\centering
\includegraphics[scale=0.8]{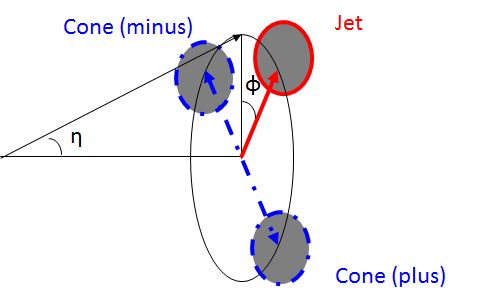}
\caption{The illustration of two off-axis cones relative to a jet. }
\label{fig:cones}
\end{figure}

The off-axis cone method is a method to study underlying event on the level of jet by jet, instead of on the level of event. First for a reconstructed jet draw two off-axis cones, each of which is centered at the same \(\eta\) as the jet but \(\pm \frac{\pi}{2}\) away in \(\phi\) from the jet \(\phi\) as shown in Figure \ref{fig:cones}. Then collect particles falling inside the two cones. The particle candidate pool is the exact same input as used for the jet finding algorithm. The off-axis cone radius is chosen to be the same as the jet parameter of the anti-\(k_T\) algorithm used in this analysis, \(0.5\). The \(p_T\) of the off-axis cone is defined as the scalar sum of the all the particles inside the cone, denoted as \(p_{T,ue}\). The energy density, \(\sigma_{ue, cone}\) is defined as the off-axis cone \(p_T\) divided by the cone area which is \(\pi R^2\). The multiplicity of the off-axis cone is the number of particles inside the cone. Finally the average density of the two cones is taken as the estimate to the underlying event energy density, \(\sigma_{ue} = \frac{1}{2}(\sigma_{ue,1} + \sigma_{ue,2})\). The correction on jet \(p_T\),  is therefore \(dp_T = \sigma_{ue}\times A_{jet}\) where \(A_{jet}\) is the jet area. The jet area is given by the anti-\(k_T\) jet finding algorithm, by using the technique of ghost particles \cite{fastjet}.


Given that the physics of the underlying event is perceived to be evenly distributed over the \(\eta\)-\(\phi\) space, the observed underlying event energy density in the \(\eta\)-\(\phi\) space is approximately uniform. However detector acceptance and efficiency is usually not uniform in \(\eta\)-\(\phi\) space. For example, at STAR there is a gap between the BEMC and EEMC and the TPC tracking efficiency degrades drastically at \(1.0 < |\eta| < 1.5\). Fortunately the STAR detector has good symmetry in \(\phi\) and the two off-axis cones are centered at the same \(\eta\) as the jet, therefore it is well applicable here. The underlying event estimation method doesn't require boost-invariance, so it can also be applied in proton and ion collisions.

\begin{figure}[H]
\centering
\includegraphics[scale=1.0]{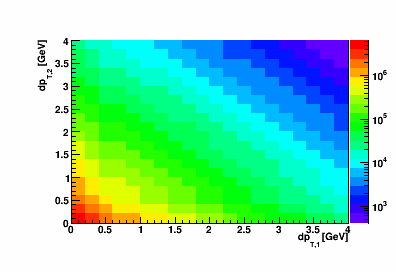}
\caption{The 2-D distribution of the summed track and tower $p_T$ from the two cones. }
\label{fig:run12uept}
\end{figure}

It is important to note that particles produced by the beam remnants and pile-up effects are also approximately uniform over the mid-rapidity STAR detectors. Therefore, this procedure also provides a first-order subtraction of those background. Figure \ref{fig:run12uept} shows 2-D distribution of the summed track and tower \(p_T\) from the two cones in the 2012 data.

\section{Counting the Number of Jets and Its Statistical Uncertainty}
To calculate the inclusive jet \(A_{LL}\), the number of inclusive jets for the four spin states \(++\), \(+-\), \(-+\) and \(--\) are counted for each run. While counting inclusive jets, a trigger category algorithm is applied. During the data taking, both JP0 and JP1 events were pre-scaled heavily to accommodate the band-with of the data acquisition system. JP2 events are not pre-scaled. Obtaining a mutually exclusive sample among triggers will help to understand the effects contributed by the individual trigger. This will play an important role in terms of weighting the individual trigger ensemble in the simulation to compare with the trigger combined measurements in the data. Only the JP0, JP1 and JP2 triggers are used in this analysis. 

For each event, the event record provides the information if a trigger was actually fired, "didfire" in the process of data acquisition. If a trigger was fired, the didfire would be one. Otherwise, the didfire is zero. In this analysis, an offline trigger simulator was applied where the trigger decision was simulated by imitating the online trigger system. The same input data was used, the raw ADC of the BEMC and EEMC towers. The same trigger algorithm was implemented. If a trigger was fired in the trigger simulator, likewise to the online system, a "shouldfire" bit is set to one. Otherwise the "shouldfire" is zero. The main purpose of the offline trigger simulator is to select events in the simulation which will be discussed in the following sections. However in the analysis process of the real data,  requiring both didfire and shouldfire provides consistent event selection between data and simulation. This is particularly important for calorimeter towers that were included in the on-line trgger, but were found to be bad during off-line QA. Shouldfire provides a mechanism to remove those towers from the trigger during off-line analysis. Also in cases where there was malfunction in the online trigger system, the trigger simulator will provide a second proof to classify the event. Another benefit is for those events skipped by pre-scaling, the offline trigger simulator will be able to tag them, which can be used to promote a more heavily pre-scaled trigger category to a less heavily pre-scaled trigger category. The performance of the offline trigger simulator is well matched with the online trigger system.

The JP2 trigger is not pre-scaled. The JP2 threshold is approximately equivalent to a deposition energy of 14.4 GeV inside the BEMC and EEMC towers covered by a full jet patch. Giving a little room for the track contributions to the jet momentum, as long as a reconstructed jet \(p_T\) is greater than 15.3 GeV and both JP2 didfire and shouldfire are one in the event, this jet will be classified as a JP2 jet. For a jet that was not able to be classified as a JP2 jet, if the jet \(p_T\) is greater than 8.2 GeV which is a little higher than the corresponding energy deposition of 7.3 GeV in the jet patch for JP1 threshold, and both the didfire and should fire are one, then the jet will be classified as a JP1 jet. In addition, if the jet satisfies the JP1 jet \(p_T\) requirement, the didfire for JP0 is one and shouldfire for JP1 is one, then the jet is promoted from JP0 jet to JP1. For a jet not satisfying either of JP2 and JP1 requirements, as long as the jet \(p_T\) satisfies the minimum kinematic limit, 6 GeV, and both didfire and shouldfire for JP0 are one, the jet will be classified as a JP0 jet. For each jet patch trigger, a geometric match is imposed to identify the jet candidate is the jet that fires the jet patch. The match requires the jet \(\eta\) and \(\phi\) are both within the range of \(\pm 0.6\) of the \(\eta\) and \(\phi\) of the center of the jet patch, and the jet patch sum ADC should be above the specific jet threshold. The details of this algorithm are described in Algorithm \ref{algo:trg}. In this analysis, all the candidate jets reconstructed in an event are sorted by their jet \(p_T\), then they are fed into this algorithm. The algorithm ends when no more candidate jets left or two triggered jets have been found in the event.

\begin{algorithm}[H]
\setstretch{1.0}
\begin{algorithmic}[1]
\If {jet $p_T$ $>$ 15.3 GeV}
	\If {didfire(JP2) $\land$ shouldfire(JP2) $\land$ GeoMatch(JP2)}
		\State jet $\gets$ JP2
	\ElsIf {(didfire(JP0) $\lor$ didfire(JP1)) $\land$ shouldfire(JP1) $\land$ GeoMatch(JP1)}
		\State jet $\gets$ JP1
	\ElsIf {(didfire(JP0) $\land$ shouldfire(JP0) $\land$ GeoMatch(JP0)}
		\State jet $\gets$ JP0
	\Else
		\State skip
	\EndIf
\ElsIf {jet $p_T > $ 8.2 GeV}
	\If {(didfire(JP0) $\lor$ didfire(JP1)) $\land$ shouldfire(JP1) $\land$ GeoMatch(JP1)}
		\State jet $\gets$ JP1
	\ElsIf {didfire(JP0) $\land$ shouldfire(JP0) $\land$ GeoMatch(JP0)}
		\State jet $\gets$ JP0
	\Else
		\State skip
	\EndIf
\ElsIf {jet $p_T >$ 6.0 GeV}
	\If {didfire(JP0) $\land$ shouldfire(JP0) $\land$ GeoMatch(JP0)}
		\State jet $\gets$ JP0
	\Else
		\State skip
	\EndIf
\Else
	\State skip	
\EndIf
\end{algorithmic}
\caption{Algorithm used to categorize jets into three jet patch JP0, JP1 and JP2 triggers.} 
\label{algo:trg}
\end{algorithm}

The two jet correlation due to two inclusive jets within the same event falling into one or two jet \(p_{T}\) bins is considered when calculating the uncertainties on the number of inclusive jets found. Events contain two jets that pass all cuts 4.7\% of the time. The two jet correlation is made in such way that assuming every event has at most two inclusive jets found, the number of jets found is \(N = N_{1} + 2 \times N_{2}\) where \(N_{1}\) is the number of events having one jet and \(N_{2}\) is the number of events having two jets, and the statistical uncertainty on the number of jets is \(\sigma^{2} = N_{1} + 4N_{2}\)  by taking the number of events obeying Poisson distribution. The same calculation can be applied to the case where counting jets within the jet \(p_T\) bins. The number of jets found in jet \(p_{T}\) bin \(i\) is \(N^{i} = N_{1}^{i} + 2 \times N_{2}^{ii}+ \sum_{q\neq i} N_{2}^{qi}\), where \(N^{i}\) is the number of jets in the \(i\)-th \(p_{T}\) bin, \(N_{1}^{i}\) is the number of events having only one jet with the jet falling in the \(i\)-th \(p_{T}\) bin, \(N_{2}^{ii}\) is the number of events having two jets with both jets falling in the \(i\)-th \(p_{T}\) bin and \(N_{2}^{qi}\) is the number of events having two jets with one jet falling in the \(q\)-th \(p_{T}\) bin and the other falling in the \(i\)-th \(p_{T}\) bin. Therefore the statistical uncertainty of number of jets found in the \(i\)-th \(p_{T}\) bin is \(\sigma_{i} = \sqrt{N_{1}^{i} +4\times N_{2}^{ii} + \sum_{q \neq i}N_{2}^{qi}} \).

The statistical uncertainties of inclusive jet \(A_{LL}\) are estimated based on the following equation \ref{eq:aLLstat}.
\begin{align}
\Delta{A_{LL}} &= \frac{\sqrt{\sum_{run} P_A^2P_B^2(((\Delta N^{++})^2 + (\Delta N^{--})^2) + R^2((\Delta N^{+-})^2+ (\Delta N^{-+})^2))}}{\sum_{run} P_A^2P_B^2((N^{++}+N^{--})+ R (N^{+-}+N^{-+}))} \label{eq:aLLstat}
\end{align}
The formula is an approximation to the the statistical uncertainty for the jet \(A_{LL}\). The corresponding error in \(\Delta A_{LL}\) is negligible.

The way two jets in the same events are falling into bins according to their \(p_T\) also introduces correlations between two \(p_T\) bins. The correlation can be expressed as the covariance of the two bins divided by the square root of the product of the variance of the two bins, for example the correlation between \(i\)-th and \(j\)-th bin can be expressed as \( \rho_{ij} = \text{cov}(i, j)/ \sqrt{\sigma_i^2 \sigma_j^2}\). The covariance for two distinct bins is the number of events that have two jets that fall into the two bins \(N_2^{ij}\) where \(i \neq j\), based from the assumption of the Poisson distribution mentioned above. The covariance of the two jets falling into the same \(p_T\) bin is exactly the variance of the \(p_T\) bin. In summary, the correlation can be expressed in the following equations \ref{eq:jetcorr}.

\begin{align}
\rho_{ij} = \frac{\text{cov}(i, j)}{\sqrt{\sigma_i^2 \sigma_j^2}} = 
	\begin{cases}
	1 & \quad \text{if } i = j \\
	 \frac{N_2^{ij}}{\sqrt{(N_{1}^{i} +4\times N_{2}^{ii} + \sum_{q \neq i}N_{2}^{qi})\times (N_{1}^{j} +4\times N_{2}^{jj} + \sum_{q \neq j}N_{2}^{qj})}}& \quad \text{if } i \neq j \\
	\end{cases}
\label{eq:jetcorr}
\end{align}

The two jet correlation matrix among all jet \(p_T\) bins from the data are calculated as the following matrix.
\newpage
\input{chapter4_matrix}.

%

%
%
%
%

%% file: chapter4_rellum.tex

\section{Relative Luminosities for the RHIC 2012 Longitudinally Polarized $pp$ Run}

The relative luminosities account for different numbers of collisions for the four helicity combinations of the two beams, \(++\), \(+-\), \(-+\) and \(--\). Six of them are defined by the following Equations \ref{eq:r1} - \ref{eq:r6}. \(R_1\) and \(R_2\) are associated with longitudinal single-spin asymmetry measurements for the yellow and blue beams respectively. \(R_3\) is required to measure the inclusive jet \(A_{LL}\).

\begin{align}
R_{1} &= \frac{N^{++} +  N^{-+}}{N^{+-} + N^{--}} \label{eq:r1} \\
R_{2} &= \frac{N^{++} + N^{+-}}{N^{-+} + N^{--}} \label{eq:r2} \\
R_{3} &= \frac{N^{++} + N^{--}}{N^{+-} + N^{-+}} \label{eq:r3} \\
R_{4} &= \frac{N^{++}}{N^{--}} \label{eq:r4} \\
R_{5} &= \frac{N^{-+}}{N^{--}} \label{eq:r5} \\
R_{6} &= \frac{N^{+-}}{N^{--}} \label{eq:r6}
\end{align}

The relative luminosities are calculated on a run by run basis. The scaler boards are used to record numbers of events that produce signals in the STAR relative luminosity detectors BBC, ZDC and VPD. The scaler board is a VME module with histogramming functionality. It has 24 input bits. These 24 bits make up a 24-bit address which corresponds to one of the \(2^{24}\) memory locations. Each address has a 40 bit content. When the VME module receives a 24 bit input, it finds the memory location based on the 24 bin input address and then increments its content. The scaler board is operated under the RHIC bunch crossing frequency, also called the RHIC strobe. The bunch crossing frequency is about 9.38 MHz. At each RHIC bunch crossing, the scaler board receives input bits sent from the STAR trigger system that specify hits in the relative luminosity monitoring detectors. Seven of the 24 input bits are assigned to hold the RHIC bunch crossing ID from 0 to 119. During a certain period of run time, one scaler board is designed to collect detector responses for the relative luminosity calculation.

\subsection{Relative Luminosities}
A \(pp\) collision at very high energy produces a large number of final particles. The majority of the produced particles tend to be closer to the beam line. Luminosity detectors are therefore installed near the beam line at both sides of the collision center. Sitting near the beam line they sample a different mix of physics processes rather than the physics that generates the jet in the mid-rapidity region. A single hit detected on one side of the collision center or two simultaneous hits detected on both sides, also known as coincidence hits, can signal a real collision. At STAR, the two sides are defined by their geometrical locations, east and west. Three binary bits are used to flag the east hit, west hit and coincident hits. These bits are a part of the 24 input bits sent to the scaler board at every bunch crossing.

A collision can produce a single hit on one side of the detector, or coincident hits on both sides of the detector. Under perfect conditions a hit implies a real collision. However in reality the two simultaneous hits detected on both sides of the detector can be two individual collisions that produce single hits that hit both sides. These types of hits are classified as random coincidences. As the performance of the accelerator has enhanced over the past decades, collision rates can achieve a very high level such as the collision rates in the 2012 RHIC \(pp\) run at \(\sqrt{s}=\) 510 GeV. It is possible at each bunch crossing, there are multiple collisions happening at the very short amount of time and the detector only records hits from one of them. In this case, multiple collisions could be disguised as one single collision. Therefore two corrections are applied to the relative luminosity calculation, one is the accidental correction and the other is the multiple correction \cite{rellumcdf2000}. In other words, the accidental correction corrects for over-counting and the multiple correction corrects for under-counting.

The random corrections are made in such way. Assume there are three independent probabilities \(P_{A}\), \(P_{B}\), and \(P_{C}\) for processes where a collision produces an east single hit, a collision produces a west single hit, and a collision produces a coincident hit. Then the probabilities to observe a hit on the east side of the detector, a hit on the west side of the detector and two simultaneous hits on both side of the detector, \(P_E\), \(P_W\), and \(P_{EW}\) respectively, can be expressed as the following equations:
\begin{align}
P_{E} &= P_{A} + P_{C} - P_{A}P_{C}, \\
P_{W} &= P_{B} + P_{C} - P_{B}P_{C}, \\
P_{EW} &= P_{C} + P_{A}P_{B} - P_{A}P_{B}P_{C}.
\end{align}
The \(P_{A}\), \(P_{B}\), and \(P_{C}\) can be solved as shown in the following equations:
\begin{align}
P_{A} &= \frac {P_{E}-P_{EW}} {1 - P_{W}}, \label{eq:pa}\\
P_{B} &= \frac {P_{W}-P_{EW}} {1 - P_{E}}, \label{eq:pb}\\
P_{C} &= \frac {P_{EW}-P_{E}P_{W}} {1 +  P_{EW} - P_{E} - P_{W}}. \label{eq:pc}
\end{align}
They will be used in the relative luminosity calculation.

The multiple correction is made by assuming the number of collisions that happened during a bunch crossing obey Poisson distributions. The average number of collisions per bunch crossing, \(\mu\), is used to estimate the real number of collisions that happened at a bunch crossing. The probability of \(k\) number of collisions during a bunch crossing is \(P_{k} = \frac{e^{-\mu}\mu^{k}}{k!}\), where \(\mu\) is the average number of collisions. Therefore the probability that at least one collision happened is \(P = 1 - P_{0} = 1 - e^{-\mu}\) and \(\mu = -\ln(1-P)\). By plugging the probabilities calculated in equations \eqref{eq:pa} \eqref{eq:pb} and \eqref{eq:pc}, the average number of collisions per bunch crossing can be calculated for the single hits and the coincident hits. The total number of collisions at a particular bunch crossing during a period can then be calculated as \(N=N_{BC} \times \mu\), where \(N_{BC}\) is the total number of bunch crossings during the period. The number of collisions that happened for the singles and coincidences \(N_{A}\), \(N_{B}\), and \(N_{C}\), can be expressed as the following equations, where \(N_{E}\), \(N_{W}\), \(N_{EW}\), and \(N_{BC}\) are the number of bunch crossings observed for east single hits, west single hits, coincident hits and the total bunch crossing number.

\begin{align}
N_{A} &= - N_{BC} \times \ln( \frac{N_{E} - N_{EW}}{N_{BC} - N_{W}} ) \label{eq:na} \\ 
N_{B} &= - N_{BC} \times \ln( \frac{N_{W} - N_{EW}}{N_{BC} - N_{E}} ) \label{eq:nb} \\
N_{C} &= - N_{BC} \times \ln( \frac{N_{EW} - \frac{N_{E}N_{W}}{N_{BC}}}{N_{BC} + N_{EW} - N_{E} - N_{W}} ) \label{eq:nc}
\end{align}

\subsection{Counting East and West Singles Hits and Coincidence Hits}
There are three bits, east ADC sum greater than its threshold, west ADC sum greater than its threshold and TAC difference within a certain window, as a part of the 24 scaler input bits for three relative monitoring detectors, BBC, ZDC and VPD. For BBC and VPD, both have 16 PMT channels read out to the trigger system at the east and west sides (where three of 19 VPD tiles are not read out). Each PMT has one 12 bit ADC value and a 12 bit TAC value. Two QT boards hold all 16 PMT ADC and TAC values for the east and west sides. The trigger system receives information from the two boards at level 0 and calculates the sum ADC and maximal TAC value for each board. The maximal TAC is corresponding to the earliest hit to the detector. During the calculation, only PMT channels that have ADC greater than the threshold and TAC within a certain range are considered. The outputs of these two boards, two sets of a 16 bit ADC sum and a 12 bit maximal TAC, are sent to level 1. At level 1 the two ADCs are compared to the ADC sum thresholds. The TAC difference is calculated and checked if the value is within a certain TAC window. The TAC difference is calculated as 4096 + east TAC – west TAC to guarantee its value is positive. If the east ADC sum is greater than its sum ADC threshold, the east ADC sum bit is set to 1 and sent to the scaler system. So does the west ADC sum bit. If the TAC difference is greater than its lower limit and less than its upper limit, then the TAC difference bit is set to 1 and sent to the scaler system.

For the ZDC there are three PMT channels corresponding to three ZDC modules at the east and west sides. Each PMT also has a 12 bit ADC value and a 12 bit TAC value. Different from BBC and VPD, the information from both sides of the detector is sent to one QT board at level 0 in the trigger system. The ADC sums from the front, middle and back module at both sides are calculated and are compared with ADC sum thresholds. If the ADC sum is greater than its threshold, the output bit is set to 1. The leftmost 10 bits of the TAC values from the front module at both sides are sent to the output. During the ADC sum calculation and output of TAC values, only those PMT channels with ADC values greater than a threshold and TAC value within a certain window are included. The output of the QT board is sent to the level 1 to calculate the TAC difference which is defined as 1024 \(+\) east TAC \(-\) west TAC. If the TAC difference is within a certain window, the TAC difference is set to 1 and sent to the scaler system. The ADC sum threshold bits are passed through the level 1 straight to the scaler system.

The following table \ref{tab:tacadcthresholds} shows the nominal thresholds for PMT ADC thresholds and PMT TAC limits, ADC sum thresholds and TAC difference limits for BBC, ZDC and VPD during the 2012 510 GeV run.

\begin{table}[H]
\centering
\scalebox{0.8}{
\begin{tabular}{|c|c|c|c|c|c|c|}
\hline
 \multirow{2}{*}{} & \multicolumn{2}{c|}{BBC} & \multicolumn{2}{c|}{ZDC} & \multicolumn{2}{c|}{VPD}\\
\cline{2-7}
& east & west & east & west & east & west \\
\hline
PMT ADC & 5 & 20 & 25 & 25 & 10 & 10 \\
\hline
PMT TAC & (100, 2300) & (100, 2300) & (100, 3000) & (100, 3000) & (100, 3000) & (100, 3000) \\
\hline
ADC Sum & 20 & 20 & 25 & 25 & 10 & 10\\
\hline
TAC Diff & \multicolumn{2}{c|}{(3267, 4933)} & \multicolumn{2}{c|}{(50, 1300)} & \multicolumn{2}{c|}{(3883, 4083)}\\
\hline
\end{tabular}
}
\caption{The nominal TAC and ADC thresholds in the trigger system for the east and west side of BBC, ZDC and VPD. The fact that the BBC east PMT ADC and ADC sum thresholds differed was a configuration error that was found during this analysis.}
\label{tab:tacadcthresholds}
\end{table}
To be consistent with what is defined in Equations \eqref{eq:na}, \eqref{eq:nb}, and \eqref{eq:nc}, the number of observed bunch crossings for the east single hits, the west single hits, and the coincident hits \(N_{E}\), \(N_{W}\) and \(N_{EW}\), are defined in the following way:
\begin{align}
N_{E} &= C(1, 0, 0) + C(1, 1, 1) + C(1, 0, 1) + C(1, 1, 0), \label{eq:neast}\\
N_{W} &= C(0, 1, 0) + C(1, 1, 1) + C(0, 1, 1) + C(1, 1, 0), \label{eq:nwest}\\
N_{EW} &= C(1, 1, 1) + C(1, 1, 0), \label{eq:nx}
\end{align}
where \(C(E,W,X)\) is the content of the scaler board corresponding to the east ADC sum bit (E), the west ADC sum bit (W) and the TAC difference bit (X) of the scaler input. The net effect of this definition is to disregard the TAC difference bit. There are other definitions for \(N_{E}\), \(N_{W}\) and \(N_{EW}\), in the previous relative luminosity studies at STAR \cite{rellum2009}, however this definition is the most internally consistent with the random correction. It is also worthy to note that the TAC difference bit is discussed only for the purpose to compare with other definitions used in the previous studies.

\subsection{Bunch Crossing Distributions}
\begin{figure}[H]
\centering
\includegraphics[scale=0.45]{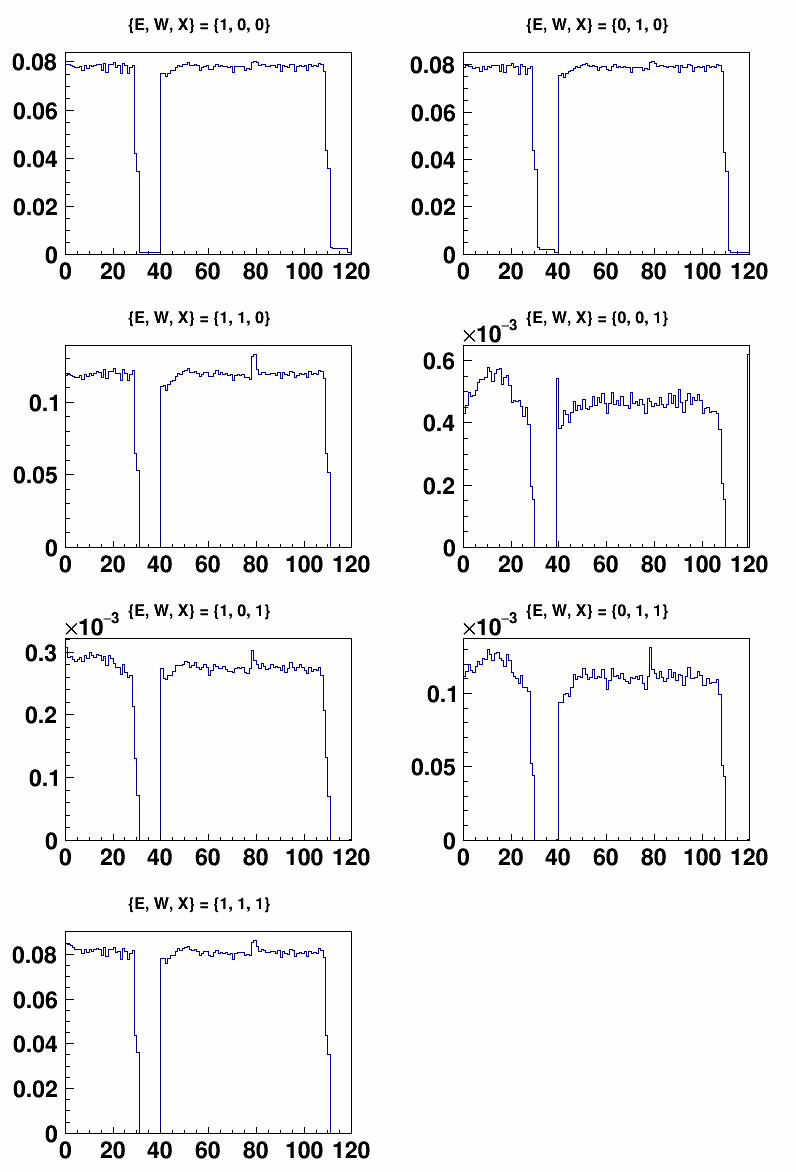}
\caption{Probabilities to find the various hit combinations from the VPD.}
\label{fig:vpdcmb}
\end{figure}

The Figure \ref{fig:vpdcmb} shows the bunch crossing distributions for all possible combinations of the three bits from the VPD. The bunch crossing is numbered from 0 to 119. The data used here cover all the good candidate runs from 2012 510 GeV longitudinal \(pp\) collisions.

In Figure \ref{fig:vpdcmb}, the two abort gaps can be seen clearly, bunch 31 to 39 is the yellow abort gap and bunch 111 to 119 is the blue abort gap. The two empty bunches in blue beam, 38 and 39, overlap with the yellow abort gap and the two empty bunches in the yellow beam 78 and 79 overlap with the blue abort gaps. Normally each bunch crossing has very similar beam intensity, so all the bunch crossing distributions should be more or less uniform with small fluctuations except the two abort gaps. However a few bunches right after the two abort gaps show a climbing effect and the possible reasons for this effect could be a portion of a previous bunch leaking through to the next bunch, a ringing effect in the detectors and the likes. In this analysis, this effect is corrected by removing the first a few bunches right after the two aborts. Bunches 78 and 79 systematically have higher  counts relative to other nearby bunches. The reason for this is that blue beam bunches 78 and 79  and their colliding partners yellow beam bunches 38 and 39 only collide once at STAR not at PHENIX. At PHENIX the blue beam bunches 78 and 79 meet with the empty yellow beam bunches 78 and 79, and the yellow beam bunches 38 and 39 meet with the empty blue beam bunches 38 and 39. All the other blue beam bunches collide with yellow beam bunches at both STAR and PHENIX. In this analysis, for certain runs bunch crossing 78 and 79 are removed only if they have severely larger counts than the other normal bunches.

\begin{figure}[H]
\centering
\includegraphics[scale=0.45]{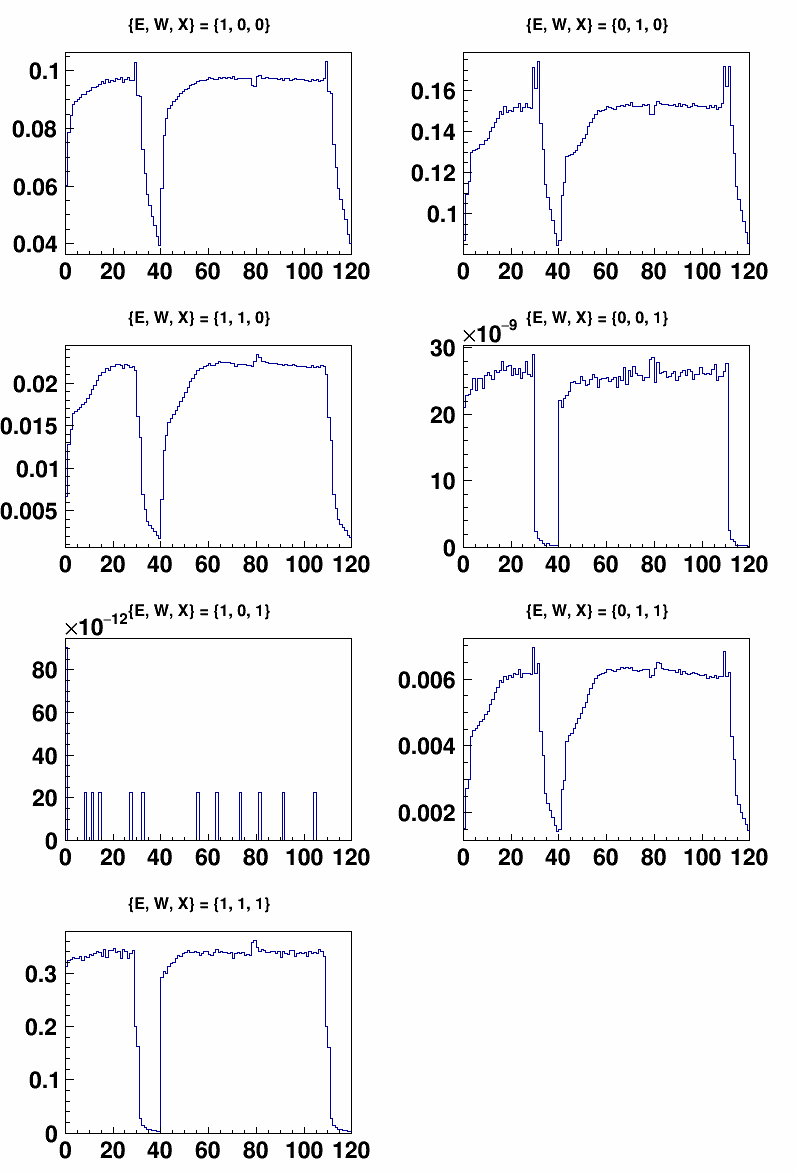}
\caption{Probabilities to find the various hit combinations from the BBC.}
\label{fig:bbccmb}
\end{figure}

In Figures \ref{fig:bbccmb}, the combinations (\({0, 0, 1}\)), (\({1, 0, 1}\)), or (\({0, 1, 1}\)) do not seem logical, however under some circumstances they in fact happen. For the BBC frequencies of illegal combinations (\({0, 0, 1}\)), and (\({1, 0, 1}\)) is at the level of \(10^{-8}\) or below and therefore they are negligible. For the combination (\({0, 1, 1}\)), this is in fact allowed in the scaler system. The BBC individual channel ADC threshold at level 0 is different than the sum ADC threshold at level 1, as seen in Table \ref{tab:tacadcthresholds}. At level 0 only if a good hit requirement is satisfied, where a channel ADC is greater than a threshold and a channel TAC is within a limit, the corresponding TAC value is passed to level 1. At the level 1, the TAC values from both sides could produce a good TAC difference. However if the summed ADC is less than the sum ADC threshold, then in this case the system would produce the illogical combination. In general the sum ADC threshold is set to the same value as the channel threshold, therefore a valid TAC value from level 0 would guarantee a pass to the sum ADC threshold, and there would not be illegal combinations. Since at the level 0 and the level 1, the BBC individual channel thresholds are the same. The combination (\({1, 0, 1}\)) happens very rarely. 

\begin{figure}[H]
\centering
\includegraphics[scale=0.45]{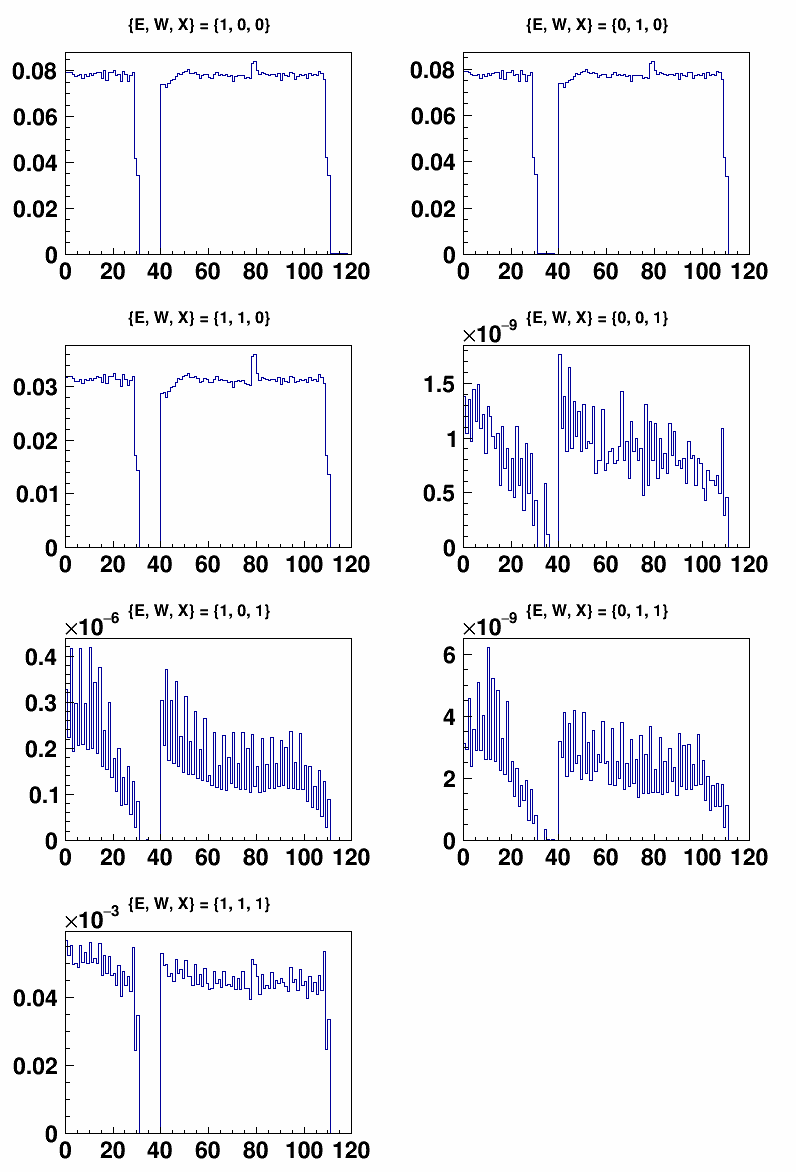}
\caption{Probabilities to find the various hit combinations from the ZDC.}
\label{fig:zdccmb}
\end{figure}

For the ZDC, all thresholds are set at the identical value at both level 0 and level 1. The three illegal combinations mentioned aove are negligible as seen in Figure \ref{fig:zdccmb}. However the frequency of the combination, (\({1, 1, 1}\)), is around three order of magnitude less than that of the other three logical combinations (\({1, 0, 0}\)), (\({0, 1, 0}\)), and (\({1, 1, 0}\)). The cause for this problem is not understood, it could be due to hardware or parameter setting errors during the data taking. In this analysis, the ZDC coincidence event is not used for the relative luminosity calculation.

For the VPD, as seen in Figure \ref{fig:vpdcmb} all three illegal combinations the combinations (\({1, 0, 0}\)), (\({1, 1, 0}\)), and (\({1, 0, 1}\)) happen at less frequencies than the logical combinations.The frequencies are similar, and down around two orders of magnitude. Indications are that the VPD TAC difference bit was not precisely synchronized with the bunch crossing clock. To minimize the side effect of the TAC bit, this bit is dropped while counting the east and west single hits and coincidence hits.

\begin{figure}[H]
\centering
\includegraphics[scale=0.5]{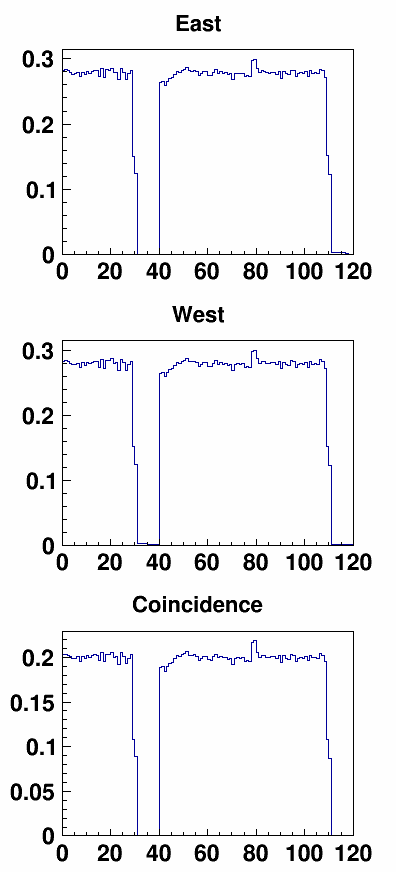}
\caption{Bunch crossing distributions for the defined VPD east, west and coincidence hits.}
\label{fig:vpdhit}
\end{figure}

Figure \ref{fig:vpdhit} shows the bunch crossing distribution for the east single hits, the west single hits, and the coincidence hits. There hits are added up according to equation \eqref{eq:neast} \eqref{eq:nwest} and \eqref{eq:nx}. Similar features have shown up as the previous bunch crossing distribution for all the possible bit combinations.

\subsection{Application of Accidental and Multiple Corrections}
\begin{figure}[H]
\centering
\includegraphics[scale=0.5]{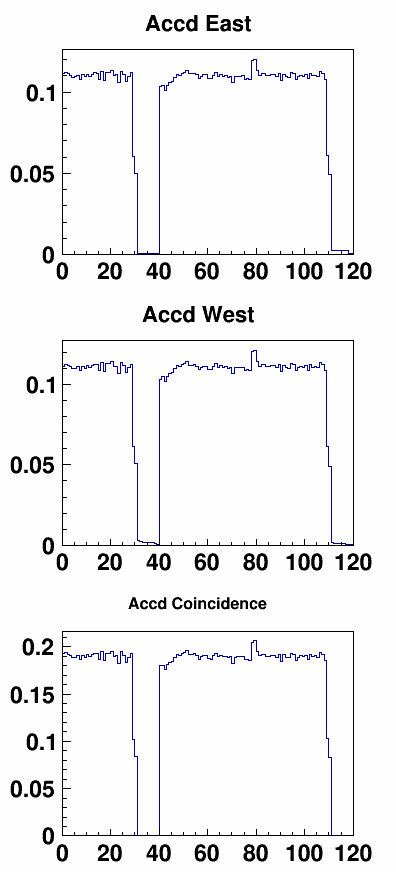}
\caption{Bunch crossing distributions for the defined VPD east, west and coincidence hits after accidental corrections.}
\label{fig:vpdcmbaccd}
\end{figure}

\begin{figure}[H]
\centering
\includegraphics[scale=0.5]{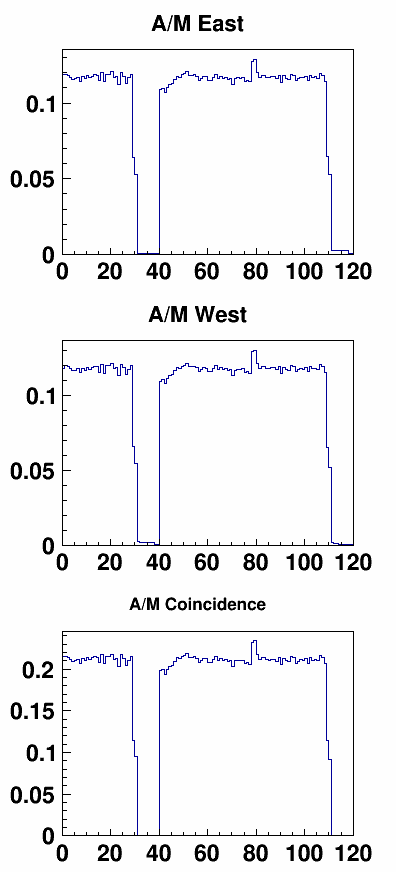}
\caption{Bunch crossing distributions for the defined VPD east, west and coincidence hits after accidental and multiple corrections.}
\label{fig:vpdcmbmult}
\end{figure}

The accidental and multiple corrections are applied to the east single hits, the west single hits,  and the coincident hits on a basis of bunch by bunch. Figures \ref{fig:vpdcmbaccd} and \ref{fig:vpdcmbmult} show the bunch crossing distributions for the east and west singles and coincidences while applying the accidental and multiple correction step by step. It is easy to see that the accidental correction removes the east-west coincident events from the observed single hits to identify the real single events. Also the multiple correction corrects for the under-counting due to the high collision rate and the finite detector response. After accidental and multiple corrections the shape of the distributions become more uniformly distributed except the abort gaps.


%

\subsection{Relative Luminosity Calculations and Their Uncertainties}
Given the number of the accidental and multiple corrected single and coincidence hits at each bunch crossing and the spin configuration of bunch crossing, the relative luminosities can be calculated by using the single and coincidence hits from equations \eqref{eq:r1} - \eqref{eq:r6} on a run-by-run bases. 

%

%
One way to characterize the quality of relative luminosity calculation is to compare the results calculated by single hits with those by coincidence hits. Since the hits in BBC, ZDC and VPD are minimum biased by physics scattering, the number of hits detected in those detectors has little dependency on the spin orientations of colliding bunches and therefore the difference is expected to be very small, ideally zero. In other words, \(\Delta\) is defined as the difference between \(R_c\) and \(R_s\), the relative luminosities observed by coincidence and single hits, normalized by the sum. The less spin dependency observed by the detector, the smaller \(\Delta\) and closer between \(R_{c}\) and \(R_{s}\) should be. By comparing among BBC, ZDC and VPD, VPD shows the smallest difference between \(R_{c}\) and \(R_{s}\), so VPD is chosen to be used to calculate the relative luminosities in this analysis. The coincident hits are chosen over the single hits to calculate the relative luminosities, because coincidence hits are less sensitive to backgrounds and more reliable to represent real collisions.

\begin{align}
\Delta 
&= R_{c} - R_{s} \label{eq:delta}
\end{align}

As analogous to Equation \eqref{eq:delta}, \(\Delta\) can be defined to reflect the relative luminosity difference among detectors such as \(\Delta ^{BBC,VPD}\), \(\Delta^{ZDC,VPD}\), and \(\Delta^{BBC,ZDC}\). The three detectors are sensitive to different physics processes in different kinematic regions. In the absence of systematic difference \(\Delta^{BBC,VPD}\),\(\Delta^{ZDC,VPD}\), and \(\Delta^{BBC,ZDC}\) are expected to be small. 

%
\begin{align}
\Delta^{BBC,VPD} &= R^{BBC} - R^{VPD} \\
\Delta^{ZDC,VPD} &= R^{ZDC} - R^{VPD} \\
\Delta^{BBC,ZDC} &= R^{BBC} - R^{ZDC}
\end{align}

%

%
The difference of relative luminosities calculated by BBC and ZDC compared to those by VPD is also reduced by removing several bad bunches on a fill by fill basis. Bad bunches include the gradually climbing bunches right after the two abort gaps and abnormal counts at certain bunch crossings. The bad bunches are identified by finding bunches that have too large or too small counts than the average bunch counts for all three detectors.

\begin{figure}[H]
\centering
\includegraphics[scale=0.5]{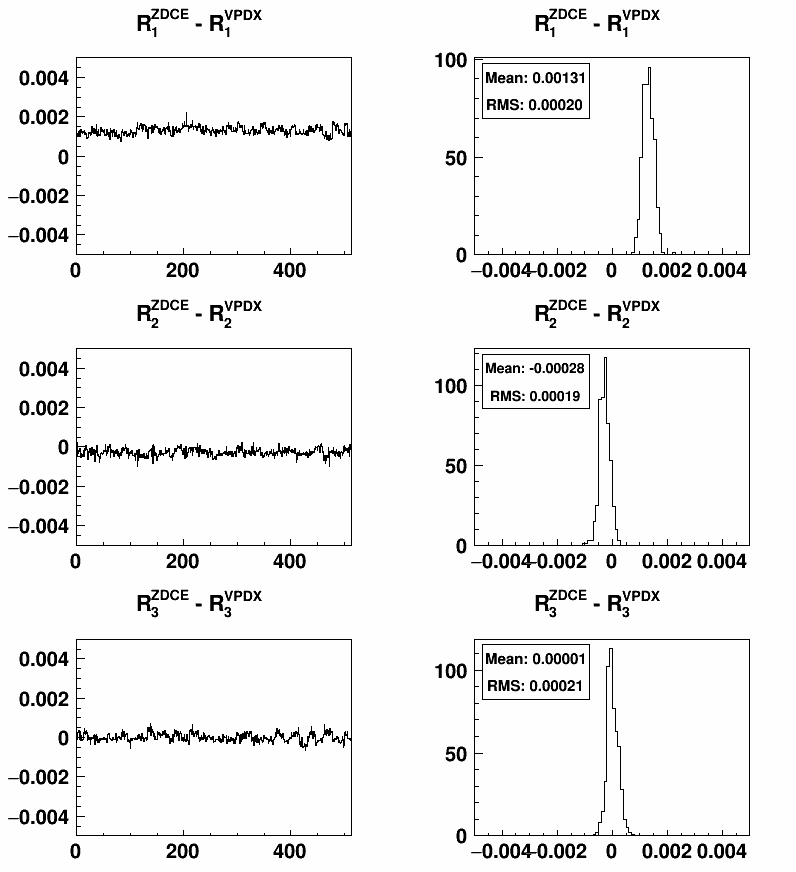}
\caption{$\Delta R_{1,2,3}$ calculated by ZDC east hits and VPD coincidence hits vs. run index (left) and the associated distributions (right).}
\label{fig:deltaRzdceVvpdx}
\end{figure}

\begin{figure}[H]
\centering
\includegraphics[scale=0.5]{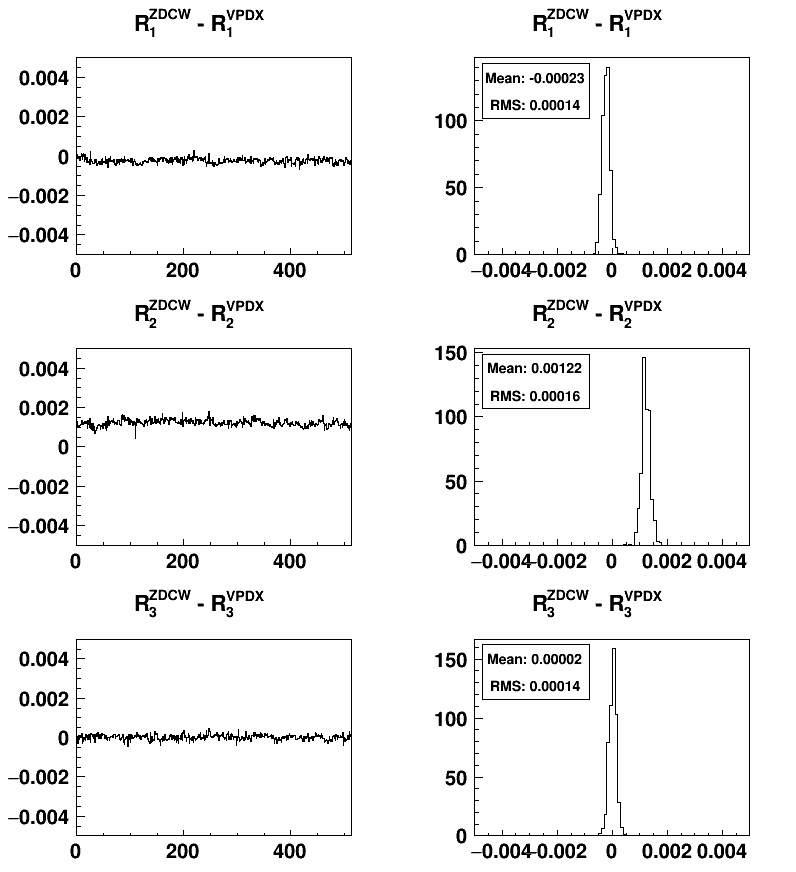}
\caption{$\Delta R_{1,2,3}$ calculated by ZDC west hits and VPD coincidence hits vs. run index (left) and the associated distributions (right).}
\label{fig:deltaRzdcwVvpdx}
\end{figure}

\begin{figure}[H]
\centering
\includegraphics[scale=0.5]{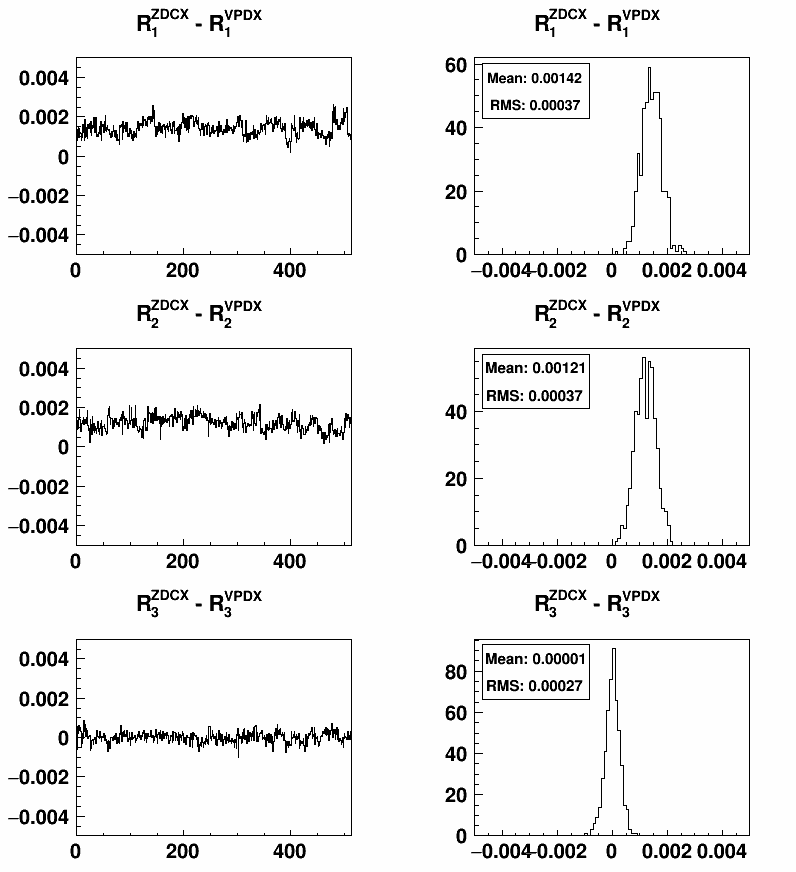}
\caption{$\Delta R_{1, 2, 3}$ calculated by ZDC coincidence hits and VPD coincidence hits vs. run index (left) and the associated distributions (right).}
\label{fig:deltaRzdcxVvpdx}
\end{figure}

\begin{figure}[H]
\centering
\includegraphics[scale=0.5]{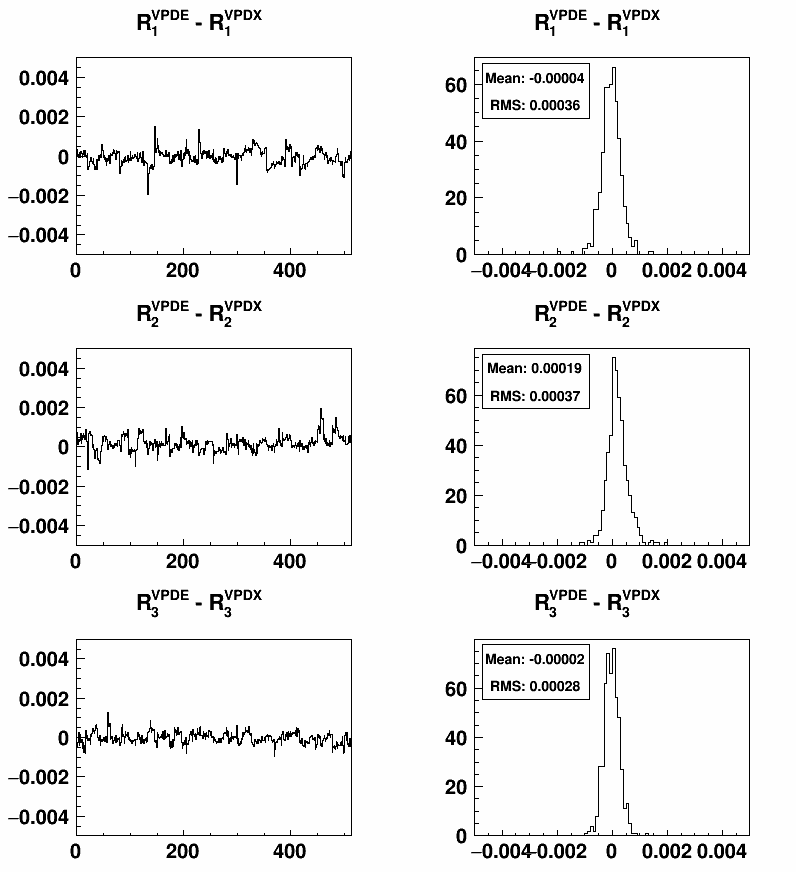}
\caption{$\Delta R_{1,2,3}$ calculated by VPD east hits and VPD coincidence hits vs. run index (left) and the associated distributions (right).}
\label{fig:deltaRvpdeVvpdx}
\end{figure}

\begin{figure}[H]
\centering
\includegraphics[scale=0.5]{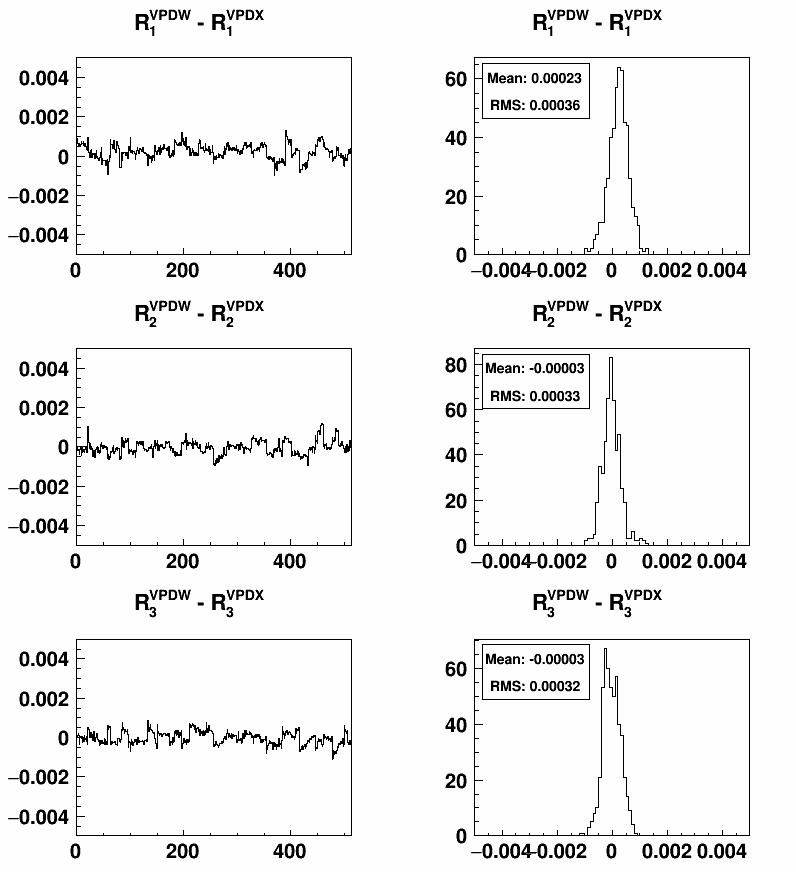}
\caption{$\Delta R_{1,2,3}$ calculated by VPD west hits and VPD coincidence hits vs. run index (left) and the associated distributions (right).}
\label{fig:deltaRvpdwVvpdx}
\end{figure}

The relative luminosity differences among three detectors on a run-by-run basis are calculated after removing all the bad bunches.  Figure 
\ref{fig:deltaRzdceVvpdx}, \ref{fig:deltaRzdcwVvpdx} \ref{fig:deltaRzdcxVvpdx} \ref{fig:deltaRvpdeVvpdx} and \ref{fig:deltaRvpdwVvpdx}, show the relative luminosity difference calculated among ZDC and VPD single and coincidence hits. The means and RMSs of the relative luminosity differences are also calculated.Table \ref{tab:allDeltaRs}  summarizes the means and RMSs of \(R_{1}\) to \(R_{6}\) between ZDC and VPD and between VPD singles and VPD coincidences.

\begin{figure}[H]
\centering
\includegraphics[scale=0.8]{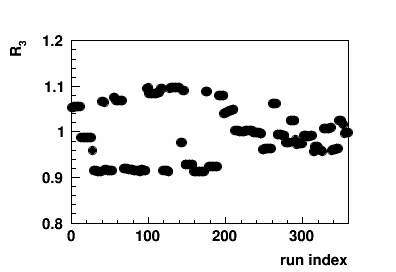}
\caption{$R_3$ vs. run number where runs in this plot are those run selected in this analysis.}
\label{fig:run12r3}
\end{figure}

Since single events and coincidence events are triggered by different physics processes, to maximize the usage of all the measurements, the five \(\Delta R\)s are grouped into three category based on the underlying physics processes. They are difference between ZDC singles and VPD coincidences, difference between VPD singles and VPD coincidences, and difference between ZDC coincidences and VPD coincidence. Within each group, the grouped mean and RMS of \(\Delta R\) is the linear average of each measurement. For the final mean and RMS of \(\Delta R\), the weighed averages of the grouped mean and RMS are taken. For \(R_3\) the weighed mean RMS, \(0.00013\),  is taken as the systematic uncertainty.
Figure \ref{fig:run12r3} shows the relative luminosity \(R_{3}\) for the final selected runs. The impact of the relative luminosity uncertainty on the uncertainty of inclusive jet \(A_{LL}\) is discussed in the next section.

\begin{table}[H]
\centering

\rotatebox{90}{
\begin{varwidth}{\textheight}
\scalebox{0.64}{
\begin{tabular}{|c|c|c|c|c|c|c|c|c|c|}
\hline
 R & ZDCE$-$VPDX & ZDCW$-$VPDX & Linear average & VPDE$-$VPDX & VPDW$-$VPDX & Linear average & ZDCX$-$VPDX & Linear average & Weighted average \\
\hline
$R_1$ & 0.00131$\pm$0.00020 & -0.00023$\pm$0.00014 & 0.00054$\pm$0.00017 & -0.00005$\pm$0.00036 & 0.00023$\pm$0.00036 & 0.00009$\pm$0.00036 & 0.00142$\pm$0.00037 & 0.00142$\pm$0.00037 & 0.00060$\pm$0.00014 \\ \hline
$R_2$ & -0.00028$\pm$0.00019 & 0.00122$\pm$0.00016 & 0.00047$\pm$0.00018 & 0.00019$\pm$0.00037 & -0.00003$\pm$0.00033 & 0.00008$\pm$0.00035 & 0.00121$\pm$0.00037 & 0.00121$\pm$0.00037 & 0.00052$\pm$0.00014 \\ \hline
$R_3$ & 0.00001$\pm$0.00021 & 0.00002$\pm$0.00013 & 0.00001$\pm$0.00017 & -0.00002$\pm$0.00028 & -0.00003$\pm$0.00032 & -0.00003$\pm$0.00030 & 0.00001$\pm$0.00027 & 0.00001$\pm$0.00027 & 0.00001$\pm$0.00013 \\ \hline
$R_4$ & 0.00104$\pm$0.00032 & 0.00099$\pm$0.00021 & 0.00102$\pm$0.00026 & 0.00015$\pm$0.00055 & 0.00020$\pm$0.00060 & 0.00017$\pm$0.00058 & 0.00264$\pm$0.00056 & 0.00264$\pm$0.00056 & 0.00114$\pm$0.00022 \\ \hline
$R_5$ & 0.00132$\pm$0.00031 & -0.00025$\pm$0.00019 & 0.00053$\pm$0.00025 & -0.00001$\pm$0.00047 & 0.00028$\pm$0.00046 & 0.00014$\pm$0.00046 & 0.00143$\pm$0.00046 & 0.00143$\pm$0.00046 & 0.00063$\pm$0.00020 \\ \hline
$R_6$ & -0.00028$\pm$0.00026 & 0.00121$\pm$0.00020 & 0.00047$\pm$0.00023 & 0.00023$\pm$0.00051 & 0.00002$\pm$0.00047 & 0.00012$\pm$0.00049 & 0.00121$\pm$0.00045 & 0.00121$\pm$0.00045 & 0.00054$\pm$0.00019 \\ \hline
\end{tabular}
}
\caption{Mean and RMS of $\Delta$R among the ZDC and VPD and the weighted average of the linear averages.} \label{tab:allDeltaRs}
\end{varwidth}
}

\end{table}

%% file: chapter4_matrix.tex
\begin{align}
\left (
\scalebox{0.6}{\mbox{\ensuremath{\displaystyle
\arraycolsep = 1.2pt
\begin{array}{@{}*{14}{c}@{}}
1 & 0.0091 & 0.0055 & 0.0047 & 0.0039 & 0.0031 & 0.0018 & 0.0011 & 0.00058 & 0.00033 & 0.00018 & 8.1\times10^{-5} & 3.7\times10^{-5} & 1.2\times10^{-5} \\
0.0091 & 1 & 0.0062 & 0.0057 & 0.0051 & 0.0041 & 0.0025 & 0.0015 & 0.00083 & 0.00046 & 0.00024 & 0.00012 & 4.6\times10^{-5} & 1.6\times10^{-5} \\
0.0055 & 0.0062 & 1 & 0.0094 & 0.0097 & 0.0092 & 0.0064 & 0.0042 & 0.0027 & 0.0015 & 0.00081 & 0.00045 & 0.0002 & 9.5\times10^{-5} \\
0.0047 & 0.0057 & 0.0094 & 1 & 0.012 & 0.012 & 0.0089 & 0.0062 & 0.0041 & 0.0023 & 0.0013 & 0.00065 & 0.00025 & 0.00012 \\
0.0039 & 0.0051 & 0.0097 & 0.012 & 1 & 0.016 & 0.012 & 0.0091 & 0.0062 & 0.0037 & 0.0021 & 0.0011 & 0.00043 & 0.00021 \\
0.0031 & 0.0041 & 0.0092 & 0.012 & 0.016 & 1 & 0.016 & 0.013 & 0.0093 & 0.0059 & 0.0034 & 0.0018 & 0.0007 & 0.00037 \\
0.0018 & 0.0025 & 0.0064 & 0.0089 & 0.012 & 0.016 & 1 & 0.017 & 0.015 & 0.012 & 0.0083 & 0.0053 & 0.0029 & 0.0016 \\
0.0011 & 0.0015 & 0.0042 & 0.0062 & 0.0091 & 0.013 & 0.017 & 1 & 0.02 & 0.018 & 0.014 & 0.0093 & 0.0053 & 0.0031 \\
0.00058 & 0.00083 & 0.0027 & 0.0041 & 0.0062 & 0.0093 & 0.015 & 0.02 & 1 & 0.024 & 0.021 & 0.016 & 0.0097 & 0.0055 \\
0.00033 & 0.00046 & 0.0015 & 0.0023 & 0.0037 & 0.0059 & 0.012 & 0.018 & 0.024 & 1 & 0.03 & 0.025 & 0.017 & 0.0099 \\
0.00018 & 0.00024 & 0.00081 & 0.0013 & 0.0021 & 0.0034 & 0.0083 & 0.014 & 0.021 & 0.03 & 1 & 0.035 & 0.027 & 0.018 \\
8.1\times10^{-5} & 0.00012 & 0.00045 & 0.00065 & 0.0011 & 0.0018 & 0.0053 & 0.0093 & 0.016 & 0.025 & 0.035 & 1 & 0.039 & 0.029 \\
3.7\times10^{-5} & 4.6\times10^{-5} & 0.0002 & 0.00025 & 0.00043 & 0.0007 & 0.0029 & 0.0053 & 0.0097 & 0.017 & 0.027 & 0.039 & 1 & 0.042 \\
1.2\times10^{-5} & 1.6\times10^{-5} & 9.5\times10^{-5} & 0.00012 & 0.00021 & 0.00037 & 0.0016 & 0.0031 & 0.0055 & 0.0099 & 0.018 & 0.029 & 0.042 & 1 
\\
\end{array}
}}}
\right ) \nonumber
\end{align}

%% file: chapter5.tex
\chapter{\uppercase{Monte Carlo Simulation}}
To study the systematics due to the hadronization process and the detector response, a Monte Carlo (M/C) sample is necessary to quantify the possible distortion. PYTHIA \cite{pythia2006} is used to simulate the proton-proton collisions including parton scatterings, hadronization processes, and decays from unstable particles. The final stable particles are then fed into GEANT \cite{geant4} to simulate the detector response. The detector responses are then mixed with zero-bias events collected during the run. The mixed responses are then used to compare with what was measured in the data. This process is called the embedding process, and the final produced sample is called the embedding sample. The reason to choose zero-bias events is that those events have no trigger requirement and were randomly taken, therefore they are a good approximation of the collision backgrounds recorded during data-taking.

\section{Event Generator and Tunes}
PYTHIA 6.4.28 is the event generator used in the M/C simulation. The \(2 \rightarrow 2\) hard QCD jet process is turned on along with initial state radiation, final state radiation, beam remnants and underlying event activities.Table \ref{tab:qcdjets} lists the subprocesses generated by PYTHIA. The PYTHIA QCD jet calculation is different from the NLO jet calculation in a sense that the existence of initial state and final state radiations mimic an all-order QCD calculation of jet production. In this analysis, the subprocesses are grouped only by the incoming hard-scattered partons, therefore it ends up with three subprocesses \(qq\), \(qg\) and \(gg\).

\begin{table}[H]
\centering
\begin{tabular}{|c|c|}
\hline
PYTHIA ID & subprocess \\
\hline
11 & $q_i q_j  \rightarrow q_i q_j $\\
\hline
12 & $q_i \bar{q}_i \rightarrow q_k \bar{q}_k $\\
\hline
13 & $q_i \bar{q}_i \rightarrow g g $\\
\hline
28 & $q_i g \rightarrow q_i g $\\
\hline
53 & $g g \rightarrow q_k \bar{q}_k $\\
\hline
68 & $g g \rightarrow g g$\\
\hline
96 & semihard QCD 2 $\rightarrow 2$ \\
\hline
\end{tabular}
\caption{QCD jet processes generated by PYTHIA by setting $MSEL = 1$.}
\label{tab:qcdjets}
\end{table}

The parameters to control all those processes are chosen from the Perugia 2012 tune \cite{perugia2010}. The Perugia 2012 tune parameters are from \(e^+e^-\) annihilation, DIS, Tevatron and LHC data. The PDF used in this tune is CTEQ6L1 \cite{cteq6l2002}. However by comparing with previous published STAR \(\pi ^{\pm}\) data at \(\sqrt{s} = 200\) GeV \cite{starpipm2006, starpipm2012}, it overestimates the underlying event contribution. Therefore the exponential parameter that controls the underlying event behavior has been further modified to match the published STAR \(\pi^{\pm}\) data as described below.

The final stable particles generated from PYTHIA are reconstructed by the same jet finding algorithm used in the data analysis. These clustered final particles are called particle jets. To study the effects of hadronization, jets are also reconstructed at the parton level. At the parton level, only the hard-scattered partons plus those from initial state radiation and final state radiation, are fed into the jet finder. Partons from beam remnants and underlying events are excluded. This formulation approximates jets in a NLO calculation roughly. However this is certainly not an exact correlation, so some difference between NLO jets and PYTHIA parton jets are expected.

The \(p_{T,0}\) in PYTHIA is introduced to avoid the divergence of the jet cross-section for low-\(p_T^2\) scatterings, since the jet cross-section is roughly proportional to the \(\frac{1}{p_T^4}\). PYTHIA makes the jet cross-section proportional to \(\frac{1}{(p_T^2 + p_{T,0}^2)^2}\). Since the underlying event involves a large number of parton scatterings with low \(p_T\), the larger the \(p_{T,0}\), the less the partonic cross-section. Ultimately it leads to smaller underlying events. The value of the \(p_{T,0}\) depends on the center-of-mass energy of the collision, as expressed in equation \ref{eq:pT0}. For the Perugia 2012 tune, \(\sqrt{s_{ref}}\) is 7000 GeV, \(p_{T,ref}\) is 2.65 GeV and \(P_{90}\) is 0.24. By reducing the exponent \(P_{90}\) from 0.24 to 0.213, the \(p_{T,0}\) is increased to 1.51 GeV from 1.41 GeV at 510 GeV, while leaving the underlying event unchanged at 7000 GeV where Perugia 2012 was tuned.

\begin{align}
p_{T,0}(s) = p_{T,ref} \times (\frac{\sqrt{s}}{\sqrt{s_{ref}}})^{P_{90}} \label{eq:pT0}
\end{align}

The smaller exponent value generates less underlying event contributions, which also improves the matching probabilities from particle jet to parton jet. A particle jet is matched to a parton jet only if the smallest distance from the particle jet to all the parton jets is less than 0.5, min(\(\Delta R = \sqrt{\Delta \phi ^2 + \Delta \eta ^2}\) ) \( < \) 0.5 where \( \Delta \phi = \phi_{particle} - \phi_{parton}\) and  \( \Delta \eta = \eta_{particle} - \eta_{parton}\).

An event weighting technique is also applied in this simulation. Since the parton scattering cross-section decreases approximately exponentially as a function of momentum transfer \(Q\), also known as partonic \(p_T\), and jet \(p_T\) is proportional to \(Q\), the high \(p_T\) jet yields are substantially smaller than the low jet \(p_T\) yields. In the data, due to high cross-sections at low jet \(p_T\), triggering on the low jet \(p_T\) events was highly pre-scaled. In simulation, to guarantee enough statistics in the high jet \(p_T\) region, events are generated from various partonic \(p_T\), or momentum transfer \(Q\),  bins. Then a proper weight is applied to each partonic \(p_T\) bin to make up the whole partonic \(p_T\) spectrum. The weight is calculated as the partonic cross-section given by PYTHIA divided by the number of events generated for the partonic \(p_T\) bin, \(w_{p_T} = \frac{\sigma_{p_T}}{N_{p_T}}\).

Note that PYTHIA tends to report a jet cross-section which is larger than the inelastic, non-diffractive cross-section near a few GeV of the cutoff \(p_{T, 0}\) \cite{pythia2006}. This is due to when PYTHIA generates both the hard scattering and the multi-parton interactions, if a multi-parton interaction has higher momentum transfer \(Q\) than the hard scattering, the generated event will be rejected. However, this is not factored into the reported cross-section. Therefore PYTHIA over-estimates the parton scattering cross-sections near the multi-parton interaction threshold. At high partonic \(p_T\) which is much larger than the multi-parton threshold, PYTHIA reports the parton scattering cross-section quite accurately.

To overcome the over-estimate of the parton scattering cross-section at low partonic \(p_T\), the "fudge factors" are introduced to get proper weights from the individual partonic \(p_T\) bins. The way to calculate fudge factors is by applying these factors, a smooth partonic \(p_T\) spectrum will be obtained while connecting the \(p_T\) distribution from all the partonic \(p_T\) bins. The partonic \(p_T\) spectrum from each partonic \(p_T\) bin is fitted by an exponential decay function with a quadratic-polynomial in the exponent. The yields at the two ends of each bin can be calculated from the fitted functions. The fudge factors can be assigned recursively starting from the highest partonic \(p_T\) bin, by equating the yields at the boundaries. The reason to start from the highest partonic \(p_T\) bin is PYTHIA gives reliable partonic cross-section at high partonic \(p_T\) values. As seen from Table \ref{tab:partonicbins}, for the first a few partonic \(p_T\) bins, the corresponding fudge factors are less than 1.

To simulate where the real collisions happen and account for the detector acceptance, a Gaussian \(z\) vertex distribution is generated with mean at zero and \(\sigma\) at 45 cm, chosen to match the width of the vertex distribution seen in the data. The \(x\) and \(y\) positions of the collision vertex are obtained using the measured location with a Gaussian smear with \(\sigma\) at 0.015 cm. The partonic \(p_T\) bins and their number of events, total cross-section and fudge factor are listed in the following table \ref{tab:partonicbins}.

\begin{table}[H]
\centering
\begin{tabular}{|c|c|c|c|c|}
\hline
 $\hat{p}_T$ bin (GeV) & $\sigma$($mb^{-1}$) &  No. of evts. gen. & No. of evts req. & Fudge factor \\
\hline
(2, 3) & 28.7 & 486442 & 490000 & 0.73\\
\hline
(3, 4) & 5.87 & 318342 & 320000 & 0.89\\
\hline
(4, 5) & 1.69 & 297264 & 300000& 0.95\\
\hline
(5, 7) & 0.859 & 504850 & 510000 & 0.97\\
\hline
(7, 9) & 0.178 & 336350 & 340000 & 1.0\\
\hline
(9, 11) & 0.0509 & 318115 & 320000 & 0.99\\
\hline
(11, 15) & 0.0251 & 523236 & 530000 & 0.99\\
\hline
(15, 20) & 0.00532 & 366573 & 370000 &0.99\\
\hline
(20, 25) & 0.00106 & 238804 & 250000 & 1.0\\
\hline
(25, 35) & 0.000371 & 269903 & 290000 & 1.0\\
\hline
(35, 45) & 4.58$\times 10^{-5}$ & 113973 & 120000 & 1.0\\
\hline
(45, 55) & 8.31$\times 10^{-6}$ & 69696 & 73000 & 0.99\\
\hline
(55, infinity) & 2.59$\times 10^{-6}$ & 65343 & 69000 & 1.0\\
\hline
\end{tabular}
\caption{Partonic $p_T$ bins and the number of events requested, the number of events simulated, total cross-section from PYTHIA and fudge factor.}
\label{tab:partonicbins}
\end{table}

\section{Pure Pythia Study}

\begin{figure}[H]
\centering
\includegraphics[scale=0.8]{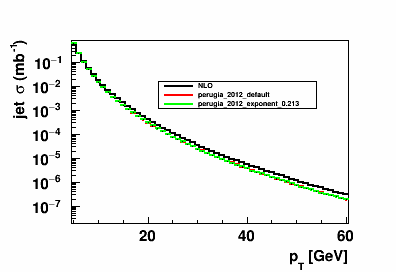}
\nocitecaption{Parton  jet cross-sections from PYTHIA from the default  Perugia 2012 tune and the Perugia 2012 tune with reduced exponent of 0.213 compared to NLO calculations \cite{nlojet2012}.}
\label{fig:pythianlocrs}
\end{figure}

\begin{figure}[H]
\centering
\includegraphics[scale=0.8]{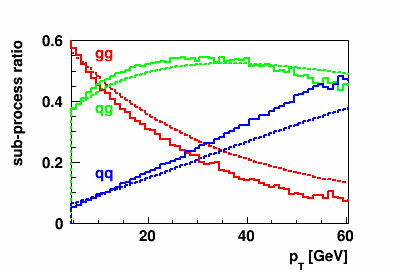}
\caption{Sub process ratios from PYTHIA (solid line) are compared to those from the NLO calculations (dashed line). }
\label{fig:pythianlosub}
\end{figure}

Figure \ref{fig:pythianlocrs} shows the comparison between the parton jet cross-section from the PYTHIA to NLO calculations using the CT10 PDF \cite{CT10}. The parton jet cross-section provides an acceptable match with the NLO calculations. Figure \ref{fig:pythianlosub} shows the sub-process ratio comparison between PYTHIA and the NLO. The subprocess ratios agreement is quite good at low jet \(p_T\). This region is particularly important to simulate trigger bias between quark and gluon jets correctly.

\begin{figure}[H]
\centering
\includegraphics[scale=0.8]{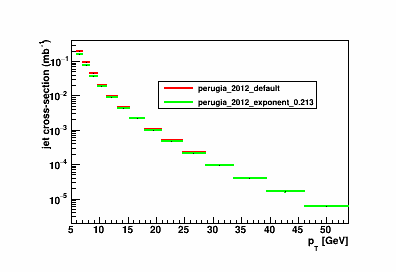}
\caption{Particle jet cross-section vs. jet $p_T$ for the default Perugia 2012 tune and the Perugia 2012 tune with reduced exponent value at 0.213.}
\label{fig:pythiaparticle}
\end{figure}

\begin{figure}[H]
\centering
\includegraphics[scale=0.8]{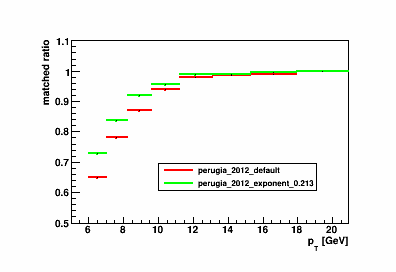}
\caption{Particle jet to parton jet matching ratio vs. jet $p_T$ for the default Perugia 2012 tune and the Perugia 2012 tune with reduced exponent value at 0.213. Note that these matching fractions are before the off-axis cone correction is applied.}
\label{fig:pythiaparticlematch}
\end{figure}

\begin{figure}[H]
\centering
\includegraphics[scale=0.8]{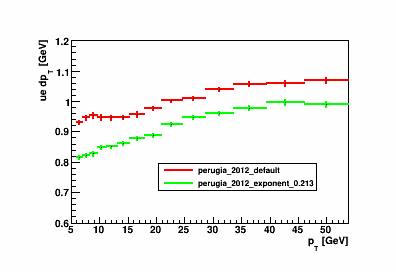}
\caption{Off-aixs cone underlying event correction $dp_T$ vs. jet $p_T$ for the default Perugia 2012 tune and the Perugia 2012 tune with reduced exponent value at 0.213.}
\label{fig:pythiaparticleuedpt}
\end{figure}

Figure \ref{fig:pythiaparticle} shows the particle jet cross-section for the two tunes, the tune chosen in this analysis with reduced underlying event contribution and the default Perugia 2012 tune.  Note the particle jet cross-sections here are the raw jet cross-sections without the off-axis cone underlying event correction. The reduced exponent has less particle jet cross-section at low jet \(p_T\). This is indeed expected because the reduced exponent reduces the underlying event contributions. The particle jets are also matched to the parton jets and their matching ratios are shown in Figure \ref{fig:pythiaparticlematch}. Though the matching ratios are close to each other, the modified Perugia 2012 tune helps the matching probabilities at low jet \(p_T\). The off-axis cone underlying event corrections ,\(dp_{T}\), vs. the reconstructed particle jet \(p_T\) for the two tunes are also compared. The tune of the reduced exponent generates less underlying event activities, which results in less underlying event correction \(dp_{T}\) as shown in Figure \ref{fig:pythiaparticleuedpt}. 

\begin{figure}[H]
\centering
\includegraphics[scale=0.8]{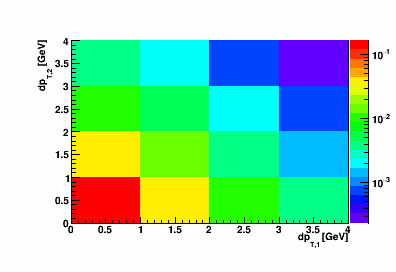}
\caption{The 2D distributions of the cone sum $p_T$ from the two off-axis cones $dp_T$ vs. jet $p_T$ for the Perugia 2012 tune with reduced exponent value at 0.213.}
\label{fig:pythiaparticleuedpt2d}
\end{figure}

\begin{figure}[H]
\centering
\includegraphics[scale=0.8]{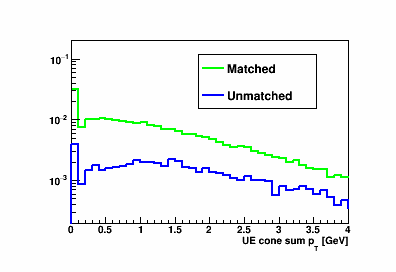}
\caption{The distribution of the sum of the two cone sum $p_T$ from the two off-axis cones for the matched and un-matched particle jets by using the Perugia 2012 tune with reduced exponent value at 0.213.}
\label{fig:pythiaparticleuedptdis}
\end{figure}

\begin{figure}[H]
\centering
\includegraphics[scale=0.8]{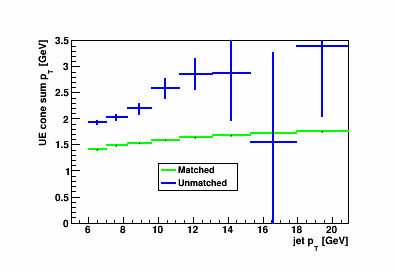}
\caption{The sum of the two cone sum $p_T$ from the two off-axis cones vs. the reconstructed jet $p_T$ for the matched and un-matched particle jets by using the Perugia 2012 tune with reduced exponent value at 0.213.}
\label{fig:pythiaparticleuedptprof}
\end{figure}

\begin{figure}[H]
\centering
\includegraphics[scale=0.8]{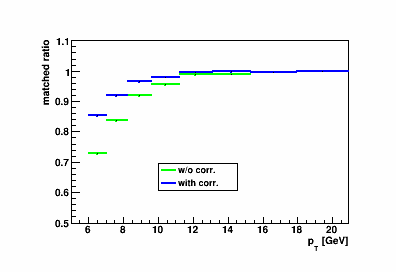}
\caption{Particle jet to parton jet matching ratio vs. jet $p_T$ without and with the underlying event correction on jet $p_T$ by using the Perugia 2012 tune with reduced exponent value at 0.213.}
\label{fig:pythiaparticlematchcrr}
\end{figure}

The underlying event quantities are also studied for the Perugia 2012 tune with the reduced exponent at 0.213. The sum \(p_T\) inside the two off-axis cones are plotted against each other as shown in Figure \ref{fig:pythiaparticleuedpt2d}. As the asymmetric feature of the sum \(p_T\) of the two cones, it is a reasonable estimate of underlying event correction by using the averaged sum \(p_T\)  of the two cones. The sum cone \(p_T\) distributions are also investigated for particle jets that have parton jets matched to them and do not have parton jets matched to them as shown in Figure \ref{fig:pythiaparticleuedptdis}. The un-matched particle jets have a harder spectrum compared to the matched particle jets, which implies a larger energy deposition inside the two cones for un-matched particle jets. The averaged cone sum \(p_T\) is also inspected inside each particle jet \(p_T\) bin for matched particle jets and un-matched particle jets as seen in Figure \ref{fig:pythiaparticleuedptprof}. Since the matching proability for jet with \(p_T\) above 20 GeV is essentially 1, only the bins with jet \(p_T\) less than 20 GeV are plotted. It is clear that the un-matched particle jets exhibit larger underlying event activities. This is the reason that there could not be a parton jet found matched to these particle jets. This characteristic also implies applying this underlying event correction would help trim the un-matched particle jets more than the matched particle and in consequence improve the matching probabilities at low jet \(p_T\) bins, where the un-matched probabilities are significant, as shown in Figure \ref{fig:pythiaparticlematchcrr}

\section{Embedding and Data Comparison}

\begin{figure}[H]
\centering
\includegraphics[scale=0.5]{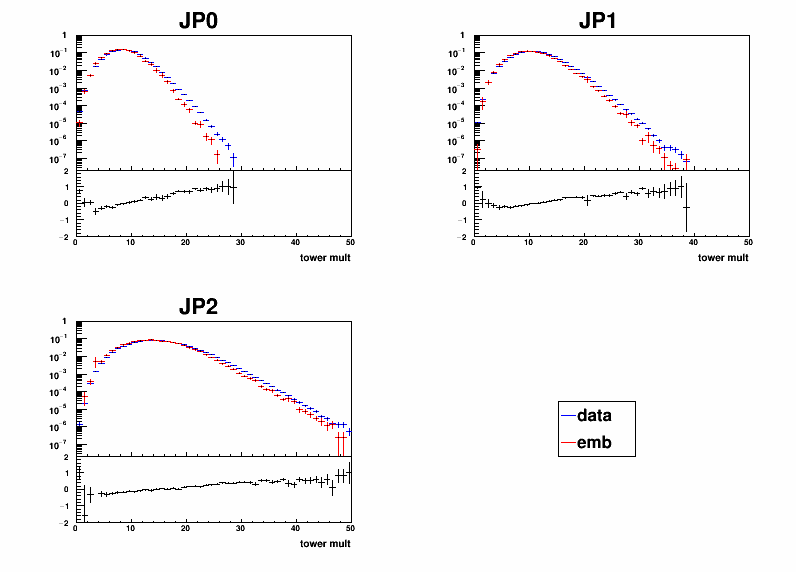}
\caption{JP0, JP1 and JP2 jet tower multiplicity distribution comparisons between data and embedding. }
\label{fig:twrmultembed}
\end{figure}

\begin{figure}[H]
\centering
\includegraphics[scale=0.5]{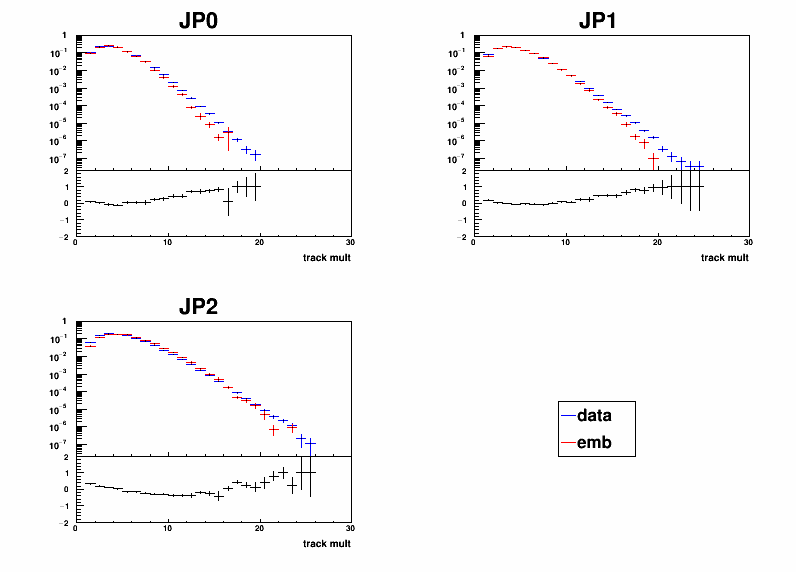}
\caption{JP0, JP1 and JP2 jet track multiplicity distribution comparisons between data and embedding. }
\label{fig:trkmultembed}
\end{figure}

\begin{figure}[H]
\centering
\includegraphics[scale=0.5]{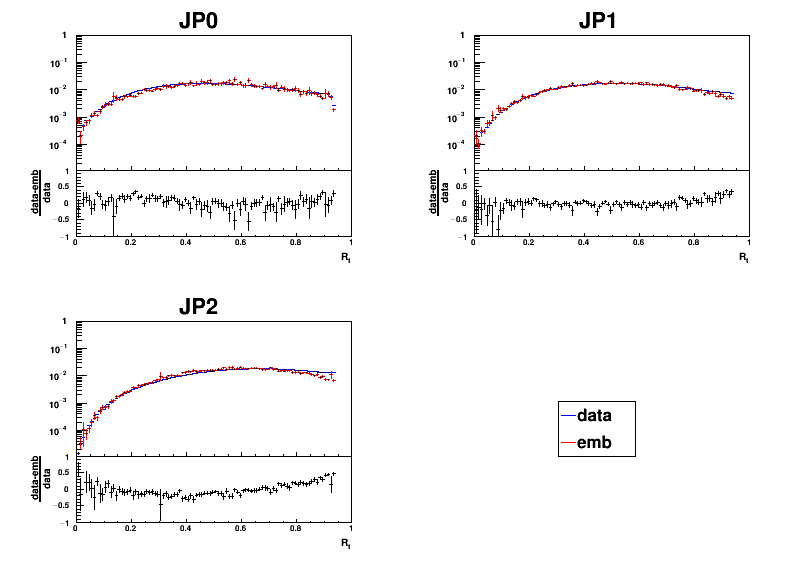}
\caption{JP0, JP1 and JP2 jet $R_t$ distribution comparisons between data and embedding. }
\label{fig:rtembed}
\end{figure}

\begin{figure}[H]
\centering
\includegraphics[scale=0.5]{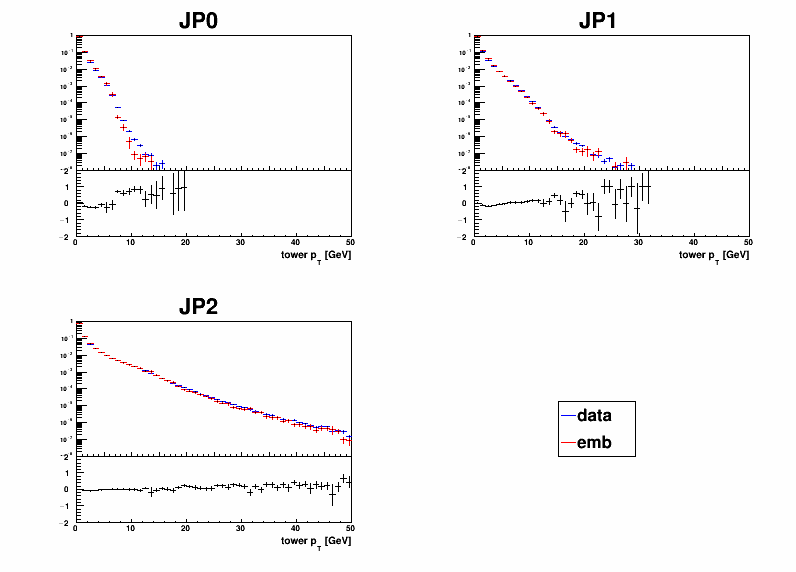}
\caption{JP0, JP1 and JP2 jet tower $p_T$ distribution comparisons between data and embedding. }
\label{fig:twrptembed}
\end{figure}

\begin{figure}[H]
\centering
\includegraphics[scale=0.5]{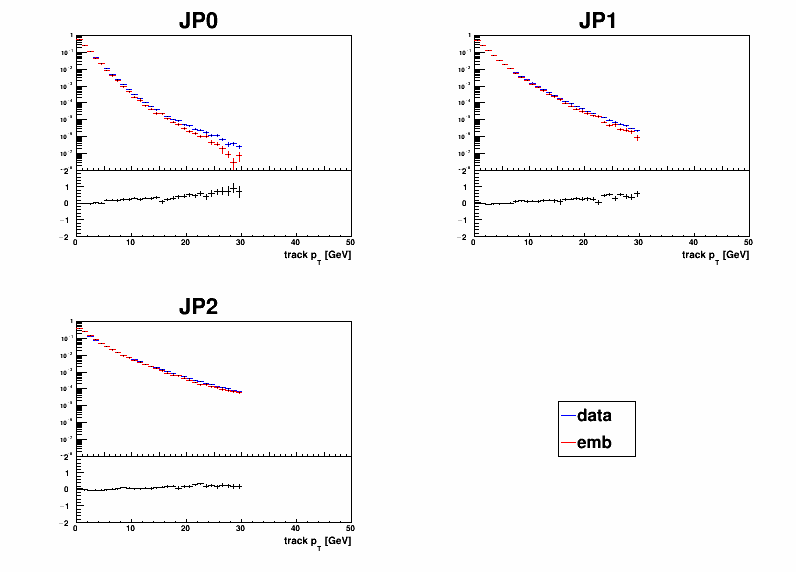}
\caption{JP0, JP1 and JP2 jet track $p_T$ distribution comparisons between data and embedding. }
\label{fig:trkptembed}
\end{figure}

\begin{figure}[H]
\centering
\includegraphics[scale=0.5]{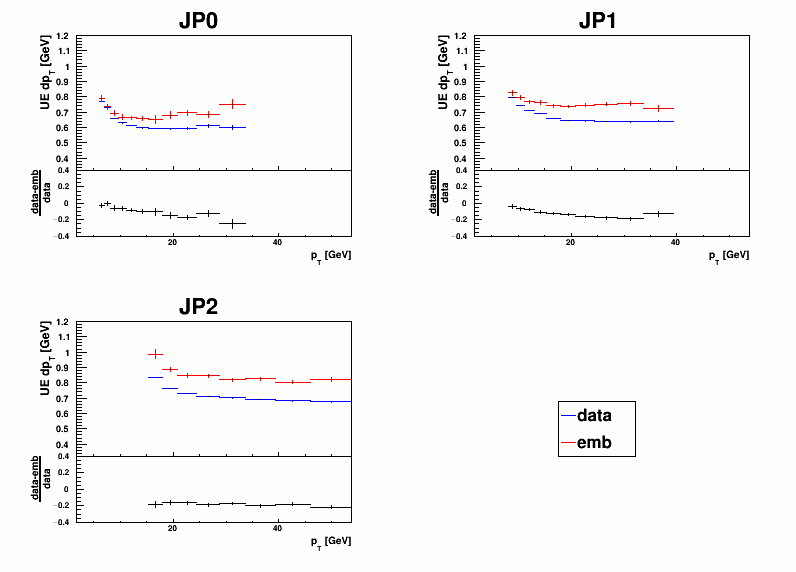}
\caption{JP0, JP1 and JP2 jet off-axis correction $dp_T$ profile vs. jet $p_T$ comparisons between data and embedding. }
\label{fig:dptprofembed}
\end{figure}

\begin{figure}[H]
\centering
\includegraphics[scale=0.5]{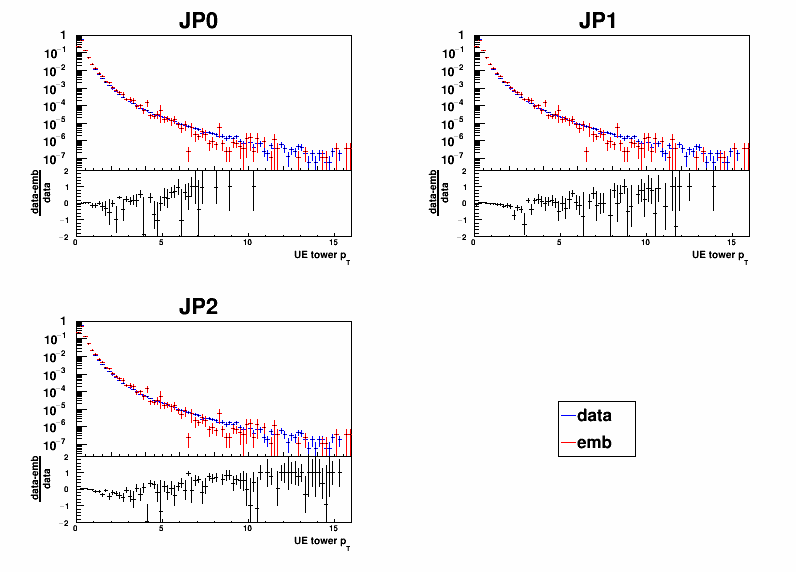}
\caption{The distributions of tower $p_T$ inside the two off-axis cones are compared between data and embedding for JP0, JP1 and JP2 jets. }
\label{fig:uetwrptembed}
\end{figure}

\begin{figure}[H]
\centering
\includegraphics[scale=0.5]{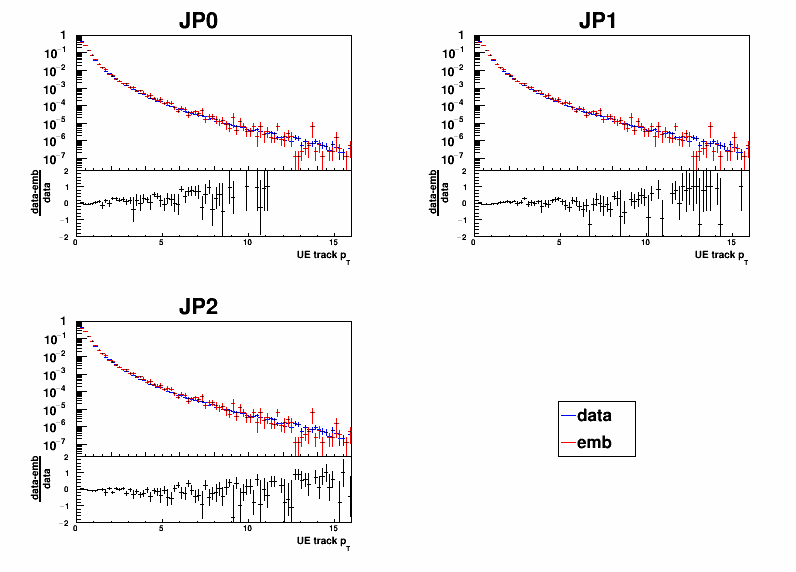}
\caption{The distributions of track $p_T$ inside the two off-axis cones are compared between data and embedding for JP0, JP1 and JP2 jets. }
\label{fig:uetrkptembed}
\end{figure}

\begin{figure}[H]
\centering
\includegraphics[scale=0.5]{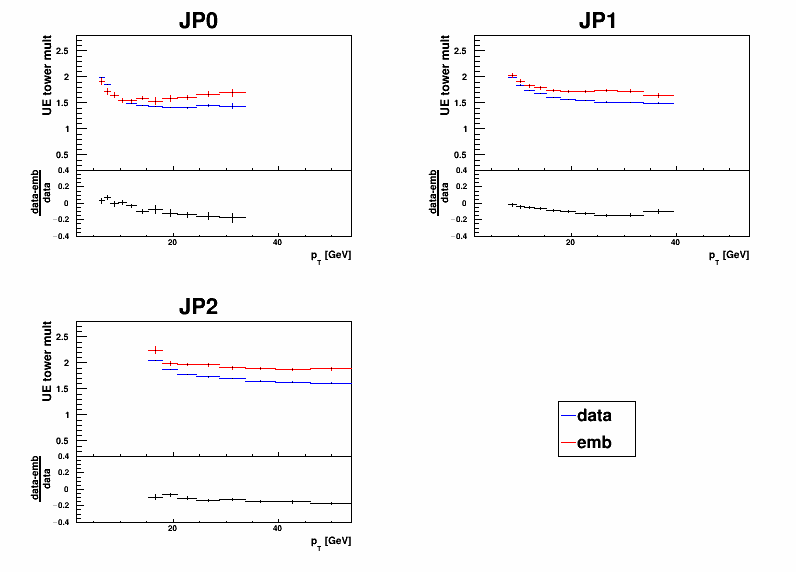}
\caption{The sum of tower multiplicities from the two off-axis cones profile vs. jet $p_T$ comparisons between data and embedding for JP0, JP1 and JP2 jets. }
\label{fig:uetwrmultembed}
\end{figure}

\begin{figure}[H]
\centering
\includegraphics[scale=0.5]{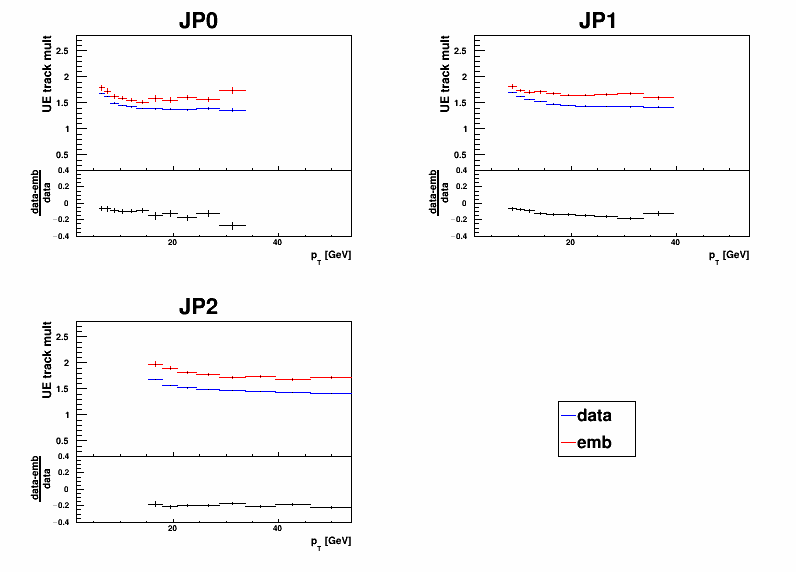}
\caption{The sum of track multiplicities from the two off-axis cones profile vs. jet $p_T$ comparisons between data and embedding for JP0, JP1 and JP2 jets. }
\label{fig:uetrkmultembed}
\end{figure}

The following Figures \ref{fig:ptembed}, \ref{fig:etaembed}, \ref{fig:phiembed}, \ref{fig:twrmultembed}, \ref{fig:trkmultembed} and \ref{fig:rtembed} show the comparison of the jet  \(p_T\) \(\eta\), \(\phi\), tower multiplicity, and track multiplicity and \(R_t\) distribution between the embedding sample and the real data after applying the underlying event correction on the jet \(p_T\). The tower multiplicity is somewhat higher in the data. The high multiplicity tail of the track multiplicity is also a bit higher than the data. Otherwise, the matches are quite good. The individual tower and track \(p_T\) inside the jet also match well between embedding and data as seen in Figures \ref{fig:twrptembed} and \ref{fig:trkptembed}. However the underlying event \(dp_T\) does not match well with the data at high jet \(p_T\) as show in Figure \ref{fig:dptprofembed}. There is a good agreement at low jet pt regions for JP0 and JP1 triggered jets. However high \(p_T\) jets triggering JP1 and JP2 exhibit about 15\% to 20 \% differences. After a deeper look, good agreement is found for the individual track and tower \(p_T\) distributions inside the off-axis cones as seen from Figures \ref{fig:uetwrptembed} and \ref{fig:uetrkptembed}. The \(dp_{T}\) discrepancy comes from the discrepancy in the multiplicity of the two off-axis cones. This is verified by the sum tower and track multiplicity of the two off-axis cones for JP0, JP1 and JP2 jets as shown in Figures \ref{fig:uetwrmultembed} and \ref{fig:uetrkmultembed}. Additional PYTHIA  simulations found that a further increase in \(p_{T,0}\) could match the M/C underlying event multiplicities. But this caused a degradation of the data vs. M/C match for the primary jet quantities.

\begin{figure}[H]
\centering
\includegraphics[scale=0.5]{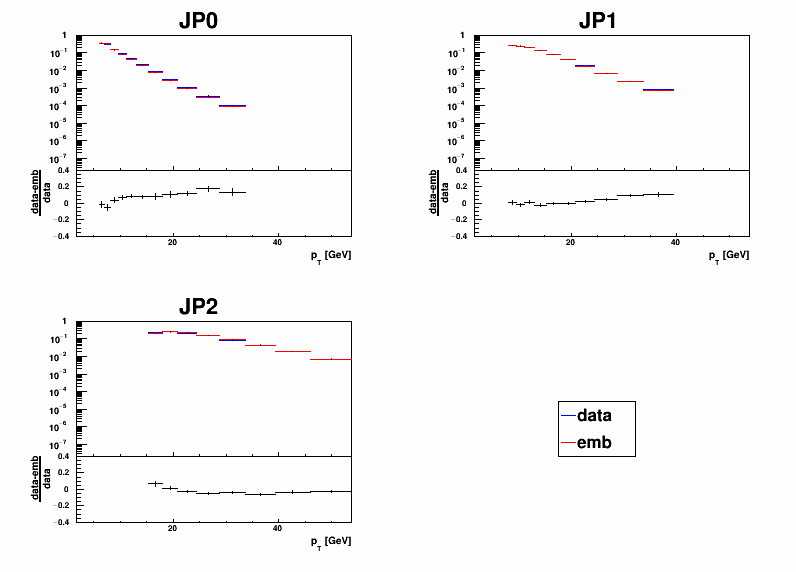}
\caption{JP0, JP1 and JP2 jet $p_T$ distribution comparisons between data and embedding. }
\label{fig:ptembed}
\end{figure}

\begin{figure}[H]
\centering
\includegraphics[scale=0.5]{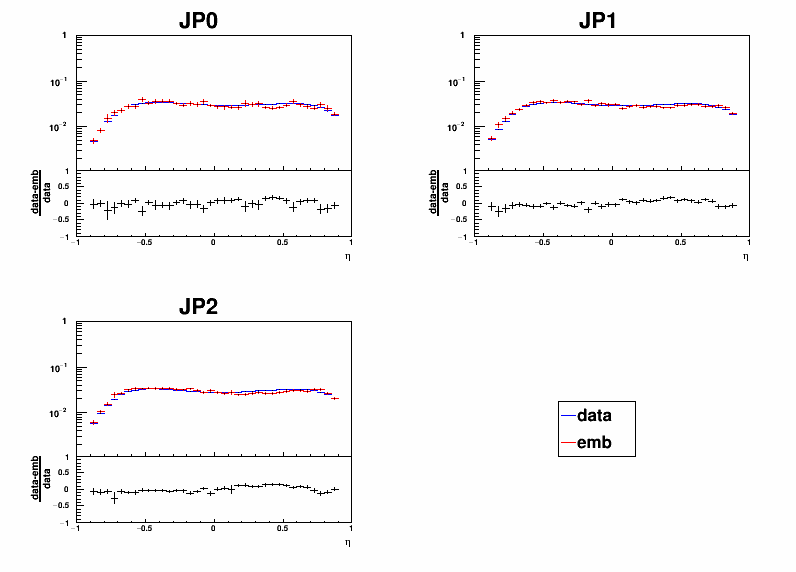}
\caption{JP0, JP1 and JP2 jet $\eta$ distribution comparisons between data and embedding. }
\label{fig:etaembed}
\end{figure}

\begin{figure}[H]
\centering
\includegraphics[scale=0.5]{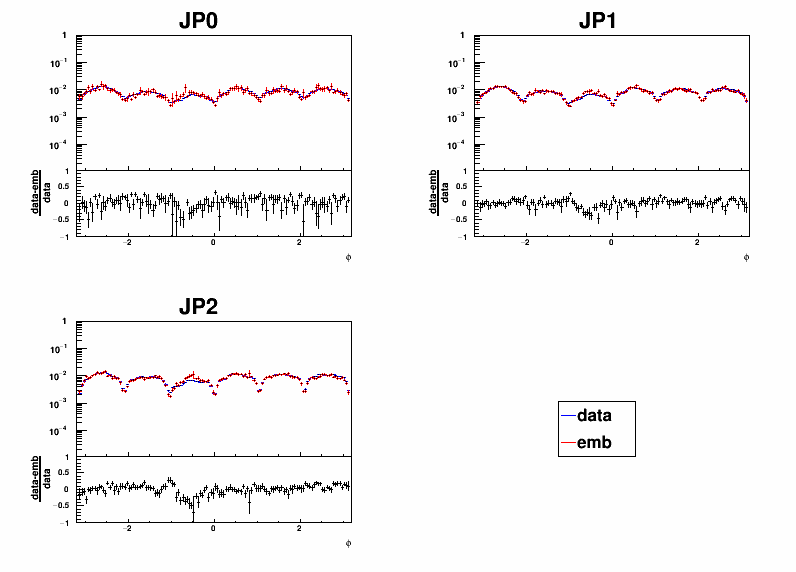}
\caption{JP0, JP1 and JP2 jet $\phi$ distribution comparisons between data and embedding. }
\label{fig:phiembed}
\end{figure}

%% file: chapter6.tex
\chapter{\uppercase{Systematic Uncertainties}}

The systematic uncertainties of the 2012 510 GeV inclusive jet \(A_{LL}\) are considered to be contributed by the following aspects: relative luminosity uncertainty, underlying event systematic uncertainty, trigger bias and reconstruction uncertainty, transverse residual double spin asymmetry uncertainty, non-collision background uncertainty, and jet energy scale uncertainties. The first five uncertainties contribute to the uncertainty on the inclusive jet \(A_{LL}\) value and the last contributes to the uncertainty on the inclusive jet \(p_{T}\) for the \(A_{LL}\) measurements.

\section{Relative Luminosity Uncertainty}
As in equation \ref{eq:aLLexp}, the inclusive jet \(A_{LL}\) can be written as,
\begin{align}
A_{LL} = \frac{1}{P_{A}P_{B}} \frac{N^{+} - R_{3}N^{-}}{N^{+}+R_{3}N^{-}}
\end{align}

where \(N^{+} = N^{++} + N^{--}\)  and \(N^{-} = N^{+-} + N^{-+}\), then

\begin{align}
A_{LL} &= \frac{1}{P_{A}P_{B}} \frac{\frac{N^{+}}{N^{-}} - R_{3}}{\frac{N^{+}}{N^{-}}+R_{3}} \nonumber \\
&= \frac{1}{P_{A}P_{B}} \frac{R_{N} - R_{3}}{R_{N} + R_{3}}, \label{eq:aLLasRn}
\end{align}
where \(R_{N} = \frac{N^{+}}{N^{-}}\).  The uncertainty on \(A_{LL}\) due to the uncertainty on \(R_{3}\) can be expressed as
\begin{align}
\Delta A_{LL} = |\frac{\partial A_{LL}}{\partial R_{3}} | \Delta R_{3} = \frac{1}{P_{A}P_{B}} \frac{2R_{N}}{(R_{N}+R_{3})^{2}} \Delta R_{3} \label{eq:daLL}.
\end{align}

From Equation \ref{eq:aLLasRn}, \(R_{N}\) can be written as,
\begin{align}
R_{N} = \frac{1 + A_{LL}P_{A}P_{B}}{1 - A_{LL}P_{A}P_{B}} R_{3} \label{eq:rn}
\end{align}
Plug Equation \ref{eq:rn} into Equation \ref{eq:daLL}, the \(\Delta A_{LL}\) can be written as,
\begin{align}
\Delta A_{LL} = \frac{1}{P_{A}P_{B}} \frac{1 - (P_{A}P_{B}A_{LL})^{2}}{2} \frac{\Delta R_{3}}{R_{3}}
\end{align}

Taking \(P_{A} = 0.54 \) and \(P_{B} = 0.55\) and ignoring the second order term of \(A_{LL}\) since \(A_{LL}\) is at the order of \(10^{-2}\), the estimated \(\Delta A_{LL}\) can be written as,

\begin{align}
\Delta A_{LL} =  \frac{1}{2 P_A P_B}\frac{\Delta R_{3}}{R_{3}}
\end{align}
In the previous section, \(\frac{\Delta R}{R}\) is estimated as \(1.3 \times 10^{-4}\), therefore \(\Delta A_{LL}\) due to the systematic uncertainty of relative luminosity \(R_{3}\) is \(2.2 \times 10^{-4}\).

\section{False Asymmetries}

Four false asymmetries are a good measure to cross-check of the relative luminosity determinations. The four false asymmetries- blue beam single spin asymmetry, \(A_L^B\), yellow beam single spin asymmetry, \(A_L^Y\), like-sign double spin asymmetry, \(A_{LL}^{l.s.}\), and unlike-sign double spin asymmetry, \(A_{LL}^{u.s.}\)- are expressed in the following equations \eqref{eq:aLb} - \eqref{eq:aLLus} where \(P_B\) and \(P_Y\) are blue and yellow beam polarizations and \(R_1\), \(R_2\), \(R_4\), \(R_5\) and \(R_6\) are relative luminosities.

\begin{align}
A_L^B = \frac{\sum_{run}P_B((N^{++} + N^{+-}) - R_2(N^{-+} + N^{--}))}{\sum_{run}P_B^2( (N^{++} + N^{+-}) + R_2 (N^{-+} + N^{--}) )} \label{eq:aLb}
\end{align}
\begin{align}
A_L^ Y= \frac{\sum_{run}P_Y((N^{++} + N^{-+}) - R_1(N^{+-} + N^{--}))}{\sum_{run}P_Y^2((N^{++} + N^{-+}) + R_1(N^{+-} + N^{--}))} \label{eq:aLy}
\end{align}
\begin{align}
A_{LL}^{l.s.} = \frac{\sum_{run}P_B P_Y ( N^{++} - R_4 N^{--})}{\sum_{run}P_B^2 P_Y^2 ( N^{++} + R_4 N^{--} )} \label{eq:aLLls}
\end{align}
\begin{align}
A_{LL}^{u.s.} = \frac{\sum_{run}P_B P_Y (R_5 N^{+-} - R_6 N^{-+}) }{\sum_{run}P_B^2 P_Y^2 (R_5 N^{+-} + R_6 N^{-+}) } \label{eq:aLLus}
\end{align}

\begin{figure}[H]
\centering
\includegraphics[scale=0.5]{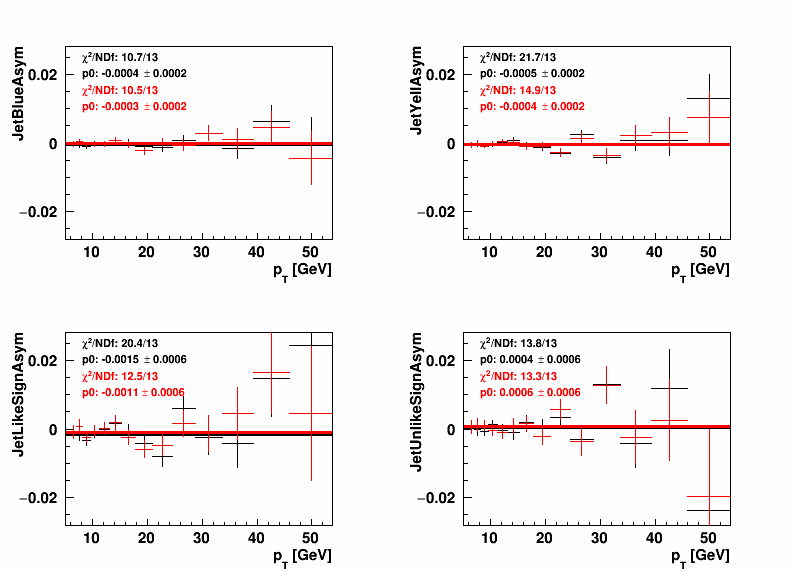}
\caption{Falses asymmetries measured in the data before (black) and after (red) the underlying event correction.}
\label{fig:falseasym}
\end{figure}

The blue beam single spin asymmetry, \(A_L^B\), the yellow beam single spin asymmetry, \(A_L^Y\), and the like-sign double spin asymmetry \(A_{LL}^{l.s.}\) deviate from zero by the parity-violating interactions. However the current precision is not good enough to observe this effect. The un-like sign double spin asymmetry, \(A_{LL}^{u.l.}\), is expected to be zero because of the geometric symmetry of the two spin orientations. As seen in Figure \ref{fig:falseasym}, before the making underlying event correction, the false asymmetries are slightly deviated from zero. However after the underlying event correction, all four false asymmetries are consistent with zero. The non-zero false asymmetries may indicate instrumental effects of the collider.

%

%
\section{Underlying Event Correction and Its Contribution}

\begin{figure}[H]
\centering
\includegraphics[scale=0.6]{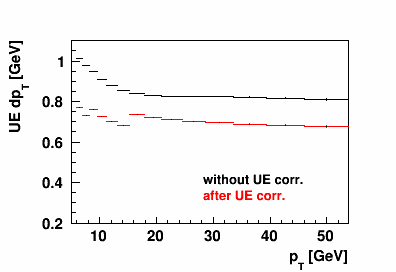}
\caption{Underlying event $dp_T$ in each jet $p_T$ bin before and after the underlying event correction.}
\label{fig:dptrun12}
\end{figure}

The underlying event contribution to inclusive jet \(A_{LL}\) is considered by measuring the averaged underlying event correction \(dp_T\) and the \(dp_T\) double spin asymmetry in the data. Figure \ref{fig:dptrun12} shows the averaged \(dp_T\) and its statistical uncertainty for all jet \(p_T\) bins before and after the underling event correction, where the corrected jet \(p_T\) is the uncorrected jet \(p_T\) minus the underlying event correction, \(p_{T,crr} = p_{T,uncrr} - dp_{T}\).

The \(dp_T\) double spin asymmetry, \(A_{LL}^{dp_T}\) is defined by
\begin{align}
A_{LL}^{dp_T} = \frac{1}{P_AP_B} \frac{(<dp_T>^{++} +<dp_T>^{--}) - (<dp_T>^{+-} +<dp_T>^{-+})}{(<dp_T>^{++} +<dp_T>^{--}) + (<dp_T>^{+-} +<dp_T>^{-+})},
\end{align} 
where \(<dp_T>^{++}\) is the measured mean underlying event correction \(dp_T\) for the spin state \(++\) and similar definition for the other three spin states. Since the beams are not 100\% polarized, beam polarizations are included in this calculation. Also the beam polarization varies from run to run, therefore the \(A_{LL}^{dp_T}\) is measured run by run individually and so is its statistical uncertainty \(\sigma_{A_{LL}^{dp_T}}\). The final measured \(A_{LL}^{dp_T}\) is the weighed average of \(A_{LL}^{dp_T}\).

\begin{figure}[H]
\centering
\includegraphics[scale=0.6]{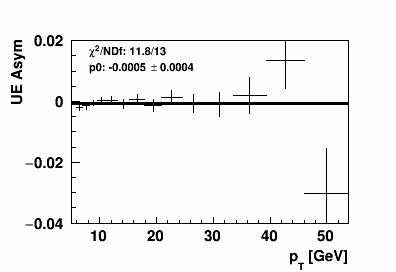}
\caption{Underlying event $dp_T$ asymmetry measured in the data with a $p0$ fit in each jet $p_T$ bin after the jet $p_T$ corrected from the underlying events.}
\label{fig:dptALL}
\end{figure}

Figure \ref{fig:dptALL} shows the final measured \(A_{LL,exp}^{dp_T}\) for all the jet \(p_T\) bins and the results from a zero-th order polynomial fit where the jet \(p_T\) is the underlying event corrected jet \(p_T\). The measured \(A_{LL,exp}^{dp_T}\) comes directly from the data and has not been corrected for effect from finite detector efficiencies etc. The fitted result shows the measured \(A_{LL,exp}^{dp_T}\) is consistent with zero across all jet \(p_T\) bin.

To estimate the underlying event contribution, it is assumed that the underlying event adds extra energy to the jet, which effectively shifts the jet \(p_T\) spectrum to the positive \(p_T\) direction. Therefore for the jet cross-section with the same helicity \(++\), the jet cross-section is expressed as \(\sigma_{jet}^{++} = \int\limits_{p_{T,min}-dp_{T}^{++}}^{p_{T,max}-dp_{T}^{++}} \frac{d\sigma^{++}}{dp_T}dp_T\), where \(p_{T, min}\) and \(p_{T, max}\) are the lower and upper limits of the jet \(p_T\) bin, \(\frac{\sigma^{++}}{dp_T}\) is the jet spin dependent differential cross-section for the same helicity. Similar for the jet cross-section with the opposite helicity, \(\sigma_{jet}^{+-} = \int\limits_{p_{T,min}-dp_{T}^{+-}}^{p_{T,max}-dp_{T}^{+-}} \frac{d\sigma^{+-}}{dp_T}dp_T\). With these, the uncertainty due to the underlying event correction can be estimated by recognizing that the spin-independent piece of \(dp_T^{++}\) and \(dp_T^{+-}\) is accounted for with the "\(p_T\) shift" correction, which is discussed later. Therefore the small spin-dependent difference between \(dp_T^{++}\) and \(dp_T^{+-}\) is considered. This can be calculated with the following equation,
%
\begin{align}
\delta A_{LL} = \frac{\int\limits_{p_{T,min}-<dp_T>\times A_{LL}^{dp_T}}^{p_{T,max}-<dp_T>\times A_{LL}^{dp_T}} \frac{d\sigma}{dp_T}dp_T - \int\limits_{p_{T,min}+<dp_T>\times A_{LL}^{dp_T}}^{p_{T,max}+<dp_T>\times A_{LL}^{dp_T}} \frac{d\sigma}{dp_T}dp_T}{\int\limits_{p_{T,min}-<dp_T>\times A_{LL}^{dp_T}}^{p_{T,max}-<dp_T>\times A_{LL}^{dp_T}} \frac{d\sigma}{dp_T}dp_T+ \int\limits_{p_{T,min}+<dp_T>\times A_{LL}^{dp_T}}^{p_{T,max}+<dp_T>\times A_{LL}^{dp_T}}\frac{d\sigma}{dp_T}dp_T}, \label{eq:dptsyst}
\end{align}
where \(\frac{d\sigma}{dp_T}\) is the un-polarized jet cross-section and \(<dp_T>\) is the mean spin independent underlying event correction. Here the NLO pQCD calculation using the MSTW PDF set \cite{nlojet2012,MSTW} is used as the un-polarized jet cross-section. 

\begin{figure}[H]
\centering
\includegraphics[scale=0.8]{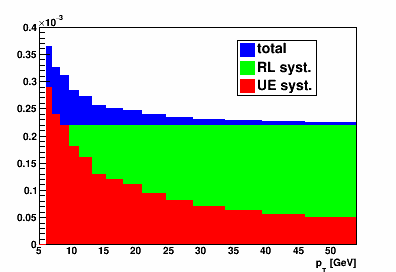}
\caption{Underlying event systematic (red), relative luminosity systematic (green) and their quadrature sum (blue).}
\label{fig:uerlsyst}
\end{figure}

The underlying event and relative luminosity systematics uncertainties are correlated across all the jet \(p_T\) bins, so their uncertainty will be added in quadrature to constitute the final relative luminosity and underlying event uncertainty. Figure \ref{fig:uerlsyst} shows the numerical values of relative luminosity and underlying event uncertainties and Table \ref{tab:uerlsyst} gives the numerical values.

\begin{table}[H]
\centering
\begin{tabular}{|c|c|c|}
\hline
$p_T$ & UE syst. & UE/RL syst. \\
\hline
(6.0, 7.0) & 0.00029 & 0.00036 \\
\hline
(7.0, 8.2) & 0.00024 & 0.00033 \\
\hline
(8.2, 9.6) & 0.00022 & 0.00031 \\
\hline
(9.6, 11.2) & 0.00018 & 0.00028 \\
\hline
(11.2, 13.1) & 0.00016 & 0.00027 \\
\hline
(13.1, 15.3) & 0.00013 & 0.00026 \\
\hline
(15.3, 17.9) & 0.00012 & 0.00025 \\
\hline
(17.9, 20.9) & 0.00011 & 0.00025 \\
\hline
(20.9, 24.5) & 9.4e-05 & 0.00024 \\
\hline
(24.5, 28.7) & 8.1e-05 & 0.00023 \\
\hline
(28.7, 33.6) & 7.1e-05 & 0.00023 \\
\hline
(33.6, 39.3) & 6.3e-05 & 0.00023 \\
\hline
(39.3, 46.0) & 5.6e-05 & 0.00023 \\
\hline
(46.0, 53.8) & 5e-05 & 0.00023 \\
\hline
\end{tabular}
\caption{Underlying event and relative luminsoity systematics.}
\label{tab:uerlsyst}
\end{table}

\section{Trigger Bias and Reconstruction Uncertainty}
The trigger bias and reconstruction uncertainty accounts for the systematic uncertainty caused by selecting events of interest by triggers and jet reconstruction based on the detector response in the experiment. Selecting events by requiring BEMC and EEMC jet patch ADC above certain thresholds may prefer jets that fragment in ways that maximize the calorimeter response, which imposes a bias on the event selection, however in theory the calculation knows nothing about the detector responses. Therefore it introduces uncertainties due to the trigger bias. In the experiment, the jets are reconstructed based on the experimental measurement; on contrary, in NLO pQCD calculation the inclusive jet cross-section and double spin asymmetry are calculated at the parton level before the hadronization process. It is easy to imagine that the reconstruction procedure inevitably introduces an uncertainty. Since both sources originate because of the need to trace the experimental measurements back to the theoretical calculations, these two uncertainties are studied together.

The way to estimate this uncertainty uses the PYTHIA simulation and embedding method mentioned in the previous section. The same jet \(p_{T}\) bins are used in this part of analysis as in the analysis of the real data. Jets reconstructed at the detector level are compared with both jets reconstructed at the particle and parton level. Since most of the theoretical calculations use unpolarized and polarized partonic cross-sections to predict the inclusive jet \(A_{LL}\), the systematic uncertainty is estimated mainly by comparing results obtained from the detector level and parton level.

For jet \(p_{T}\), due to the hadronization and underlying events in the process of producing the final particles and the reality of detector system, the measured jet \(p_{T}\) at the detector would not truely represent the real jet \(p_{T}\) after the \(2 \rightarrow 2\) parton scattering. The difference of the jet \(p_{T}\) between the detector level and parton level is defined as the \(p_{T}\) shift from the parton level. In this analysis, the \(p_{T}\) shift is calculated by taking the difference between the detector jet \(p_{T}\)  and the parton jet \(p_T\) where the parton is deemed to be matched with the detector jet.

\begin{align}
p_{T,shift} = p_{T,parton} - p_{T, particle}. \label{eq:ptshift}
\end{align}

\begin{figure}[H]
\centering
\includegraphics[scale=0.8]{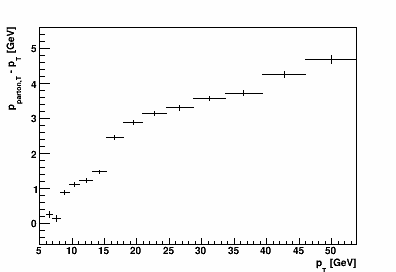}
\caption{Jet $p_T$ shift from the detector leve to the parton level.}
\label{fig:ptshift}
\end{figure}

A detector jet is matched to a parton jet only if the minimum distance from the detector jet to all parton jets is less than 0.5, 
\begin{align}
\Delta R_{min} = min(\sqrt{(\eta_{detector} - \eta_{parton})^{2} + (\phi_{detector}-\phi_{parton})^{2}}) < 0.5.\nonumber
\end{align}
The parton jet that has the minimum \(\Delta R\) is the parton jet that the detector jet  is matched to. The 0.5 is chosen because of the \(R\) parameter of the jet finding algorithm used in this analysis is also 0.5. Table \ref{tab:partonmatch} shows matching ratios from the detector jets to the parton jets vs. jet \(p_{T}\) where the jet \(p_{T}\) has been corrected for underlying the event effect. Figure\ref{fig:ptshift} shows the averaged \(p_{T}\) shift with its statistical errors.

\begin{table}[H]
\centering
\begin{tabular}{|c|c|}
\hline
$p_T$ & Matching ratio \\
\hline
(6.0, 7.0) & 0.8 \\
\hline
(7.0, 8.2) & 0.87 \\
\hline
(8.2, 9.6) & 0.96 \\
\hline
(9.6, 11.2) & 0.99 \\
\hline
(11.2, 13.1) & 1 \\
\hline
(13.1, 15.3) & 0.99 \\
\hline
(15.3, 17.9) & 0.99 \\
\hline
(17.9, 20.9) & 1 \\
\hline
(20.9, 24.5) & 1 \\
\hline
(24.5, 28.7) & 0.99 \\
\hline
(28.7, 33.6) & 1 \\
\hline
(33.6, 39.3) & 1 \\
\hline
(39.3, 46.0) & 1 \\
\hline
(46.0, 53.8) & 1 \\
\hline
\end{tabular}
\caption{Matching ratios from the detector jets to the parton jets in the embedding sample. Note the detector jet $p_T$ is after the underlying event correction.}
\label{tab:partonmatch}
\end{table}

The trigger bias and reconstruction uncertainty is estimated by comparing the \(A_{LL}\) found in every detector jet \(p_{T}\) bin with the \(A_{LL}\) found at the parton jet \(p_{T}\) corresponding to the detector jet \(p_{T}\) plus the \(p_{T}\) shift obtained in the previous step. The \(A_{LL}\) is calculated by first extracting the parton scattering kinematics, the Mandelstam variables \(u\),\(s\), and \(t\) of the parton to parton scattering process and the parton flavors of the \(2 \rightarrow 2\) scattering process in order to calculate the ratio of the polarized partonic cross-section and unpolarized partonic cross-section at the leading order. This is multiplied by the ratio of the polarized PDF to unpolarized PDF of the two incoming partons at their respective \(x_{1}\) and \(x_{2}\) and \(Q^{2}\) as in the following equation
\begin{align}
A_{LL} = \frac{\Delta \sigma}{\sigma} \times \frac{\Delta f_{1}(x_{1},Q^{2})}{f_{1}(x_{1},Q^{2})} \frac{\Delta f_{2}(x_{2},Q^{2})}{f_{2}(x_{2},Q^{2})},
\end{align}
where \(\Delta \sigma\) is the polarized partonic cross-section, \(\sigma\) is the un-polarized partonic cross-section, \(\Delta f(x,Q^{2})\) is the polarized PDF and  \(f(x,Q^{2})\) is the un-polarized PDF. The un-polarized and polarized partonic cross-sections due to different processes at leading order are listed in table \ref{tab:subprocsigma} \cite{locrssection1977}. The polarized PDFs used here are the 100 replicas from the NNPDF group for their NNPDFpol1.1 global fit \cite{NNPDFpol}. The unpolarized PDF is the un-polarized PDF from the same group NNPDF2.3 with \(\alpha_s = 0.119\) \cite{NNPDFunp}. The inclusive jet \(A_{LL}\) predictions from these 100 replicas provide good description of both the previous published STAR 200 GeV inclusive jet \(A_{LL}\) results and the measurements here at the 510 GeV. In addition the \(A_{LL}\) and its statistical uncertainties from the best fit of NNPDFpol 1.1 are calculated. The best fit from NNPDFpol1.1 is approximately the equally weighted average from the 100 replicas. With the 100 calculated \(A_{LL}\)s at both the detector jet level and parton jet level, it's easy to calculate the mean and the error on the mean for the difference in \(A_{LL}\) between the detector level and the parton level. It turns out that the error on the mean from the 100 \(A_{LL}\) replicas is impressively small at the level of \(10^{-5}\). Those values are completely negligible compared to the statistical uncertainties of the best fit, which are fully correlated in the replica companions. Finally, the mean of the difference in the detector \(A_{LL}\) and the parton \(A_{LL}\) is taken as a correction on the \(A_{LL}\) due to trigger bias and reconstruction, and the statistical uncertainties on the detector \(A_{LL}\) from the best fit is taken as the systematic uncertainty due to trigger and reconstruction bias.

\begin{figure}[H]
\centering
\includegraphics[scale=0.5]{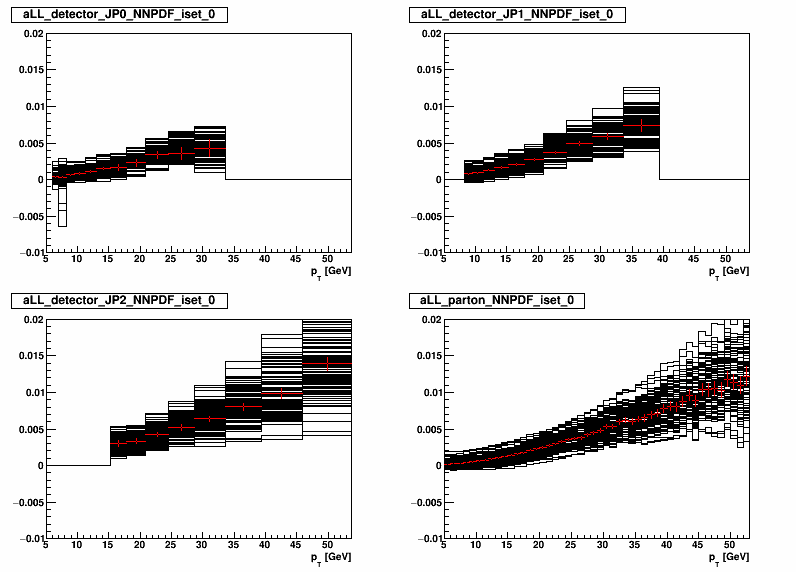}
\caption{Inclusive jet $A_{LL}$ from embedding for JP0, JP1, JP2 triggered jets at the detector level and all the jets at the parton level.}
\label{fig:aLLrepembed}
\end{figure}

\begin{table}[H]
\centering
\begin{tabular}{|c|c|c|}
\hline Sub-process & Un-polarized $|M|^2$ & Polarized $ |M|^2$\\
\hline
$q_1 q_2 (\bar{q_2})\rightarrow q_1 q_2 (\bar{q_2}) $ & 
$\frac{4}{9}\frac{s^2+u^2}{t^2}$ &
$\frac{4}{9}\frac{s^2 - u^2}{t^2}$ \\
\hline
$q_1q_1 \rightarrow q_1q_1$ &
$\frac{4}{9}(\frac{s^2+u^2}{t^2} + \frac{s^2+t^2}{u^2}) - \frac{8}{27}\frac{s^2}{ut}$ &
$\frac{4}{9}(\frac{s^2 - u^2}{t^2} + \frac{s^2 - t^2}{u^2}) - \frac{8}{27}\frac{s^2}{ut}$ \\
\hline
$q_1 \bar{q_1} \rightarrow q_2 \bar{q_2}$ &
$\frac{4}{9}\frac{t^2+u^2}{s^2}$ &
$-\frac{4}{9}\frac{t^2+u^2}{s^2}$\\
\hline
$q_1 \bar{q_1} \rightarrow q_1 \bar{q_1}$ &
$\frac{4}{9}(\frac{s^2+u^2}{t^2} + \frac{t^2+u^2}{s^2}) - \frac{8}{27}\frac{u^2}{st}$ &
$\frac{4}{9}(\frac{s^2 - u^2}{t^2} - \frac{t^2+u^2}{s^2}) + \frac{8}{27}\frac{u^2}{st}$ \\
\hline
$q \bar{q} \rightarrow g g$ &
$\frac{32}{27}\frac{u^2+t^2}{ut} - \frac{8}{3}\frac{u^2+t^2}{s^2}$ &
$ - \frac{32}{27}\frac{u^2+t^2}{ut} + \frac{8}{3}\frac{u^2+t^2}{s^2}$ \\
\hline
$ g g\rightarrow q \bar{q}$ &
$\frac{1}{6}\frac{u^2+t^2}{ut} - \frac{8}{3}\frac{u^2+t^2}{s^2}$ &
$ - \frac{1}{6}\frac{u^2+t^2}{ut} + \frac{8}{3}\frac{u^2+t^2}{s^2}$ \\
\hline
$ q g\rightarrow q g$ &
$-\frac{4}{9}\frac{u^2+s^2}{us} + \frac{u^2+s^2}{t^2}$ &
$-\frac{4}{9}\frac{ - u^2 + s^2}{us} + \frac{- u^2+s^2}{t^2}$ \\
\hline
$ g g\rightarrow g g$ &
$\frac{9}{2}(3 - \frac{ut}{s^2} - \frac{us}{t^2} - \frac{st}{u^2})$ &
$\frac{9}{2}( - 3 + \frac{ut}{s^2} + \frac{s^2}{ut})$ \\
\hline
\end{tabular}
\nocitecaption{Un-polarized and polarized partonic cross-sections for various sub-processes at the leading order, $\frac{d\sigma}{d\hat{t}} = \frac{\pi \alpha^2}{s^2}\times |M|^2$, where $M$ is the corresponding matrix element \cite{locrssection1977}.}
\label{tab:subprocsigma}
\end{table}
Figure \ref{fig:aLLrepembed} shows the spectra of the detector level \(A_{LL}\) and parton level \(A_{LL}\) vs. jet \(p_{T}\). The value of the parton level \(A_{LL}\) is estimated from the TSpline interpolation at the shifted \(p_{T}\) based on the spectrum. Table \ref{tab:trgbias} shows the difference of the \(A_{LL}\) between those two levels and the statistical uncertainty of \(A_{LL}\) from the best fit at the detector level and the final trigger bias and reconstruction uncertainty in each jet \(p_{T}\) bin measured in the data.

\begin{table}[H]
\centering
\begin{tabular}{|c|c|c|c|c|}
\hline
$p_T$ & model corr. & model error & stat.  & total \\
\hline
(6.0, 7.0) & -9.9e-05 & 2e-05 & 7.3e-05 & 7.6e-05 \\
\hline
(7.0, 8.2) & 4.9e-05 & 9.4e-05 & 9.7e-05 & 0.00014 \\
\hline
(8.2, 9.6) & -0.00022 & 7.3e-06 & 7.9e-05 & 7.9e-05 \\
\hline
(9.6, 11.2) & -0.00015 & 7.0e-06 & 8.3e-05 & 8.3e-05 \\
\hline
(11.2, 13.1) & -0.00025 & 8.4e-06 & 9.3e-05 & 9.3e-05 \\
\hline
(13.1, 15.3) & -0.00027 & 8.8e-06 & 0.00012 & 0.00012 \\
\hline
(15.3, 17.9) & -0.00033 & 1.1e-05 & 0.00019 & 0.00019 \\
\hline
(17.9, 20.9) & -0.00028 & 1.9e-05 & 0.00022 & 0.00022 \\
\hline
(20.9, 24.5) & -0.00041 & 1.3e-05 & 0.0002 & 0.0002 \\
\hline
(24.5, 28.7) & -0.0003 & 1.9e-05 & 0.0003 & 0.0003 \\
\hline
(28.7, 33.6) & -0.00027 & 2.3e-05 & 0.00034 & 0.00034 \\
\hline
(33.6, 39.3) & -4.3e-05 & 1.8e-05 & 0.00045 & 0.00045 \\
\hline
(39.3, 46.0) & 0.00037 & 5.5e-05 & 0.00066 & 0.00066 \\
\hline
(46.0, 53.8) & 7.3e-05 & 6e-05 & 0.00089 & 0.0009 \\
\hline

\end{tabular}
\caption{Trigger bias and reconstruction uncertainty.}
\label{tab:trgbias}
\end{table}

\section{Dilution due to Vertex Finding}
The possible distortion on inclusive jet asymmetry due to the vertex finding algorithm is also studied in the analysis. A dilution uncertainty is introduced to account for this effect.	
\begin{align}
\Delta A_{LL, dilu} = A_{LL} \times (\frac{1}{r} - 1),
\end{align}
where \(r\) is the vertex finding efficiency. It is estimated by comparing the reconstructed most probable primary vertex to the vertex thrown at the beginning of the simulation. The \(r\) is calculated as the matching ratio between the reconstructed vertex at the detector level and the true vertex thrown in the simulation, where the matching requirement is that the difference between the two vertices is less than 2 cm. Table \ref{tab:vtxmatchingratio} shows the matching ratio for each of the jet \(p_{T}\) bins. Since the vertex matching ratio is above 90\% beyond the first two jet \(p_{T}\) bins, the dilution effect is expected to be tiny for the rest of the larger jet \(p_{T}\) bins.

\begin{table}[H]
\centering
\begin{tabular}{|c|c|}
\hline
$p_T$ & Vertex matching ratio \\
\hline
(6.0, 7.0) & 0.75 \\
\hline
(7.0, 8.2) & 0.80 \\
\hline
(8.2, 9.6) & 0.94 \\
\hline
(9.6, 11.2) & 0.94 \\
\hline
(11.2, 13.1) & 0.96 \\
\hline
(13.1, 15.3) & 0.95 \\
\hline
(15.3, 17.9) & 0.99 \\
\hline
(17.9, 20.9) & 0.99 \\
\hline
(20.9, 24.5) & 1 \\
\hline
(24.5, 28.7) & 1 \\
\hline
(28.7, 33.6) & 0.99 \\
\hline
(33.6, 39.3) & 1 \\
\hline
(39.3, 46.0) & 1 \\
\hline
(46.0, 53.8) & 1 \\
\hline
\end{tabular}
\caption{Vertex matching fraction in each jet $p_T$ after the underlying event correction from the embedding sample.}
\label{tab:vtxmatchingratio}
\end{table}
\section{Transverse Residual Double Spin Asymmetry Uncertainty}
The longitudinally polarized beams are not always 100\% perfectly polarized along the beam moving direction. The asymmetry measured, \(\varepsilon_{LL}\), therefore has contributions not only from the true double longitudinal double spin asymmetry, \(A_{LL}\), but also from the residual transverse asymmetry, \(A_\Sigma\), such as,
\begin{align}
\varepsilon_{LL} &= P_{b}P_{y}A_{LL} + \vec{P_{bT}} \cdot \vec{P_{yT}}A_{\Sigma} \nonumber \\
&= P_{b}P_{y}(A_{LL} + \frac{P_{bT}P_{yT}}{P_{b}P_{y}}A_{\Sigma}),
\end{align}
where \(P_{b(y)}\) is the longitudinal beam polarization for blue (yellow) beam and \(P_{b(y)T}\) is the transverse beam polarization for blue (yellow) beam. The transverse component fraction is estimated from the local transverse asymmetry measured by the ZDC. At the beginning of the 2012 510 GeV running period, the local transverse asymmetry was about 5.5\% and 5.0\% for the yellow and blue beams before the spin rotators were turned on. After the spin rotators were turned on, the value was 0.3\% and 0.3\% apiece for both beams. The transverse polarization fractions are \(\frac{P_{bT}}{P_{b}} \sim \frac{0.3\%}{5.5\%} \sim 0.05\) and \(\frac{P_{yT}}{P_{y}} \sim \frac{0.3\%}{5.0\%} \sim 0.06\). Based on the residual transverse asymmetry, \(A_{\Sigma}\), that was found to be less than 0.008 for jet \(p_{T} < 15\) GeV in the \(pp\) collisions at \(\sqrt{s} = 200\) GeV, \(A_{\Sigma}\) is expected to be less than 0.008 for jet \(p_{T} < 38\) GeV at \(\sqrt{s} = 510\) GeV by assuming \(x_{T} = \frac{2P_{T}}{\sqrt{s}}\) scaling holds. So the systematic uncertainty caused by the residual transverse asymmetry is at the level of \( 0.008 \times 0.05 \times 0.06 < 3 \times 10^{-5}\) in the jet \(p_{T}\) range measured in this analysis, which is negligible compared to other sources of systematical uncertainty.

\section{Non-collision Background Uncertainty}

\begin{figure}[H]
\centering
\includegraphics[scale=0.8]{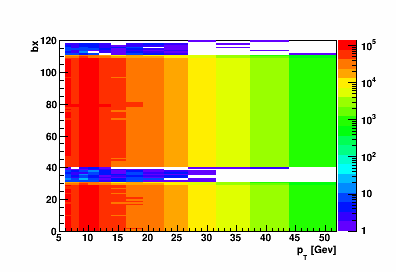}
\caption{Jet yields a function of jet $p_T$ and bunch crossing numbers from zero-bias events.}
\label{fig:noncollbkg}
\end{figure}

\begin{figure}[H]
\centering
\includegraphics[scale=0.5]{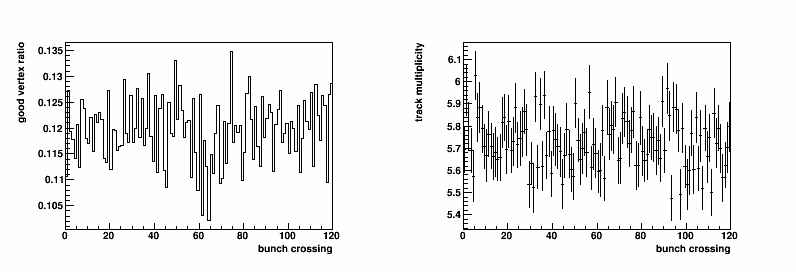}
\caption{Fractions of good vertex found vs. bunch crossings (left) and track multiplicity vs. bunch crossings (right) from zero-bias events. Note the offset vertical axes in each case.}
\label{fig:noncollbkgbx}
\end{figure}

The asymmetry from the non-collision background is unknown yet, however it is easy to estimate the fraction of the non-collision background contributions to the real jet signals. It was estimated by utilizing the data from the abort gaps and comparing with the data from the normal filled bunch crossings. Figure \ref{fig:noncollbkg} shows the 2-D histogram of jet \(p_{T}\) and bunch crossing number. It is easy to see that in the two abort gaps, bunch numbers from 31 to 39 and 111 to 119, the jet yields across the entire \(p_{T}\) range of the interest is down an order of \(10^{-4}\) with respect to that in the normal bunch crossings. It was also verified that all the bunching crossings including abort gaps have the same backgrounds by investigating distributions of the fraction of events that are found to have a good collision vertex and the average number of tracks in an event as a function of bunch crossing numbers. Figure \ref{fig:noncollbkgbx} show these distributions. They both turn out uniformly distributed across the bunch crossing numbers, which implies that all the bunch crossings more or less have the same non-collision backgrounds. The asymmetry for the non-collision background is expected to be small. For example, the asymmetry for the non-collision background was measured at \(\sqrt{s} = 200\) GeV during 2009. It was found to be less than 0.02 for jet \(p_T < \) 15 GeV and less than 0.08 for jet \(p_T <\) 35 GeV \cite{run9aLL2015}. However, even if the asymmetry caused by non-collision background is unity, the systematic effect on the inclusive jet \(A_{LL}\) would be small enough to be neglected because of the small contributions from the non-collision background to the inclusive jet production.

\section{Statistical Uncertainty on $p_{T}$ Shift}

The statistical uncertainty on the \(p_{T}\) shift as defined in \eqref{eq:ptshift} is considered and will be folded into the systematic uncertainty on the jet \(p_{T}\). 
%

\section{Calorimeter Energy Resolution and TPC Tracking Efficiency Uncertainty}

The energy deposited in the BEMC has its uncertainty which will contribute the jet \(p_{T}\) measured in this analysis. Since both charged tracks and neutral particles deposit energy into the BEMC towers, the systematic uncertainty on the jet \(p_{T}\) is coming from the BEMC neutral energy uncertainty and the BEMC track uncertainty, as shown in the following equation,

\begin{align}
\Delta  p_{T} = \sqrt{\Delta p_{T, BEMC, neutral}^{2} + \Delta p_{T, BEMC, track}^{2}}
\end{align}

which can also be expressed in terms of  fractional uncertainties,

\begin{align}
\Delta p_{T} = p_{T} \sqrt{\Delta f_{BEMC, neutral}^{2} + \Delta f_{BEMC, track}^{2}}
\end{align}

The fractional BEMC neutral energy uncertainty is contributed by the calibration gain uncertainty and its efficiency uncertainty, that is,

\begin{align}
\Delta f_{BEMC, neutral} = R_{t} \times \sqrt{\Delta gain^{2} + \Delta eff^{2}},
\end{align}
where \(R_{t} = \frac{\sum_{twr}p_{T}}{\sum_{trk}p_{T} + \sum_{twr} p_{T}}\) is the neutral energy fraction in the BEMC. The average \(R_t\) in each jet \(p_T\) is shown in Table \ref{tab:run12rt}. In this analysis, the gain calibration uncertainty is estimated to be 3.8\% as a conservative measure. The efficiency uncertainty is 1\%.

\begin{table}[H]
\centering
\begin{tabular}{|c|c|}
\hline
$p_T$ & $R_t$ \\
\hline
(6.0, 7.0) & 0.58 \\
\hline
(7.0, 8.2) & 0.55 \\
\hline
(8.2, 9.6) & 0.58 \\
\hline
(9.6, 11.2) & 0.54 \\
\hline
(11.2, 13.1) & 0.51 \\
\hline
(13.1, 15.3) & 0.47 \\
\hline
(15.3, 17.9) & 0.55 \\
\hline
(17.9, 20.9) & 0.55 \\
\hline
(20.9, 24.5) & 0.53 \\
\hline
(24.5, 28.7) & 0.5 \\
\hline
(28.7, 33.6) & 0.47 \\
\hline
(33.6, 39.3) & 0.45 \\
\hline
(39.3, 46.0) & 0.45 \\
\hline
(46.0, 53.8) & 0.44 \\
\hline
\end{tabular}
\caption{Average $R_t$ in each jet $p_T$ bin after the underlying event correction from the data}
\label{tab:run12rt}
\end{table}

The fractional track uncertainty is considered to be contributed by the TPC track momentum uncertainty and the BEMC track response uncertainty as in the equation

\begin{align}
\Delta f_{track} = \sqrt{[(1 - R_{t}) \times \Delta f_{trk,p}]^{2} + (1-R_{t})\times \Delta f_{BEMC, nonph}]^{2}}.
\end{align}

The TPC track momentum calibration fractional uncertainty \(\Delta f_{trk,p}\)  is conservatively estimated as 1\% based on the calibration of the TPC. \(\Delta f_{BEMC,nonph}\) is the BEMC fractional uncertainty due to non-photonic hadrons. The non-photonic hadrons include charged hadrons that were either not seen by the TPC or were seen by the TPC but did not project to the a BEMC tower and neutral hadrons that deposit a certain fraction of their energy into the BEMC towers. The total fractional possible non-phonic energy deposited in the BEMC is \( \frac{1}{\epsilon_{trk}}\times S_{hadron}\), where \(\epsilon_{trk}\) is the TPC tracking efficiency and \(S_{hadron}\) is the scale-up factor for neutral hadrons, \(\frac{E_{charged} + E_{neutral}}{E_{charged}}\). The tracking efficiency is taken as 65\%.This  is estimated from the recent STAR central \(Au + Au\) collisions at \(\sqrt{s} = 200\) GeV, which has similar TPC track densities as \(pp\) collisions at \(\sqrt{s} = 510\) GeV. The scale up factor for neutral hadrons is 1.16 \cite{MarciaET}. The 100\% track momentum subtraction from the projected tower is applied in the jet reconstruction, therefore the track energy that was measured from the TPC and projected to a BEMC tower should be subtracted away. Conservatively assuming only 50\% of a track's BEMC energy appears in the projected BEMC tower, then \(f_{proj}\), the amount that is removed by the \(p_T\) subtraction process, would be 0.5. Taking the BEMC response to non-photonic hadron energy,  \(f_{nonph}\) is 30\% and the BEMC systematic uncertainty due to non-photonic energy,  \(\Delta f_{nonph}\) is 9\% \cite{MarciaET}, then systematic uncertainty on jet \(p_T\) can be expressed as,

\begin{align}
\Delta f_{BEMC, nonph} = (\frac{1}{\epsilon_{trk}}S_{hadron} - f_{proj}) \times f_{nonph} \times \Delta f_{nonph}.
\end{align}

With all these numbers set, the track momentum fractional uncertainty \(\Delta f_{trk,p}\)  is estimated to be 3.5\%.

\section{Total Tracking Efficiency Uncertainty}
\begin{figure}[H]
\centering
\includegraphics[scale=0.8]{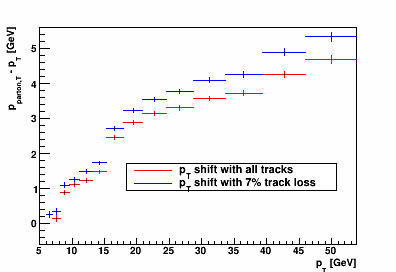}
\caption{Jet $p_T$ shift for 7\% track loss (blue) and no track loss (red).}
\label{fig:ptshift7pct}
\end{figure}

The uncertainty on jet \(p_{T}\) is due to tracking efficiency in TPC is also considered in this analysis. It is estimated by taking the difference the averaged jet \(p_{T}\) shift from the detector jet to parton jet between by using the full set of reconstructed tracks from the TPC and by using a partial set of reconstructed tracks from TPC. The partial set of reconstructed tracks from TPC is usually chosen by randomly rejecting a certain percent of tracks from the full set. Considering the luminosity in the 2012 RHIC 510 GeV \(pp\) run and general performance of the STAR TPC, the rejection fraction was chosen to be 7\%, which means rejecting seven tracks in every 100 tracks. The \(p_T\) shift for the 7\% percent track loss compared with the \(p_T\) shift with no track loss is shown in Figure \ref{fig:ptshift7pct}. The difference of the averaged jet \(p_T\) shift for each jet \(p_{T}\) bin is shown in Table \ref{tab:ptsyst}.

\section{Jet $p_T$ Shift Uncertainties due to Tune Variations}
The jet \(p_T\) shift uncertainties due to the PYTHIA is estimated in the analysis by utilizing the possible variants provided for Perugia 2012 in the PYTHIA version of 6.4.28. There are 13 other variants for Perugia 2012 tunes except the default tune. As shown in the Table \ref{tab:perugia2012vars}, some variants change the same set of parameters to control common activities for example the pair of tunes, 371 and 372, and the pair of 376 and 377. Some of them are related with underlying events. Since the underlying event systematic on jet \(p_T\) is estimated by the disagreement in underlying event correction, \(dp_T\), between the data and embedding sample, the tunes 373, 380, 381, and 382 are dropped. Both tunes 374 and 375 are related with color reconnections, so only tune 374 is kept. The tunes 378 and 379 replace the PDF in the default tune with MSTW 2008 and MRST PDFs. The newer MSTW 2008 PDF is kept. In total, 7 tunes, 371, 372, 374, 376, 377, 378 and 383 are chosen to compare with the default tune. Note that in the M/C simulation used in the embedding sample the exponent \(PARP(90)\) is reduced to 0.213; in this tune study, the \(p_{T,0}\), \(PARP(82)\), is increased by 7.3\% which corresponds the same effect as reducing the exponent. The \(p_{T,0}\) is controlled by several parameters beside the exponent parameter. These parameters vary in those variants. In order to avoid the complexity of changing multiple parameters in a tune to achieve the same effect, a constant up scale is applied for each tune to reduce the underlying event effect.
\begin{table}[H]
\centering
\begin{tabular}{|c|c|}
\hline
Tune number & Description \\
\hline
370 & Default \\
\hline
371 &  radHi, $\alpha_s(\frac{1}{2}p_{\perp})$ for ISR and FSR\\
\hline
372 & radLo, $\alpha_s(p_{\perp})$ for ISR and FSR\\
\hline
373 & mpiHi, $\lambda_{QCD} = 0.26$ GeV for MPI\\
\hline 
374 & loCR, less color reconnections\\
\hline
375 & noCR, no color reconnections\\
\hline
376 & FL, more longitudinal fragmentation\\
\hline
377 & FT, more transverse fragmentation\\
\hline
378 & MSLO, MSTW 2008 LO PDFs\\
\hline
379 & LO**, MRST LO** PDFs\\
\hline
380 & mb2, PARP(87) = 0\\
\hline
381 & ueHi, higher UE (lower $p_{T.0}$)\\
\hline
382 & ueLo, lower UE(higher $p_{T.0}$)\\
\hline
383 & IBK, Innsbruck hadronization parameters \\
\hline
\end{tabular}
\caption{The default Perugia 2012 tune and its variants.}
\label{tab:perugia2012vars}
\end{table}

\begin{figure}[H]
\centering
\includegraphics[scale=0.8]{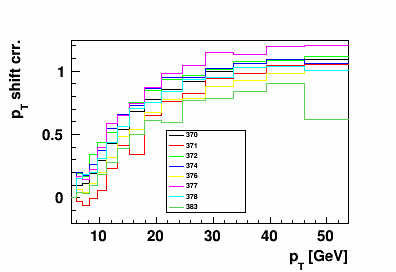}
\caption{Jet $p_T$ shift vs. particle jet $p_T$ from the default Perugia 2012 tune and its variants}
\label{fig:tuneptshift}
\end{figure}

Both particle and parton jets are reconstructed from the various tunes by using the same algorithm as used in this analysis. The particle jets are then matched to the parton jets. The averaged \(p_T\) shifts are calculated in each particle jet \(p_T\) bin where the bin scheme is the same as used in the data analysis. Figure \ref{fig:tuneptshift} shows the \(p_T\) shift vs particle jet \(p_T\) among all the tunes. The 371 and 372 are bracketing the default tune. This is reasonable since those tunes differ in the same set of parameters. The same situation applies for tunes 376 and 377.

\begin{figure}[H]
\centering
\includegraphics[scale=0.8]{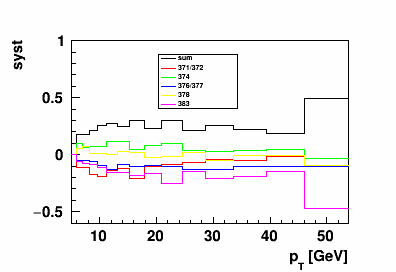}
\caption{Jet $p_T$ shift vs. particle jet $p_T$ from the default Perugia 2012 tune and its variants}
\label{fig:tunedpt}
\end{figure}

The systematic uncertainty due to tune variation is estimated by taking the square root of quadrature sum of the difference of \(dp_T\) for various tunes relative to the default tune. In exception, for the pair, 371 and 372, and the pair 376 and 377, the half of the absolute difference between the pairs is taken. For the other three tunes, 374, 378 and 383, the difference relative to the default tune is taken. Figure \ref{fig:tunedpt} shows the total \(p_T\) shift uncertainty that are contributed by these tunes.

Table \ref{tab:ptsyst} summarizes all the contributions to the jet \(p_T\) shift systematic uncertainties. Table \ref{tab:run12pt} shows the averaged jet \(p_T\) in each jet \(p_T\) bin and its corresponding \(p_T\) shifts.

\begin{table}[H]
\centering
\rotatebox{90}{
\begin{varwidth}{\textheight}
\begin{tabular}{|c|c|c|c|c|c|c|c|}
\hline
$p_T$ & $p_T$ shift syst. & BEMC track & BEMC neutral & syst. 7\% track loss & UE syst. & Tune syst. & Total \\
\hline
(6.0, 7.0) & 0.084 & 0.098 & 0.15 & -0.0062 & -0.016 & 0.18 & 0.26 \\
\hline
(7.0, 8.2) & 0.083 & 0.12 & 0.16 & -0.21 & -0.0017 & 0.18 & 0.35 \\
\hline
(8.2, 9.6) & 0.062 & 0.14 & 0.2 & -0.21 & -0.031 & 0.22 & 0.39 \\
\hline
(9.6, 11.2) & 0.053 & 0.17 & 0.22 & -0.13 & -0.045 & 0.25 & 0.41 \\
\hline
(11.2, 13.1) & 0.053 & 0.22 & 0.24 & -0.27 & -0.053 & 0.27 & 0.51 \\
\hline
(13.1, 15.3) & 0.045 & 0.27 & 0.26 & -0.26 & -0.071 & 0.24 & 0.52 \\
\hline
(15.3, 17.9) & 0.059 & 0.27 & 0.36 & -0.28 & -0.11 & 0.3 & 0.62 \\
\hline
(17.9, 20.9) & 0.054 & 0.31 & 0.42 & -0.36 & -0.11 & 0.23 & 0.68 \\
\hline
(20.9, 24.5) & 0.051 & 0.38 & 0.47 & -0.4 & -0.11 & 0.3 & 0.80 \\
\hline
(24.5, 28.7) & 0.059 & 0.48 & 0.52 & -0.48 & -0.13 & 0.21 & 0.89 \\
\hline
(28.7, 33.6) & 0.065 & 0.59 & 0.57 & -0.53 & -0.12 & 0.26 & 1.0 \\
\hline
(33.6, 39.3) & 0.075 & 0.71 & 0.64 & -0.53 & -0.13 & 0.22 & 1.1 \\
\hline
(39.3, 46.0) & 0.094 & 0.84 & 0.74 & -0.63 & -0.12 & 0.19 & 1.3 \\
\hline
(46.0, 53.8) & 0.11 & 0.99 & 0.85 & -0.65 & -0.14 & 0.49 & 1.6 \\
\hline
\end{tabular}
\caption{Jet $p_T$ systematics.}
\label{tab:ptsyst}
\end{varwidth}
}

\end{table}

\begin{table}[H]
\centering
\begin{tabular}{|c|c|c|c|}
\hline
$p_T$ & $<p_T>$ & $p_T$ shift & jet $p_T$ \\
\hline
(6.0, 7.0) & 6.48 & 0.26 & 6.74 \\
\hline
(7.0, 8.2) & 7.56 & 0.14 & 7.70 \\
\hline
(8.2, 9.6) & 8.86 & 0.89 & 9.76 \\
\hline
(9.6, 11.2) & 10.35 & 1.12 & 11.47 \\
\hline
(11.2, 13.1) & 12.07 & 1.22 & 13.29 \\
\hline
(13.1, 15.3) & 14.09 & 1.49 & 15.58 \\
\hline
(15.3, 17.9) & 16.52 & 2.45 & 18.96 \\
\hline
(17.9, 20.9) & 19.28 & 2.87 & 22.16 \\
\hline
(20.9, 24.5) & 22.52 & 3.14 & 25.66 \\
\hline
(24.5, 28.7) & 26.36 & 3.30 & 29.65 \\
\hline
(28.7, 33.6) & 30.81 & 3.56 & 34.38 \\
\hline
(33.6, 39.3) & 36.00 & 3.72 & 39.72 \\
\hline
(39.3, 46.0) & 42.06 & 4.26 & 46.32 \\
\hline
(46.0, 53.8) & 49.14 & 4.67 & 53.81 \\
\hline
\end{tabular}

\caption{Average jet $p_T$ and the $p_T$ shift for each jet $p_T$ bin.}
\label{tab:run12pt}
\end{table}

%% file: chapter7.tex
\chapter{\uppercase{Summary: Final Results and Their Impact}}
\begin{figure}[H]
\centering
\includegraphics[scale=0.6]{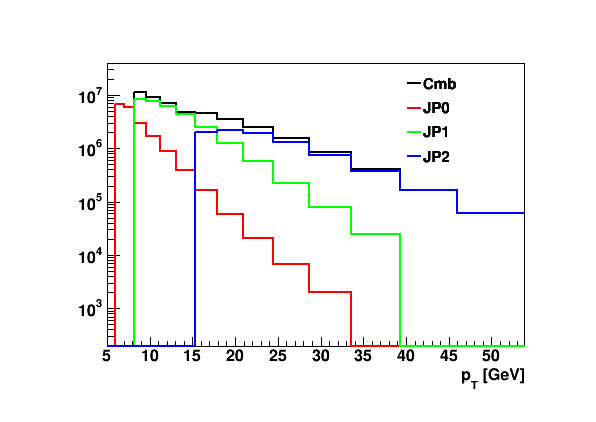}
\caption{The distribution of  the number of jets triggered by the three triggers, JP0, JP1, and JP2 found in this analysis, together with the summed number of jets distribution from the three triggers, labeled as combined.}
\label{fig:run12jetspec}
\end{figure}

Jets triggered by the three jet patch triggers, JP0, JP1 and JP2 from the 2012 RHIC run are counted towards the inclusive jet \(A_{LL}\) calculations. Figure \ref{fig:run12jetspec} shows the total number of jets triggered by the jet triggers used in this analysis. The lowest first two jet \(p_T\) bins are solely contributed by the JP0 jets. In the intermediate jet \(p_T\) range, JP1 trigger jets dominate. At high jet \(p_T\) above 20 GeV, the JP2 jets dominate.

\begin{figure}[H]
\centering
\includegraphics[scale=0.4]{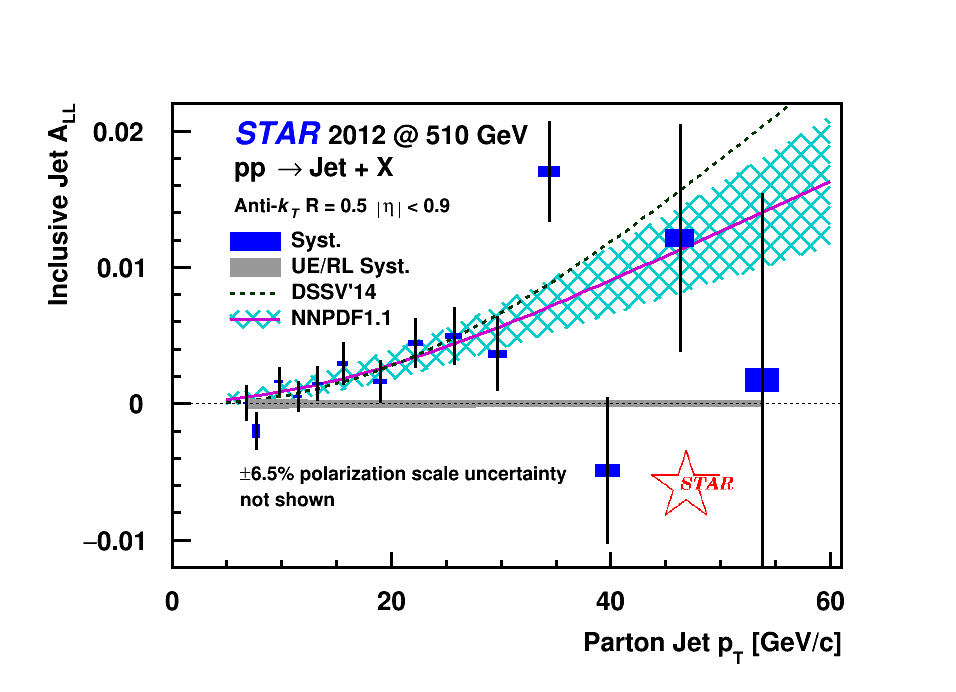}
\nocitecaption{STAR 2012 inclusive jet $A_{LL}$ with its statistical and systematical uncertainties. The results are compared with DSSV'14 \cite{DSSV14} and NNPDF predictions \cite{NNPDFpol}.}
\label{fig:run12aLLfinal}
\end{figure}

In the final results, trigger bias and reconstruction bias uncertainty and dilution are added in quadrature as the total systematic uncertainty on inclusive jet \(A_{LL}\). The jet \(p_{T}\) are plotted at the shifted average jet \(p_{T}\) value, which is equal to the average jet \(p_{T}\) in the jet \(p_{T}\) bin from the data plus the shift from the average detector level jet \(p_{T}\) to the average parton level jet \(p_{T}\) determined from the embedding sample. The uncertainty of the jet \(p_{T}\) is the quadrature sum of the BEMC energy scale uncertainty, total track efficiency uncertainty, \(p_{T}\) shift uncertainty and the PYTHIA tune uncertainties. Figure \ref{fig:run12aLLfinal} shows the inclusive jet \(A_{LL}\) vs. jet \(p_{T}\) bin with the latest theoretical predictions from those global analysis that include previous RHIC data in their fits \cite{DSSV14, NNPDFpol}. Table \ref{tab:run12aLLfinal} presents the numerical results from Figure \ref{fig:run12aLLfinal}, It shows than the measured \(A_{LL}\) is consistent with these model predictions.

\begin{figure}[H]
\centering
\includegraphics[scale=0.6]{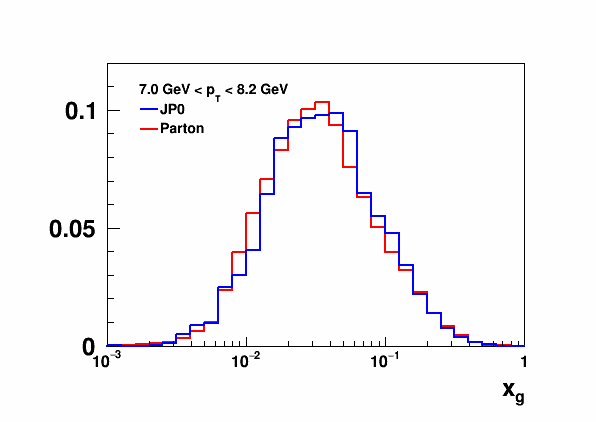}
\caption{Gluon momentum fraction $x_g$ sampled by the inclusive jet with detector jet $p_T$ between 7.0 and 8.2 GeV and by the parton jets with the same $p_T$ range.}
\label{fig:xgdis}
\end{figure}

\begin{table}[H]
\centering
\begin{tabular}{|c|c|c|c|c|c|}
\hline
$p_T$ & $p_T$ syst. & $A_{LL}$ corr. & $A_{LL}$ stat. & $A_{LL}$ syst. & UE/RL syst. \\
\hline
6.74 & 0.26 & 4.16e-05 & 0.00132 & 7.72e-05 & 0.000364 \\
\hline
7.70 & 0.35 & -0.00203 & 0.0014 & 0.000532 & 0.000326 \\
\hline
9.76 & 0.39 & 0.00162 & 0.00101 & 0.000124 & 0.000311 \\
\hline
11.5 & 0.41 & 0.000493 & 0.00112 & 8.99e-05 & 0.000284 \\
\hline
13.3 & 0.51 & 0.00147 & 0.00129 & 0.000115 & 0.000272 \\
\hline
15.6 & 0.52 & 0.00293 & 0.00159 & 0.000183 & 0.000256 \\
\hline
19.0 & 0.62 & 0.00158 & 0.00158 & 0.00019 & 0.000251 \\
\hline
22.2 & 0.69 & 0.00443 & 0.00181 & 0.000228 & 0.000246 \\
\hline
25.7 & 0.80 & 0.00494 & 0.00214 & 0.000202 & 0.000239 \\
\hline
29.7 & 0.89 & 0.00364 & 0.00273 & 0.0003 & 0.000234 \\
\hline
34.4 & 1.02 & 0.017 & 0.00372 & 0.000398 & 0.000231 \\
\hline
39.7 & 1.13 & -0.00492 & 0.00537 & 0.000453 & 0.000229 \\
\hline
46.3 & 1.31 & 0.0121 & 0.00836 & 0.000659 & 0.000227 \\
\hline
53.8 & 1.55 & 0.00173 & 0.0137 & 0.000895 & 0.000226 \\
\hline
\end{tabular}
\caption{Numerical values for the STAR 2012 inclusive jet $A_{LL}$ results, the parton jet $p_T$ with its uncertainties, the inclusive jet $A_{LL}$ with its statistical and systematic uncertainties, and the relative luminosity and underlying event uncertainties.}
\label{tab:run12aLLfinal}
\end{table}
Over this kinematic range, the inclusive jets are dominated by the \(qg\) and \(gg\) sub-processes of the two parton scattering. Therefore  the measured inclusive jets are sensitive the gluon distribution inside the proton. Figure \ref{fig:xgdis} shows the momentum fraction carried by gluons, \(x_g\), sampled by the inclusive jet with measured \(p_T\) between 7.0 and 8.2 GeV. It also shows the \(x_g\) distribution sampled by the parton jets inside the same \(p_T\) range. The difference of the two distributions reflect the \(p_T\) shift and trigger bias and reconstruction errors. For jets with \(p_T\) between 7.0 and 8.2 GeV, they sample a good fraction of gluons with \(x_g\) near 0.025.

\begin{figure}[H]
\centering
\includegraphics[scale=0.4]{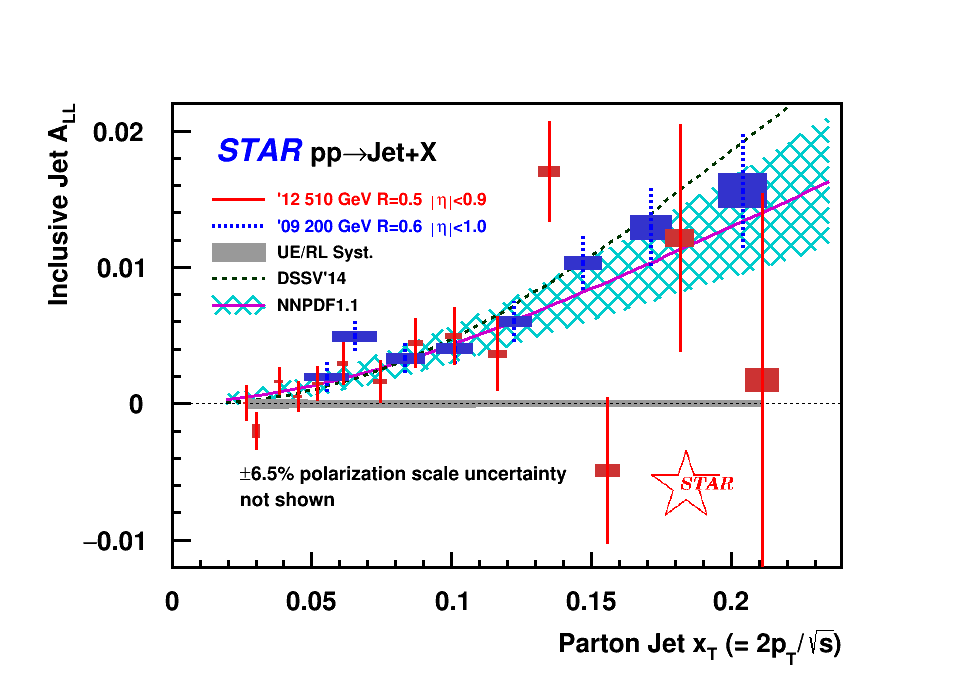}
\nocitecaption{STAR 2012 inclusive jet $A_{LL}$ vs. $x_T$ compared with STAR 2009 inclusive jet $A_{LL}$ results \cite{run9aLL2015}.}
\label{fig:run12aLLxT}
\end{figure}

The measured inclusive jet \(A_{LL}\) is compared with the STAR 2009 inclusive jet \(A_{LL}\) results. Since the two measurements have different center of mass energy, 510 GeV and 200 GeV respectively, the \(A_{LL}\) is compared on the scale of \(x_{T} = \frac{2p_{T}}{\sqrt{s}}\). The \(x_{T}\) is also an approximate proxy of \(x_g\) which is the momentum fraction carried by gluons inside the proton, since the inclusive jet production in the kinematics is dominated by the gluon contribution. The difference of inclusive jet asymmetries at the same \(x_{T}\) due to the \(x_{T}\) scaling violation is not discernible under the current experimental precision. Figure \ref{fig:run12aLLxT} shows the inclusive jet \(A_{LL}\) vs. \(x_{T}\) at \(\sqrt{s} = \)200 GeV and 510 GeV. The 510 GeV data extend inclusive \(A_{LL}\) measurements to the lower \(x_{T}\) region \(x_{T} \sim 0.02\) which is mostly dominated by gluon polarizations inside the proton with \(x_{g}\) in the approximately same range \(x_{g} \sim 0.02\). In the overlapping \(x_{T}\), the two sets of data at different center of mass energies are consistent with each other given the experimental precisions. 

\begin{figure}[H]
\centering
\includegraphics[scale=0.4]{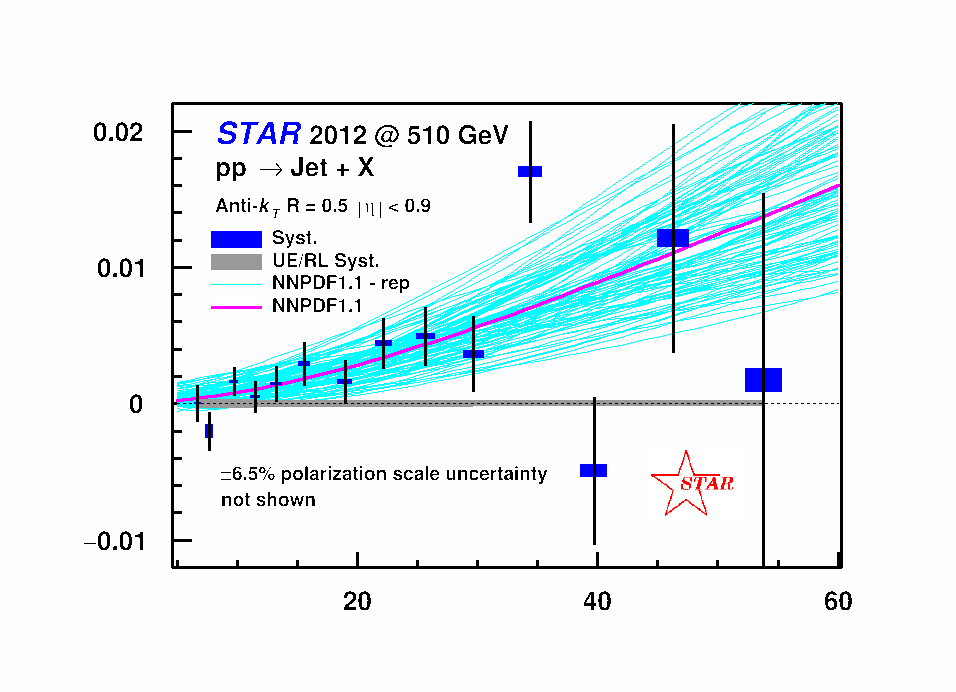}
\nocitecaption{STAR 2012 inclusive jet $A_{LL}$ vs. $p_T$ compared with the predictions from the 100 NNPDFpol1.1 replicas \cite{NNPDFpol}.}
\label{fig:run12aLLrep}
\end{figure}

In conclusion, the STAR 2012 510 GeV inclusive jet \(A_{LL}\) data explore gluon polarizations inside the proton and extend the gluon polarization measurements to lower \(x_{g}\) value in the vicinity of \(x_{g}\) around 0.02 with respect to the inclusive jet \(A_{LL}\) measurements from the STAR 2009 200 GeV data. The two data agree well in the overlapping \(x_{T}\) region by assuming the \(x_{T}\) scaling violation effect to be indiscernible. Figure \ref{fig:run12aLLrep} shows the 2012 results with the predictions from the 100 NNPDFpol1.1 equal-probability replicas \cite{NNPDFpol}. Some replicas sit below the data points and some sit above the data points. The replicas that lie far away from the data points will have smaller weights than those that lie close to the data points. Therefore a re-weighting procedure will produce smaller error bands for the \(\Delta g\) at low \(x\) than the current predictions. The results studied in this thesis will place important new constraints on the gluon polarizations in the future global NLO analysis to determine the polarized gluon PDF inside the proton, especially at low \(x_{g}\) range around \(x_{g} \sim 0.02\), where the current global data reach sparsely.

%% file: bibliography.tex
%
%
%

\phantomsection
\addcontentsline{toc}{chapter}{REFERENCES}

\renewcommand{\bibname}{{\normalsize\rm REFERENCES}}

\bibliographystyle{apsrev4-1}
\bibliography{references}

%% file: tamuthesis.bbl
\begin{thebibliography}{91}%
\makeatletter
\providecommand \@ifxundefined [1]{%
 \@ifx{#1\undefined}
}%
\providecommand \@ifnum [1]{%
 \ifnum #1\expandafter \@firstoftwo
 \else \expandafter \@secondoftwo
 \fi
}%
\providecommand \@ifx [1]{%
 \ifx #1\expandafter \@firstoftwo
 \else \expandafter \@secondoftwo
 \fi
}%
\providecommand \natexlab [1]{#1}%
\providecommand \enquote  [1]{``#1''}%
\providecommand \bibnamefont  [1]{#1}%
\providecommand \bibfnamefont [1]{#1}%
\providecommand \citenamefont [1]{#1}%
\providecommand \href@noop [0]{\@secondoftwo}%
\providecommand \href [0]{\begingroup \@sanitize@url \@href}%
\providecommand \@href[1]{\@@startlink{#1}\@@href}%
\providecommand \@@href[1]{\endgroup#1\@@endlink}%
\providecommand \@sanitize@url [0]{\catcode `\\12\catcode `\$12\catcode
  `\&12\catcode `\#12\catcode `\^12\catcode `\_12\catcode `\%12\relax}%
\providecommand \@@startlink[1]{}%
\providecommand \@@endlink[0]{}%
\providecommand \url  [0]{\begingroup\@sanitize@url \@url }%
\providecommand \@url [1]{\endgroup\@href {#1}{\urlprefix }}%
\providecommand \urlprefix  [0]{URL }%
\providecommand \Eprint [0]{\href }%
\providecommand \doibase [0]{http://dx.doi.org/}%
\providecommand \selectlanguage [0]{\@gobble}%
\providecommand \bibinfo  [0]{\@secondoftwo}%
\providecommand \bibfield  [0]{\@secondoftwo}%
\providecommand \translation [1]{[#1]}%
\providecommand \BibitemOpen [0]{}%
\providecommand \bibitemStop [0]{}%
\providecommand \bibitemNoStop [0]{.\EOS\space}%
\providecommand \EOS [0]{\spacefactor3000\relax}%
\providecommand \BibitemShut  [1]{\csname bibitem#1\endcsname}%
\let\auto@bib@innerbib\@empty
\bibitem [{\citenamefont {Zweig}(1964{\natexlab{a}})}]{ZweigOne}%
  \BibitemOpen
  \bibfield  {author} {\bibinfo {author} {\bibfnamefont {G.}~\bibnamefont
  {Zweig}},\ }\href@noop {} {\bibfield  {journal} {\bibinfo  {journal}
  {CERN-TH-401}\ } (\bibinfo {year} {1964}{\natexlab{a}})}\BibitemShut
  {NoStop}%
\bibitem [{\citenamefont {Zweig}(1964{\natexlab{b}})}]{ZweigTwo}%
  \BibitemOpen
  \bibfield  {author} {\bibinfo {author} {\bibfnamefont {G.}~\bibnamefont
  {Zweig}},\ }\href@noop {} {\bibfield  {journal} {\bibinfo  {journal}
  {CERN-TH-412}\ } (\bibinfo {year} {1964}{\natexlab{b}})}\BibitemShut
  {NoStop}%
\bibitem [{\citenamefont {Gell-Mann}(1964)}]{GMann1964}%
  \BibitemOpen
  \bibfield  {author} {\bibinfo {author} {\bibfnamefont {M.}~\bibnamefont
  {Gell-Mann}},\ }\href@noop {} {\bibfield  {journal} {\bibinfo  {journal}
  {Phys. Rev. Lett.}\ }\textbf {\bibinfo {volume} {8}},\ \bibinfo {pages} {214}
  (\bibinfo {year} {1964})}\BibitemShut {NoStop}%
\bibitem [{\citenamefont {Bloom}\ \emph {et~al.}(1969)\citenamefont {Bloom},
  \citenamefont {Coward}, \citenamefont {DeStaebler}, \citenamefont {Drees},
  \citenamefont {Miller}, \citenamefont {Mo}, \citenamefont {Taylor},
  \citenamefont {Breidenbach}, \citenamefont {Friedman}, \citenamefont
  {Hartmann},\ and\ \citenamefont {Kendall}}]{SLACiep930}%
  \BibitemOpen
  \bibfield  {author} {\bibinfo {author} {\bibfnamefont {E.~D.}\ \bibnamefont
  {Bloom}}, \bibinfo {author} {\bibfnamefont {D.~H.}\ \bibnamefont {Coward}},
  \bibinfo {author} {\bibfnamefont {H.}~\bibnamefont {DeStaebler}}, \bibinfo
  {author} {\bibfnamefont {J.}~\bibnamefont {Drees}}, \bibinfo {author}
  {\bibfnamefont {G.}~\bibnamefont {Miller}}, \bibinfo {author} {\bibfnamefont
  {L.~W.}\ \bibnamefont {Mo}}, \bibinfo {author} {\bibfnamefont {R.~E.}\
  \bibnamefont {Taylor}}, \bibinfo {author} {\bibfnamefont {M.}~\bibnamefont
  {Breidenbach}}, \bibinfo {author} {\bibfnamefont {J.~I.}\ \bibnamefont
  {Friedman}}, \bibinfo {author} {\bibfnamefont {G.~C.}\ \bibnamefont
  {Hartmann}}, \ and\ \bibinfo {author} {\bibfnamefont {H.~W.}\ \bibnamefont
  {Kendall}},\ }\href@noop {} {\bibfield  {journal} {\bibinfo  {journal} {Phys.
  Rev. Lett}\ }\textbf {\bibinfo {volume} {23}},\ \bibinfo {pages} {930}
  (\bibinfo {year} {1969})}\BibitemShut {NoStop}%
\bibitem [{\citenamefont {Breidenbach}\ \emph {et~al.}(1969)\citenamefont
  {Breidenbach}, \citenamefont {Friedman}, \citenamefont {Kendall},
  \citenamefont {Bloom}, \citenamefont {Coward}, \citenamefont {DeStaebler},
  \citenamefont {Drees}, \citenamefont {Mo},\ and\ \citenamefont
  {Taylor}}]{SLACiep935}%
  \BibitemOpen
  \bibfield  {author} {\bibinfo {author} {\bibfnamefont {M.}~\bibnamefont
  {Breidenbach}}, \bibinfo {author} {\bibfnamefont {J.~I.}\ \bibnamefont
  {Friedman}}, \bibinfo {author} {\bibfnamefont {H.~W.}\ \bibnamefont
  {Kendall}}, \bibinfo {author} {\bibfnamefont {E.~D.}\ \bibnamefont {Bloom}},
  \bibinfo {author} {\bibfnamefont {D.~H.}\ \bibnamefont {Coward}}, \bibinfo
  {author} {\bibfnamefont {H.}~\bibnamefont {DeStaebler}}, \bibinfo {author}
  {\bibfnamefont {J.}~\bibnamefont {Drees}}, \bibinfo {author} {\bibfnamefont
  {L.~W.}\ \bibnamefont {Mo}}, \ and\ \bibinfo {author} {\bibfnamefont {R.~E.}\
  \bibnamefont {Taylor}},\ }\href@noop {} {\bibfield  {journal} {\bibinfo
  {journal} {Phys. Rev. Lett}\ }\textbf {\bibinfo {volume} {23}},\ \bibinfo
  {pages} {935} (\bibinfo {year} {1969})}\BibitemShut {NoStop}%
\bibitem [{\citenamefont {Feynman}(1969)}]{Feynman1969}%
  \BibitemOpen
  \bibfield  {author} {\bibinfo {author} {\bibfnamefont {R.~P.}\ \bibnamefont
  {Feynman}},\ }\href@noop {} {\bibfield  {journal} {\bibinfo  {journal} {Phys.
  Rev. Lett.}\ }\textbf {\bibinfo {volume} {23}},\ \bibinfo {pages} {1415}
  (\bibinfo {year} {1969})}\BibitemShut {NoStop}%
\bibitem [{\citenamefont {Bjorken}\ and\ \citenamefont
  {Paschos}(1969)}]{Bjorken1969}%
  \BibitemOpen
  \bibfield  {author} {\bibinfo {author} {\bibfnamefont {J.~D.}\ \bibnamefont
  {Bjorken}}\ and\ \bibinfo {author} {\bibfnamefont {E.~A.}\ \bibnamefont
  {Paschos}},\ }\href@noop {} {\bibfield  {journal} {\bibinfo  {journal} {Phys.
  Rev. Lett.}\ }\textbf {\bibinfo {volume} {185}},\ \bibinfo {pages} {1975}
  (\bibinfo {year} {1969})}\BibitemShut {NoStop}%
\bibitem [{\citenamefont {Gross}\ and\ \citenamefont
  {Wilczek}(1973)}]{DGross1973}%
  \BibitemOpen
  \bibfield  {author} {\bibinfo {author} {\bibfnamefont {D.~J.}\ \bibnamefont
  {Gross}}\ and\ \bibinfo {author} {\bibfnamefont {F.}~\bibnamefont
  {Wilczek}},\ }\href@noop {} {\bibfield  {journal} {\bibinfo  {journal} {Phys.
  Rev. Lett.}\ }\textbf {\bibinfo {volume} {30}},\ \bibinfo {pages} {1343}
  (\bibinfo {year} {1973})}\BibitemShut {NoStop}%
\bibitem [{\citenamefont {Politzer}(1973)}]{HPolitzer1973}%
  \BibitemOpen
  \bibfield  {author} {\bibinfo {author} {\bibfnamefont {H.~D.}\ \bibnamefont
  {Politzer}},\ }\href@noop {} {\bibfield  {journal} {\bibinfo  {journal}
  {Phys. Rev. Lett.}\ }\textbf {\bibinfo {volume} {30}},\ \bibinfo {pages}
  {1346} (\bibinfo {year} {1973})}\BibitemShut {NoStop}%
\bibitem [{\citenamefont {Gribov}\ and\ \citenamefont
  {Lipatov}(1972)}]{Gribov1972}%
  \BibitemOpen
  \bibfield  {author} {\bibinfo {author} {\bibfnamefont {V.~N.}\ \bibnamefont
  {Gribov}}\ and\ \bibinfo {author} {\bibfnamefont {L.~N.}\ \bibnamefont
  {Lipatov}},\ }\href@noop {} {\bibfield  {journal} {\bibinfo  {journal} {Sov.
  J. Nucl. Phys.}\ }\textbf {\bibinfo {volume} {15}},\ \bibinfo {pages} {438}
  (\bibinfo {year} {1972})}\BibitemShut {NoStop}%
\bibitem [{\citenamefont {Dokshitzer}(1977)}]{Dokshitzer1977}%
  \BibitemOpen
  \bibfield  {author} {\bibinfo {author} {\bibfnamefont {Y.~L.}\ \bibnamefont
  {Dokshitzer}},\ }\href@noop {} {\bibfield  {journal} {\bibinfo  {journal}
  {Sov. Phys. JETP}\ }\textbf {\bibinfo {volume} {46}},\ \bibinfo {pages} {641}
  (\bibinfo {year} {1977})}\BibitemShut {NoStop}%
\bibitem [{\citenamefont {Altarelli}\ and\ \citenamefont
  {Parisi}(1977)}]{Altarelli1977}%
  \BibitemOpen
  \bibfield  {author} {\bibinfo {author} {\bibfnamefont {G.}~\bibnamefont
  {Altarelli}}\ and\ \bibinfo {author} {\bibfnamefont {G.}~\bibnamefont
  {Parisi}},\ }\href@noop {} {\bibfield  {journal} {\bibinfo  {journal} {Nucl.
  Phys. B}\ }\textbf {\bibinfo {volume} {126}},\ \bibinfo {pages} {298}
  (\bibinfo {year} {1977})}\BibitemShut {NoStop}%
\bibitem [{\citenamefont {Aaron}\ \emph {et~al.}(2010)\citenamefont {Aaron}
  \emph {et~al.}}]{h1zeus2010}%
  \BibitemOpen
  \bibfield  {author} {\bibinfo {author} {\bibfnamefont {F.~D.}\ \bibnamefont
  {Aaron}} \emph {et~al.} (\bibinfo {collaboration} {ZEUS and H1
  Collaboration}),\ }\href@noop {} {\ \textbf {\bibinfo {volume} {01}},\
  \bibinfo {pages} {109} (\bibinfo {year} {2010})},\ \Eprint
  {http://arxiv.org/abs/0911.0884} {arXiv:0911.0884 [hep-ex]} \BibitemShut
  {NoStop}%
\bibitem [{\citenamefont {Aaltonen}\ \emph {et~al.}(2008)\citenamefont
  {Aaltonen} \emph {et~al.}}]{cdfjets2008}%
  \BibitemOpen
  \bibfield  {author} {\bibinfo {author} {\bibfnamefont {T.}~\bibnamefont
  {Aaltonen}} \emph {et~al.} (\bibinfo {collaboration} {CDF Collaboration}),\
  }\href@noop {} {\bibfield  {journal} {\bibinfo  {journal} {Phys. Rev. D}\
  }\textbf {\bibinfo {volume} {78}},\ \bibinfo {pages} {052006} (\bibinfo
  {year} {2008})}\BibitemShut {NoStop}%
\bibitem [{\citenamefont {Abazov}\ \emph {et~al.}(2012)\citenamefont {Abazov}
  \emph {et~al.}}]{d0jets2012}%
  \BibitemOpen
  \bibfield  {author} {\bibinfo {author} {\bibfnamefont {V.~M.}\ \bibnamefont
  {Abazov}} \emph {et~al.} (\bibinfo {collaboration} {D0 Collaboration}),\
  }\href@noop {} {\bibfield  {journal} {\bibinfo  {journal} {Phys. Rev. D}\
  }\textbf {\bibinfo {volume} {85}},\ \bibinfo {pages} {052006} (\bibinfo
  {year} {2012})}\BibitemShut {NoStop}%
\bibitem [{\citenamefont {Acosta}\ \emph {et~al.}(2005)\citenamefont {Acosta}
  \emph {et~al.}}]{cdfw2005}%
  \BibitemOpen
  \bibfield  {author} {\bibinfo {author} {\bibfnamefont {D.}~\bibnamefont
  {Acosta}} \emph {et~al.} (\bibinfo {collaboration} {CDF Collaboration}),\
  }\href@noop {} {\bibfield  {journal} {\bibinfo  {journal} {Phys. Rev. D}\
  }\textbf {\bibinfo {volume} {71}},\ \bibinfo {pages} {051104} (\bibinfo
  {year} {2005})}\BibitemShut {NoStop}%
\bibitem [{\citenamefont {Abazov}\ \emph {et~al.}(2008)\citenamefont {Abazov}
  \emph {et~al.}}]{d0w2008}%
  \BibitemOpen
  \bibfield  {author} {\bibinfo {author} {\bibfnamefont {V.~M.}\ \bibnamefont
  {Abazov}} \emph {et~al.} (\bibinfo {collaboration} {D0 Collaboration}),\
  }\href@noop {} {\bibfield  {journal} {\bibinfo  {journal} {Phys. Rev. Lett.}\
  }\textbf {\bibinfo {volume} {101}},\ \bibinfo {pages} {211801} (\bibinfo
  {year} {2008})}\BibitemShut {NoStop}%
\bibitem [{\citenamefont {Antero}\ \emph {et~al.}(2010)\citenamefont {Antero}
  \emph {et~al.}}]{cdfz2010}%
  \BibitemOpen
  \bibfield  {author} {\bibinfo {author} {\bibfnamefont {T.}~\bibnamefont
  {Antero}} \emph {et~al.} (\bibinfo {collaboration} {CDF Collaboration}),\
  }\href@noop {} {\bibfield  {journal} {\bibinfo  {journal} {Phys. Lett. B}\
  }\textbf {\bibinfo {volume} {692}},\ \bibinfo {pages} {232} (\bibinfo {year}
  {2010})}\BibitemShut {NoStop}%
\bibitem [{\citenamefont {Abazov}\ \emph {et~al.}(2007)\citenamefont {Abazov}
  \emph {et~al.}}]{d0z2007}%
  \BibitemOpen
  \bibfield  {author} {\bibinfo {author} {\bibfnamefont {V.~M.}\ \bibnamefont
  {Abazov}} \emph {et~al.} (\bibinfo {collaboration} {D0 Collaboration}),\
  }\href@noop {} {\bibfield  {journal} {\bibinfo  {journal} {Phys. Rev. D}\
  }\textbf {\bibinfo {volume} {76}},\ \bibinfo {pages} {012003} (\bibinfo
  {year} {2007})}\BibitemShut {NoStop}%
\bibitem [{\citenamefont {Towell}\ \emph {et~al.}(2001)\citenamefont {Towell}
  \emph {et~al.}}]{e866nuesea2001}%
  \BibitemOpen
  \bibfield  {author} {\bibinfo {author} {\bibfnamefont {R.~S.}\ \bibnamefont
  {Towell}} \emph {et~al.} (\bibinfo {collaboration} {FNAL E866/NuSea
  Collaboration}),\ }\href@noop {} {\bibfield  {journal} {\bibinfo  {journal}
  {Phys. Rev. D}\ }\textbf {\bibinfo {volume} {64}},\ \bibinfo {pages} {052002}
  (\bibinfo {year} {2001})}\BibitemShut {NoStop}%
\bibitem [{\citenamefont {Abramowicz}\ \emph {et~al.}(2015)\citenamefont
  {Abramowicz} \emph {et~al.}}]{HERAPDF20}%
  \BibitemOpen
  \bibfield  {author} {\bibinfo {author} {\bibfnamefont {H.}~\bibnamefont
  {Abramowicz}} \emph {et~al.} (\bibinfo {collaboration} {H1 and ZEUS
  Collaboration}),\ }\href@noop {} {\bibfield  {journal} {\bibinfo  {journal}
  {Eur. Phys. J. C}\ }\textbf {\bibinfo {volume} {75}},\ \bibinfo {pages} {580}
  (\bibinfo {year} {2015})}\BibitemShut {NoStop}%
\bibitem [{\citenamefont {Pumplin}\ \emph {et~al.}(2002)\citenamefont
  {Pumplin}, \citenamefont {Stump}, \citenamefont {Huston}, \citenamefont
  {Lai}, \citenamefont {Nadolsky},\ and\ \citenamefont {Tung}}]{cteq6l2002}%
  \BibitemOpen
  \bibfield  {author} {\bibinfo {author} {\bibfnamefont {J.}~\bibnamefont
  {Pumplin}}, \bibinfo {author} {\bibfnamefont {D.~R.}\ \bibnamefont {Stump}},
  \bibinfo {author} {\bibfnamefont {J.}~\bibnamefont {Huston}}, \bibinfo
  {author} {\bibfnamefont {H.~L.}\ \bibnamefont {Lai}}, \bibinfo {author}
  {\bibfnamefont {P.~M.}\ \bibnamefont {Nadolsky}}, \ and\ \bibinfo {author}
  {\bibfnamefont {W.~K.}\ \bibnamefont {Tung}},\ }\href@noop {} {\bibfield
  {journal} {\bibinfo  {journal} {JHEP}\ }\textbf {\bibinfo {volume} {07}},\
  \bibinfo {pages} {012} (\bibinfo {year} {2002})},\ \Eprint
  {http://arxiv.org/abs/0201195} {arXiv:0201195 [hep-ph]} \BibitemShut
  {NoStop}%
\bibitem [{\citenamefont {Lai}\ \emph {et~al.}(2010)\citenamefont {Lai},
  \citenamefont {Guzzi}, \citenamefont {Huston}, \citenamefont {Li},
  \citenamefont {Nadolsky}, \citenamefont {Pumplin},\ and\ \citenamefont
  {Yuan}}]{CT10}%
  \BibitemOpen
  \bibfield  {author} {\bibinfo {author} {\bibfnamefont {H.-L.}\ \bibnamefont
  {Lai}}, \bibinfo {author} {\bibfnamefont {M.}~\bibnamefont {Guzzi}}, \bibinfo
  {author} {\bibfnamefont {J.}~\bibnamefont {Huston}}, \bibinfo {author}
  {\bibfnamefont {Z.}~\bibnamefont {Li}}, \bibinfo {author} {\bibfnamefont
  {P.~M.}\ \bibnamefont {Nadolsky}}, \bibinfo {author} {\bibfnamefont
  {J.}~\bibnamefont {Pumplin}}, \ and\ \bibinfo {author} {\bibfnamefont
  {C.-P.}\ \bibnamefont {Yuan}},\ }\href@noop {} {\bibfield  {journal}
  {\bibinfo  {journal} {Phys. Rev. D}\ }\textbf {\bibinfo {volume} {82}},\
  \bibinfo {pages} {074024} (\bibinfo {year} {2010})}\BibitemShut {NoStop}%
\bibitem [{\citenamefont {Dulat}\ \emph {et~al.}(2016)\citenamefont {Dulat},
  \citenamefont {Hou}, \citenamefont {Gao}, \citenamefont {Guzzi},
  \citenamefont {Huston}, \citenamefont {Nadolsky}, \citenamefont {Pumplin},
  \citenamefont {Schmidt}, \citenamefont {Stump},\ and\ \citenamefont
  {Yuan}}]{CT14}%
  \BibitemOpen
  \bibfield  {author} {\bibinfo {author} {\bibfnamefont {S.}~\bibnamefont
  {Dulat}}, \bibinfo {author} {\bibfnamefont {T.-J.}\ \bibnamefont {Hou}},
  \bibinfo {author} {\bibfnamefont {J.}~\bibnamefont {Gao}}, \bibinfo {author}
  {\bibfnamefont {M.}~\bibnamefont {Guzzi}}, \bibinfo {author} {\bibfnamefont
  {J.}~\bibnamefont {Huston}}, \bibinfo {author} {\bibfnamefont
  {P.}~\bibnamefont {Nadolsky}}, \bibinfo {author} {\bibfnamefont
  {J.}~\bibnamefont {Pumplin}}, \bibinfo {author} {\bibfnamefont
  {C.}~\bibnamefont {Schmidt}}, \bibinfo {author} {\bibfnamefont
  {D.}~\bibnamefont {Stump}}, \ and\ \bibinfo {author} {\bibfnamefont {C.-P.}\
  \bibnamefont {Yuan}},\ }\href@noop {} {\bibfield  {journal} {\bibinfo
  {journal} {Phys. Rev. D}\ }\textbf {\bibinfo {volume} {93}},\ \bibinfo
  {pages} {033006} (\bibinfo {year} {2016})}\BibitemShut {NoStop}%
\bibitem [{\citenamefont {Martin}\ \emph {et~al.}()\citenamefont {Martin},
  \citenamefont {Stirling}, \citenamefont {Thorne},\ and\ \citenamefont
  {Watt}}]{MSTW}%
  \BibitemOpen
  \bibfield  {author} {\bibinfo {author} {\bibfnamefont {A.~D.}\ \bibnamefont
  {Martin}}, \bibinfo {author} {\bibfnamefont {W.~J.}\ \bibnamefont
  {Stirling}}, \bibinfo {author} {\bibfnamefont {R.~S.}\ \bibnamefont
  {Thorne}}, \ and\ \bibinfo {author} {\bibfnamefont {G.}~\bibnamefont
  {Watt}},\ }\href@noop {} {}\Eprint {http://arxiv.org/abs/0901.0002}
  {arXiv:0901.0002 [hep-th]} \BibitemShut {NoStop}%
\bibitem [{\citenamefont {Ball}\ \emph {et~al.}(2015)\citenamefont {Ball} \emph
  {et~al.}}]{NNPDFunp}%
  \BibitemOpen
  \bibfield  {author} {\bibinfo {author} {\bibfnamefont {R.~D.}\ \bibnamefont
  {Ball}} \emph {et~al.} (\bibinfo {collaboration} {NNPDF Collaboration}),\
  }\href@noop {} {\bibfield  {journal} {\bibinfo  {journal} {JHEP}\ }\textbf
  {\bibinfo {volume} {04}},\ \bibinfo {pages} {40} (\bibinfo {year}
  {2015})}\BibitemShut {NoStop}%
\bibitem [{\citenamefont {Jaff}\ and\ \citenamefont
  {Manohar}()}]{JaffeManohar1990}%
  \BibitemOpen
  \bibfield  {author} {\bibinfo {author} {\bibfnamefont {R.~L.}\ \bibnamefont
  {Jaff}}\ and\ \bibinfo {author} {\bibfnamefont {A.}~\bibnamefont {Manohar}},\
  }\href@noop {} {\bibfield  {journal} {\bibinfo  {journal} {Nucl. Phys. B.}\
  }\textbf {\bibinfo {volume} {337}},\ \bibinfo {pages} {509}}\BibitemShut
  {NoStop}%
\bibitem [{\citenamefont {Allkofer}\ \emph {et~al.}(1981)\citenamefont
  {Allkofer} \emph {et~al.}}]{emcexp1981}%
  \BibitemOpen
  \bibfield  {author} {\bibinfo {author} {\bibfnamefont {O.~C.}\ \bibnamefont
  {Allkofer}} \emph {et~al.} (\bibinfo {collaboration} {EMC Collaboration}),\
  }\href@noop {} {\bibfield  {journal} {\bibinfo  {journal} {Nucl. Instr. and
  Meth.}\ }\textbf {\bibinfo {volume} {179}},\ \bibinfo {pages} {445} (\bibinfo
  {year} {1981})}\BibitemShut {NoStop}%
\bibitem [{\citenamefont {Bjorken}(1970)}]{bjokensum1970}%
  \BibitemOpen
  \bibfield  {author} {\bibinfo {author} {\bibfnamefont {J.~D.}\ \bibnamefont
  {Bjorken}},\ }\href@noop {} {\bibfield  {journal} {\bibinfo  {journal} {Phys.
  Lett. D}\ }\textbf {\bibinfo {volume} {1}},\ \bibinfo {pages} {1376}
  (\bibinfo {year} {1970})}\BibitemShut {NoStop}%
\bibitem [{\citenamefont {Ellis}\ and\ \citenamefont
  {Jaffe}(1974)}]{ellisjaffe1974}%
  \BibitemOpen
  \bibfield  {author} {\bibinfo {author} {\bibfnamefont {J.}~\bibnamefont
  {Ellis}}\ and\ \bibinfo {author} {\bibfnamefont {R.}~\bibnamefont {Jaffe}},\
  }\href@noop {} {\bibfield  {journal} {\bibinfo  {journal} {Phys. Lett. D}\
  }\textbf {\bibinfo {volume} {9}},\ \bibinfo {pages} {1444} (\bibinfo {year}
  {1974})}\BibitemShut {NoStop}%
\bibitem [{\citenamefont {Ashman}\ \emph {et~al.}(1988)\citenamefont {Ashman}
  \emph {et~al.}}]{emc1988}%
  \BibitemOpen
  \bibfield  {author} {\bibinfo {author} {\bibfnamefont {J.}~\bibnamefont
  {Ashman}} \emph {et~al.} (\bibinfo {collaboration} {EMC Collaboration}),\
  }\href@noop {} {\bibfield  {journal} {\bibinfo  {journal} {Phys. Lett. B}\
  }\textbf {\bibinfo {volume} {206}},\ \bibinfo {pages} {364} (\bibinfo {year}
  {1988})}\BibitemShut {NoStop}%
\bibitem [{\citenamefont {Abbon}\ \emph {et~al.}(2007)\citenamefont {Abbon}
  \emph {et~al.}}]{compassexp2007}%
  \BibitemOpen
  \bibfield  {author} {\bibinfo {author} {\bibfnamefont {P.}~\bibnamefont
  {Abbon}} \emph {et~al.} (\bibinfo {collaboration} {Compass Collaboration}),\
  }\href@noop {} {\bibfield  {journal} {\bibinfo  {journal} {Nucl. Instr. and
  Meth. A}\ }\textbf {\bibinfo {volume} {577}},\ \bibinfo {pages} {455}
  (\bibinfo {year} {2007})}\BibitemShut {NoStop}%
\bibitem [{\citenamefont {Abbon}\ \emph {et~al.}(2010)\citenamefont {Abbon}
  \emph {et~al.}}]{compassg1p2010}%
  \BibitemOpen
  \bibfield  {author} {\bibinfo {author} {\bibfnamefont {M.~G.}\ \bibnamefont
  {Abbon}} \emph {et~al.} (\bibinfo {collaboration} {Compass Collaboration}),\
  }\href@noop {} {\bibfield  {journal} {\bibinfo  {journal} {Phys. Lett. B}\
  }\textbf {\bibinfo {volume} {690}},\ \bibinfo {pages} {466} (\bibinfo {year}
  {2010})}\BibitemShut {NoStop}%
\bibitem [{\citenamefont {Alexakhin}\ \emph {et~al.}(2007)\citenamefont
  {Alexakhin} \emph {et~al.}}]{compassg1n2007}%
  \BibitemOpen
  \bibfield  {author} {\bibinfo {author} {\bibfnamefont {V.~Y.}\ \bibnamefont
  {Alexakhin}} \emph {et~al.} (\bibinfo {collaboration} {Compass
  Collaboration}),\ }\href@noop {} {\bibfield  {journal} {\bibinfo  {journal}
  {Phys. Lett. B}\ }\textbf {\bibinfo {volume} {647}},\ \bibinfo {pages} {8}
  (\bibinfo {year} {2007})}\BibitemShut {NoStop}%
\bibitem [{\citenamefont {Alexakhin}\ \emph {et~al.}(2010)\citenamefont
  {Alexakhin} \emph {et~al.}}]{compasssidisg1p2010}%
  \BibitemOpen
  \bibfield  {author} {\bibinfo {author} {\bibfnamefont {M.}~\bibnamefont
  {Alexakhin}} \emph {et~al.} (\bibinfo {collaboration} {Compass
  Collaboration}),\ }\href@noop {} {\bibfield  {journal} {\bibinfo  {journal}
  {Phys. Lett. B}\ }\textbf {\bibinfo {volume} {693}},\ \bibinfo {pages} {227}
  (\bibinfo {year} {2010})}\BibitemShut {NoStop}%
\bibitem [{\citenamefont {Alexakhin}\ \emph {et~al.}(2009)\citenamefont
  {Alexakhin} \emph {et~al.}}]{compasssidisg1n2009}%
  \BibitemOpen
  \bibfield  {author} {\bibinfo {author} {\bibfnamefont {M.}~\bibnamefont
  {Alexakhin}} \emph {et~al.} (\bibinfo {collaboration} {Compass
  Collaboration}),\ }\href@noop {} {\bibfield  {journal} {\bibinfo  {journal}
  {Phys. Lett. B}\ }\textbf {\bibinfo {volume} {680}},\ \bibinfo {pages} {217}
  (\bibinfo {year} {2009})}\BibitemShut {NoStop}%
\bibitem [{\citenamefont {Ageev}\ \emph {et~al.}(2006)\citenamefont {Ageev}
  \emph {et~al.}}]{compassphoton2006}%
  \BibitemOpen
  \bibfield  {author} {\bibinfo {author} {\bibfnamefont {E.~S.}\ \bibnamefont
  {Ageev}} \emph {et~al.} (\bibinfo {collaboration} {Compass Collaboration}),\
  }\href@noop {} {\bibfield  {journal} {\bibinfo  {journal} {Phys. Lett. B}\
  }\textbf {\bibinfo {volume} {633}},\ \bibinfo {pages} {25} (\bibinfo {year}
  {2006})}\BibitemShut {NoStop}%
\bibitem [{\citenamefont {Alekseev}\ \emph {et~al.}(2009)\citenamefont
  {Alekseev} \emph {et~al.}}]{compassdcharm2009}%
  \BibitemOpen
  \bibfield  {author} {\bibinfo {author} {\bibfnamefont {M.}~\bibnamefont
  {Alekseev}} \emph {et~al.} (\bibinfo {collaboration} {Compass
  Collaboration}),\ }\href@noop {} {\bibfield  {journal} {\bibinfo  {journal}
  {Phys. Lett. B}\ }\textbf {\bibinfo {volume} {676}},\ \bibinfo {pages} {31}
  (\bibinfo {year} {2009})}\BibitemShut {NoStop}%
\bibitem [{\citenamefont {Adolph}\ \emph
  {et~al.}(2013{\natexlab{a}})\citenamefont {Adolph} \emph
  {et~al.}}]{compasscharm2013}%
  \BibitemOpen
  \bibfield  {author} {\bibinfo {author} {\bibfnamefont {C.}~\bibnamefont
  {Adolph}} \emph {et~al.} (\bibinfo {collaboration} {Compass Collaboration}),\
  }\href@noop {} {\bibfield  {journal} {\bibinfo  {journal} {Phys. Rev. D}\
  }\textbf {\bibinfo {volume} {87}},\ \bibinfo {pages} {052018} (\bibinfo
  {year} {2013}{\natexlab{a}})}\BibitemShut {NoStop}%
\bibitem [{\citenamefont {Adolph}\ \emph
  {et~al.}(2013{\natexlab{b}})\citenamefont {Adolph} \emph
  {et~al.}}]{compasshadronpairs2013}%
  \BibitemOpen
  \bibfield  {author} {\bibinfo {author} {\bibfnamefont {C.}~\bibnamefont
  {Adolph}} \emph {et~al.} (\bibinfo {collaboration} {Compass Collaboration}),\
  }\href@noop {} {\bibfield  {journal} {\bibinfo  {journal} {Phys. Lett. B}\
  }\textbf {\bibinfo {volume} {718}},\ \bibinfo {pages} {922} (\bibinfo {year}
  {2013}{\natexlab{b}})}\BibitemShut {NoStop}%
\bibitem [{\citenamefont {Airapetian}\ \emph
  {et~al.}(2005{\natexlab{a}})\citenamefont {Airapetian} \emph
  {et~al.}}]{hermestargets2005}%
  \BibitemOpen
  \bibfield  {author} {\bibinfo {author} {\bibfnamefont {A.}~\bibnamefont
  {Airapetian}} \emph {et~al.} (\bibinfo {collaboration} {HERMES
  Collaboration}),\ }\href@noop {} {\bibfield  {journal} {\bibinfo  {journal}
  {Nucl. Instr. and Meth. A}\ }\textbf {\bibinfo {volume} {540}},\ \bibinfo
  {pages} {68} (\bibinfo {year} {2005}{\natexlab{a}})}\BibitemShut {NoStop}%
\bibitem [{\citenamefont {Ackestaff}\ \emph {et~al.}(1998)\citenamefont
  {Ackestaff} \emph {et~al.}}]{hermesexp1998}%
  \BibitemOpen
  \bibfield  {author} {\bibinfo {author} {\bibfnamefont {A.}~\bibnamefont
  {Ackestaff}} \emph {et~al.} (\bibinfo {collaboration} {HERMES
  Collaboration}),\ }\href@noop {} {\bibfield  {journal} {\bibinfo  {journal}
  {Nucl. Instr. and Meth. A}\ }\textbf {\bibinfo {volume} {417}},\ \bibinfo
  {pages} {230} (\bibinfo {year} {1998})}\BibitemShut {NoStop}%
\bibitem [{\citenamefont {Airapetian}\ \emph {et~al.}(2007)\citenamefont
  {Airapetian} \emph {et~al.}}]{hermesdis2007}%
  \BibitemOpen
  \bibfield  {author} {\bibinfo {author} {\bibfnamefont {A.}~\bibnamefont
  {Airapetian}} \emph {et~al.} (\bibinfo {collaboration} {HERMES
  Collaboration}),\ }\href@noop {} {\bibfield  {journal} {\bibinfo  {journal}
  {Phys. Rev. D}\ }\textbf {\bibinfo {volume} {75}},\ \bibinfo {pages} {012007}
  (\bibinfo {year} {2007})}\BibitemShut {NoStop}%
\bibitem [{\citenamefont {Airapetian}\ \emph
  {et~al.}(2005{\natexlab{b}})\citenamefont {Airapetian} \emph
  {et~al.}}]{hermes2005}%
  \BibitemOpen
  \bibfield  {author} {\bibinfo {author} {\bibfnamefont {A.}~\bibnamefont
  {Airapetian}} \emph {et~al.} (\bibinfo {collaboration} {HERMES
  Collaboration}),\ }\href@noop {} {\bibfield  {journal} {\bibinfo  {journal}
  {Phys. Rev. D}\ }\textbf {\bibinfo {volume} {71}},\ \bibinfo {pages} {012003}
  (\bibinfo {year} {2005}{\natexlab{b}})}\BibitemShut {NoStop}%
\bibitem [{\citenamefont {Airapetian}\ \emph {et~al.}(2010)\citenamefont
  {Airapetian} \emph {et~al.}}]{hermeshighpt2010}%
  \BibitemOpen
  \bibfield  {author} {\bibinfo {author} {\bibfnamefont {A.}~\bibnamefont
  {Airapetian}} \emph {et~al.} (\bibinfo {collaboration} {HERMES
  Collaboration}),\ }\href@noop {} {\bibfield  {journal} {\bibinfo  {journal}
  {JHEP}\ }\textbf {\bibinfo {volume} {08}},\ \bibinfo {pages} {130} (\bibinfo
  {year} {2010})}\BibitemShut {NoStop}%
\bibitem [{\citenamefont {Bl{\"{u}}mlein}\ and\ \citenamefont
  {B{\"{o}}ttcher}(2010)}]{BB10}%
  \BibitemOpen
  \bibfield  {author} {\bibinfo {author} {\bibfnamefont {J.}~\bibnamefont
  {Bl{\"{u}}mlein}}\ and\ \bibinfo {author} {\bibfnamefont {H.}~\bibnamefont
  {B{\"{o}}ttcher}},\ }\href@noop {} {\bibfield  {journal} {\bibinfo  {journal}
  {Nucl. Phys. B}\ }\textbf {\bibinfo {volume} {841}},\ \bibinfo {pages} {205}
  (\bibinfo {year} {2010})}\BibitemShut {NoStop}%
\bibitem [{\citenamefont {Leader}\ \emph {et~al.}(2010)\citenamefont {Leader},
  \citenamefont {Sidorov},\ and\ \citenamefont {Stamenov}}]{LSS10}%
  \BibitemOpen
  \bibfield  {author} {\bibinfo {author} {\bibfnamefont {E.}~\bibnamefont
  {Leader}}, \bibinfo {author} {\bibfnamefont {A.~V.}\ \bibnamefont {Sidorov}},
  \ and\ \bibinfo {author} {\bibfnamefont {D.~B.}\ \bibnamefont {Stamenov}},\
  }\href@noop {} {\bibfield  {journal} {\bibinfo  {journal} {Phys. Rev. D}\
  }\textbf {\bibinfo {volume} {82}},\ \bibinfo {pages} {1140812} (\bibinfo
  {year} {2010})}\BibitemShut {NoStop}%
\bibitem [{\citenamefont {{de Florian}}\ \emph {et~al.}(2008)\citenamefont {{de
  Florian}}, \citenamefont {Sassot}, \citenamefont {Stratmann},\ and\
  \citenamefont {Vogelsang}}]{DSSV08}%
  \BibitemOpen
  \bibfield  {author} {\bibinfo {author} {\bibfnamefont {D.}~\bibnamefont {{de
  Florian}}}, \bibinfo {author} {\bibfnamefont {R.}~\bibnamefont {Sassot}},
  \bibinfo {author} {\bibfnamefont {M.}~\bibnamefont {Stratmann}}, \ and\
  \bibinfo {author} {\bibfnamefont {W.}~\bibnamefont {Vogelsang}},\ }\href@noop
  {} {\bibfield  {journal} {\bibinfo  {journal} {Phys. Rev. Lett.}\ }\textbf
  {\bibinfo {volume} {101}},\ \bibinfo {pages} {072001} (\bibinfo {year}
  {2008})}\BibitemShut {NoStop}%
\bibitem [{\citenamefont {{de Florian}}\ \emph {et~al.}(2009)\citenamefont {{de
  Florian}}, \citenamefont {Sassot}, \citenamefont {Stratmann},\ and\
  \citenamefont {Vogelsang}}]{DSSV09}%
  \BibitemOpen
  \bibfield  {author} {\bibinfo {author} {\bibfnamefont {D.}~\bibnamefont {{de
  Florian}}}, \bibinfo {author} {\bibfnamefont {R.}~\bibnamefont {Sassot}},
  \bibinfo {author} {\bibfnamefont {M.}~\bibnamefont {Stratmann}}, \ and\
  \bibinfo {author} {\bibfnamefont {W.}~\bibnamefont {Vogelsang}},\ }\href@noop
  {} {\bibfield  {journal} {\bibinfo  {journal} {Phys. Rev. D.}\ }\textbf
  {\bibinfo {volume} {80}},\ \bibinfo {pages} {034030} (\bibinfo {year}
  {2009})}\BibitemShut {NoStop}%
\bibitem [{\citenamefont {{de Florian}}\ \emph {et~al.}(2014)\citenamefont {{de
  Florian}}, \citenamefont {Sassot}, \citenamefont {Stratmann},\ and\
  \citenamefont {Vogelsang}}]{DSSV14}%
  \BibitemOpen
  \bibfield  {author} {\bibinfo {author} {\bibfnamefont {D.}~\bibnamefont {{de
  Florian}}}, \bibinfo {author} {\bibfnamefont {R.}~\bibnamefont {Sassot}},
  \bibinfo {author} {\bibfnamefont {M.}~\bibnamefont {Stratmann}}, \ and\
  \bibinfo {author} {\bibfnamefont {W.}~\bibnamefont {Vogelsang}},\ }\href@noop
  {} {\bibfield  {journal} {\bibinfo  {journal} {Phys. Rev. Lett.}\ }\textbf
  {\bibinfo {volume} {113}},\ \bibinfo {pages} {012001} (\bibinfo {year}
  {2014})}\BibitemShut {NoStop}%
\bibitem [{\citenamefont {Ball}\ \emph {et~al.}(2014)\citenamefont {Ball} \emph
  {et~al.}}]{NNPDFpol}%
  \BibitemOpen
  \bibfield  {author} {\bibinfo {author} {\bibfnamefont {R.~D.}\ \bibnamefont
  {Ball}} \emph {et~al.} (\bibinfo {collaboration} {NNPDF Collaboration}),\
  }\href@noop {} {\bibfield  {journal} {\bibinfo  {journal} {Nucl. Phys. B}\
  }\textbf {\bibinfo {volume} {887}},\ \bibinfo {pages} {276} (\bibinfo {year}
  {2014})}\BibitemShut {NoStop}%
\bibitem [{\citenamefont {Ball}\ \emph {et~al.}(2013)\citenamefont {Ball} \emph
  {et~al.}}]{NNPDFpol10}%
  \BibitemOpen
  \bibfield  {author} {\bibinfo {author} {\bibfnamefont {R.~D.}\ \bibnamefont
  {Ball}} \emph {et~al.} (\bibinfo {collaboration} {NNPDF Collaboration}),\
  }\href@noop {} {\bibfield  {journal} {\bibinfo  {journal} {Nucl. Phys. B}\
  }\textbf {\bibinfo {volume} {874}},\ \bibinfo {pages} {36} (\bibinfo {year}
  {2013})}\BibitemShut {NoStop}%
\bibitem [{\citenamefont {Sato}\ \emph {et~al.}(2016)\citenamefont {Sato},
  \citenamefont {Melnitchouk}, \citenamefont {Kuhn}, \citenamefont {Eithier},\
  and\ \citenamefont {Accardi}}]{JAM}%
  \BibitemOpen
  \bibfield  {author} {\bibinfo {author} {\bibfnamefont {N.}~\bibnamefont
  {Sato}}, \bibinfo {author} {\bibfnamefont {W.}~\bibnamefont {Melnitchouk}},
  \bibinfo {author} {\bibfnamefont {S.~E.}\ \bibnamefont {Kuhn}}, \bibinfo
  {author} {\bibfnamefont {J.~J.}\ \bibnamefont {Eithier}}, \ and\ \bibinfo
  {author} {\bibfnamefont {A.}~\bibnamefont {Accardi}} (\bibinfo
  {collaboration} {Jefferson Lab Angular Momentum Collaboration}),\ }\href@noop
  {} {\bibfield  {journal} {\bibinfo  {journal} {Phys. Rev. D}\ }\textbf
  {\bibinfo {volume} {93}},\ \bibinfo {pages} {074005} (\bibinfo {year}
  {2016})}\BibitemShut {NoStop}%
\bibitem [{\citenamefont {Blazey}\ \emph {et~al.}()\citenamefont {Blazey},
  \citenamefont {Dittmann}, \citenamefont {Ellis}, \citenamefont {Elvira},
  \citenamefont {Frame}, \citenamefont {Grinstein}, \citenamefont {Hirosky},
  \citenamefont {Piegaia}, \citenamefont {Schellman}, \citenamefont {Snihur},
  \citenamefont {V.Sorin},\ and\ \citenamefont {Zeppenfeld}}]{conesalgo}%
  \BibitemOpen
  \bibfield  {author} {\bibinfo {author} {\bibfnamefont {G.~C.}\ \bibnamefont
  {Blazey}}, \bibinfo {author} {\bibfnamefont {J.~R.}\ \bibnamefont
  {Dittmann}}, \bibinfo {author} {\bibfnamefont {S.~D.}\ \bibnamefont {Ellis}},
  \bibinfo {author} {\bibfnamefont {V.~D.}\ \bibnamefont {Elvira}}, \bibinfo
  {author} {\bibfnamefont {K.}~\bibnamefont {Frame}}, \bibinfo {author}
  {\bibfnamefont {S.}~\bibnamefont {Grinstein}}, \bibinfo {author}
  {\bibfnamefont {R.}~\bibnamefont {Hirosky}}, \bibinfo {author} {\bibfnamefont
  {R.}~\bibnamefont {Piegaia}}, \bibinfo {author} {\bibfnamefont
  {H.}~\bibnamefont {Schellman}}, \bibinfo {author} {\bibfnamefont
  {R.}~\bibnamefont {Snihur}}, \bibinfo {author} {\bibnamefont {V.Sorin}}, \
  and\ \bibinfo {author} {\bibfnamefont {D.}~\bibnamefont {Zeppenfeld}},\
  }\href@noop {} {}\Eprint {http://arxiv.org/abs/0005012} {arXiv:0005012
  [hep-ex]} \BibitemShut {NoStop}%
\bibitem [{\citenamefont {Catani}\ \emph {et~al.}(1993)\citenamefont {Catani},
  \citenamefont {Dokshitzer}, \citenamefont {Seymour},\ and\ \citenamefont
  {Webber}}]{kt1993}%
  \BibitemOpen
  \bibfield  {author} {\bibinfo {author} {\bibfnamefont {S.}~\bibnamefont
  {Catani}}, \bibinfo {author} {\bibfnamefont {Y.~L.}\ \bibnamefont
  {Dokshitzer}}, \bibinfo {author} {\bibfnamefont {M.~H.}\ \bibnamefont
  {Seymour}}, \ and\ \bibinfo {author} {\bibfnamefont {B.~R.}\ \bibnamefont
  {Webber}},\ }\href@noop {} {\bibfield  {journal} {\bibinfo  {journal} {Nucl.
  Phys. B}\ }\textbf {\bibinfo {volume} {406}},\ \bibinfo {pages} {187}
  (\bibinfo {year} {1993})}\BibitemShut {NoStop}%
\bibitem [{\citenamefont {Ellis}\ and\ \citenamefont
  {Soper}(1993)}]{ktellis1993}%
  \BibitemOpen
  \bibfield  {author} {\bibinfo {author} {\bibfnamefont {S.~D.}\ \bibnamefont
  {Ellis}}\ and\ \bibinfo {author} {\bibfnamefont {D.~E.}\ \bibnamefont
  {Soper}},\ }\href@noop {} {\bibfield  {journal} {\bibinfo  {journal} {Phys.
  Rev. D}\ }\textbf {\bibinfo {volume} {48}},\ \bibinfo {pages} {3160}
  (\bibinfo {year} {1993})}\BibitemShut {NoStop}%
\bibitem [{\citenamefont {Dokshitzer}\ \emph {et~al.}(1997)\citenamefont
  {Dokshitzer}, \citenamefont {Leder}, \citenamefont {Moretti},\ and\
  \citenamefont {Webber}}]{cajet1997}%
  \BibitemOpen
  \bibfield  {author} {\bibinfo {author} {\bibfnamefont {Y.~L.}\ \bibnamefont
  {Dokshitzer}}, \bibinfo {author} {\bibfnamefont {G.~D.}\ \bibnamefont
  {Leder}}, \bibinfo {author} {\bibfnamefont {S.}~\bibnamefont {Moretti}}, \
  and\ \bibinfo {author} {\bibfnamefont {B.~R.}\ \bibnamefont {Webber}},\
  }\href@noop {} {\bibfield  {journal} {\bibinfo  {journal} {JHEP}\ }\textbf
  {\bibinfo {volume} {9708}},\ \bibinfo {pages} {001} (\bibinfo {year}
  {1997})}\BibitemShut {NoStop}%
\bibitem [{\citenamefont {Wobisch}\ and\ \citenamefont
  {Wengler}()}]{cajet1999}%
  \BibitemOpen
  \bibfield  {author} {\bibinfo {author} {\bibfnamefont {M.}~\bibnamefont
  {Wobisch}}\ and\ \bibinfo {author} {\bibfnamefont {T.}~\bibnamefont
  {Wengler}},\ }\href@noop {} {}\Eprint {http://arxiv.org/abs/9907280}
  {arXiv:9907280 [hep-ph]} \BibitemShut {NoStop}%
\bibitem [{\citenamefont {Abelev}\ \emph {et~al.}(2006)\citenamefont {Abelev}
  \emph {et~al.}}]{run3run4res2006}%
  \BibitemOpen
  \bibfield  {author} {\bibinfo {author} {\bibfnamefont {B.~I.}\ \bibnamefont
  {Abelev}} \emph {et~al.} (\bibinfo {collaboration} {STAR Collaboration}),\
  }\href@noop {} {\bibfield  {journal} {\bibinfo  {journal} {Phys. Rev. Lett.}\
  }\textbf {\bibinfo {volume} {97}},\ \bibinfo {pages} {252001} (\bibinfo
  {year} {2006})}\BibitemShut {NoStop}%
\bibitem [{\citenamefont {Abelev}\ \emph {et~al.}(2008)\citenamefont {Abelev}
  \emph {et~al.}}]{run6aLL2008}%
  \BibitemOpen
  \bibfield  {author} {\bibinfo {author} {\bibfnamefont {B.~I.}\ \bibnamefont
  {Abelev}} \emph {et~al.} (\bibinfo {collaboration} {STAR Collaboration}),\
  }\href@noop {} {\bibfield  {journal} {\bibinfo  {journal} {Phys. Rev. Lett.}\
  }\textbf {\bibinfo {volume} {100}},\ \bibinfo {pages} {232003} (\bibinfo
  {year} {2008})}\BibitemShut {NoStop}%
\bibitem [{\citenamefont {Adamczyk}\ \emph {et~al.}(2012)\citenamefont
  {Adamczyk} \emph {et~al.}}]{run6aLL2012}%
  \BibitemOpen
  \bibfield  {author} {\bibinfo {author} {\bibfnamefont {L.}~\bibnamefont
  {Adamczyk}} \emph {et~al.} (\bibinfo {collaboration} {STAR Collaboration}),\
  }\href@noop {} {\bibfield  {journal} {\bibinfo  {journal} {Phys. Rev. D.}\
  }\textbf {\bibinfo {volume} {86}},\ \bibinfo {pages} {032006} (\bibinfo
  {year} {2012})}\BibitemShut {NoStop}%
\bibitem [{\citenamefont {Sakuma}(2010)}]{sakumarun6}%
  \BibitemOpen
  \bibfield  {author} {\bibinfo {author} {\bibfnamefont {T.}~\bibnamefont
  {Sakuma}},\ }\emph {\bibinfo {title} {{Inclusive Jet and Dijet Production in
  Polarized proton-proton Collisions at $\sqrt{s} = $200 GeV at RHIC }}},\
  \href@noop {} {Ph.D. thesis},\ \bibinfo  {school} {{Massachusetts Institute
  of Technology}} (\bibinfo {year} {2010})\BibitemShut {NoStop}%
\bibitem [{\citenamefont {Li}(2015)}]{run9crsdis}%
  \BibitemOpen
  \bibfield  {author} {\bibinfo {author} {\bibfnamefont {X.}~\bibnamefont
  {Li}},\ }in\ \href@noop {} {\emph {\bibinfo {booktitle}
  {{PoS(DIS2015)203}}}}\ (\bibinfo {year} {2015})\BibitemShut {NoStop}%
\bibitem [{\citenamefont {Webb}(2013)}]{run9dijet500}%
  \BibitemOpen
  \bibfield  {author} {\bibinfo {author} {\bibfnamefont {G.}~\bibnamefont
  {Webb}},\ }in\ \href@noop {} {\emph {\bibinfo {booktitle}
  {{PoS(DIS2013)215}}}}\ (\bibinfo {year} {2013})\BibitemShut {NoStop}%
\bibitem [{\citenamefont {Mukherjee}\ and\ \citenamefont
  {Vogelsang}(2012)}]{nlojet2012}%
  \BibitemOpen
  \bibfield  {author} {\bibinfo {author} {\bibfnamefont {A.}~\bibnamefont
  {Mukherjee}}\ and\ \bibinfo {author} {\bibfnamefont {W.}~\bibnamefont
  {Vogelsang}},\ }\href@noop {} {\bibfield  {journal} {\bibinfo  {journal}
  {Phys. Rev. D}\ }\textbf {\bibinfo {volume} {86}},\ \bibinfo {pages} {094009}
  (\bibinfo {year} {2012})}\BibitemShut {NoStop}%
\bibitem [{\citenamefont {Adamczyk}\ \emph {et~al.}(2015)\citenamefont
  {Adamczyk} \emph {et~al.}}]{run9aLL2015}%
  \BibitemOpen
  \bibfield  {author} {\bibinfo {author} {\bibfnamefont {L.}~\bibnamefont
  {Adamczyk}} \emph {et~al.} (\bibinfo {collaboration} {STAR Collaboration}),\
  }\href@noop {} {\bibfield  {journal} {\bibinfo  {journal} {Phys. Rev. Lett.}\
  }\textbf {\bibinfo {volume} {115}},\ \bibinfo {pages} {092002} (\bibinfo
  {year} {2015})}\BibitemShut {NoStop}%
\bibitem [{\citenamefont {Harrison}\ \emph {et~al.}(2003)\citenamefont
  {Harrison}, \citenamefont {Ludlam},\ and\ \citenamefont {Ozaki}}]{rhicproj}%
  \BibitemOpen
  \bibfield  {author} {\bibinfo {author} {\bibfnamefont {M.}~\bibnamefont
  {Harrison}}, \bibinfo {author} {\bibfnamefont {T.}~\bibnamefont {Ludlam}}, \
  and\ \bibinfo {author} {\bibfnamefont {S.}~\bibnamefont {Ozaki}},\
  }\href@noop {} {\bibfield  {journal} {\bibinfo  {journal} {Nucl. Instr. and
  Meth. A}\ }\textbf {\bibinfo {volume} {499}},\ \bibinfo {pages} {235}
  (\bibinfo {year} {2003})}\BibitemShut {NoStop}%
\bibitem [{\citenamefont {Hahn}\ \emph {et~al.}(2003)\citenamefont {Hahn},
  \citenamefont {Forsyth}, \citenamefont {Foelsche}, \citenamefont {Harrison},
  \citenamefont {Kewisch}, \citenamefont {Parzen}, \citenamefont {Peggs},
  \citenamefont {Raka}, \citenamefont {Ruggiero}, \citenamefont {Stevens},
  \citenamefont {Tepikian}, \citenamefont {Thieberger}, \citenamefont
  {Trbojevic}, \citenamefont {Wei}, \citenamefont {Willen}, \citenamefont
  {Ozaki},\ and\ \citenamefont {Lee}}]{rhicdesign}%
  \BibitemOpen
  \bibfield  {author} {\bibinfo {author} {\bibfnamefont {H.}~\bibnamefont
  {Hahn}}, \bibinfo {author} {\bibfnamefont {E.}~\bibnamefont {Forsyth}},
  \bibinfo {author} {\bibfnamefont {H.}~\bibnamefont {Foelsche}}, \bibinfo
  {author} {\bibfnamefont {M.}~\bibnamefont {Harrison}}, \bibinfo {author}
  {\bibfnamefont {J.}~\bibnamefont {Kewisch}}, \bibinfo {author} {\bibfnamefont
  {G.}~\bibnamefont {Parzen}}, \bibinfo {author} {\bibfnamefont
  {S.}~\bibnamefont {Peggs}}, \bibinfo {author} {\bibfnamefont
  {E.}~\bibnamefont {Raka}}, \bibinfo {author} {\bibfnamefont {A.}~\bibnamefont
  {Ruggiero}}, \bibinfo {author} {\bibfnamefont {A.}~\bibnamefont {Stevens}},
  \bibinfo {author} {\bibfnamefont {S.}~\bibnamefont {Tepikian}}, \bibinfo
  {author} {\bibfnamefont {P.}~\bibnamefont {Thieberger}}, \bibinfo {author}
  {\bibfnamefont {D.}~\bibnamefont {Trbojevic}}, \bibinfo {author}
  {\bibfnamefont {J.}~\bibnamefont {Wei}}, \bibinfo {author} {\bibfnamefont
  {E.}~\bibnamefont {Willen}}, \bibinfo {author} {\bibfnamefont
  {S.}~\bibnamefont {Ozaki}}, \ and\ \bibinfo {author} {\bibfnamefont
  {S.}~\bibnamefont {Lee}},\ }\href@noop {} {\bibfield  {journal} {\bibinfo
  {journal} {Nucl. Instr. and Meth. A}\ }\textbf {\bibinfo {volume} {499}},\
  \bibinfo {pages} {245} (\bibinfo {year} {2003})}\BibitemShut {NoStop}%
\bibitem [{\citenamefont {Alekseev}\ \emph {et~al.}(2003)\citenamefont
  {Alekseev}, \citenamefont {Allgower}, \citenamefont {Bai}, \citenamefont
  {Batygin}, \citenamefont {Bozano}, \citenamefont {Brown}, \citenamefont
  {Bunce}, \citenamefont {Cameron}, \citenamefont {Courant}, \citenamefont
  {Erin} \emph {et~al.}}]{rhicpolarize}%
  \BibitemOpen
  \bibfield  {author} {\bibinfo {author} {\bibfnamefont {I.}~\bibnamefont
  {Alekseev}}, \bibinfo {author} {\bibfnamefont {C.}~\bibnamefont {Allgower}},
  \bibinfo {author} {\bibfnamefont {M.}~\bibnamefont {Bai}}, \bibinfo {author}
  {\bibfnamefont {Y.}~\bibnamefont {Batygin}}, \bibinfo {author} {\bibfnamefont
  {L.}~\bibnamefont {Bozano}}, \bibinfo {author} {\bibfnamefont
  {K.}~\bibnamefont {Brown}}, \bibinfo {author} {\bibfnamefont
  {G.}~\bibnamefont {Bunce}}, \bibinfo {author} {\bibfnamefont
  {P.}~\bibnamefont {Cameron}}, \bibinfo {author} {\bibfnamefont
  {E.}~\bibnamefont {Courant}}, \bibinfo {author} {\bibfnamefont
  {S.}~\bibnamefont {Erin}},  \emph {et~al.},\ }\href@noop {} {\bibfield
  {journal} {\bibinfo  {journal} {Nucl. Instr. and Meth. A}\ }\textbf {\bibinfo
  {volume} {499}},\ \bibinfo {pages} {392} (\bibinfo {year}
  {2003})}\BibitemShut {NoStop}%
\bibitem [{\citenamefont {Okada}\ \emph {et~al.}()\citenamefont {Okada},
  \citenamefont {Alekseev}, \citenamefont {Bravar}, \citenamefont {Bunce},
  \citenamefont {Dhawan}, \citenamefont {Gill}, \citenamefont {Haeberli},
  \citenamefont {Jinnouchi}, \citenamefont {Khodinov}, \citenamefont {Makdisi}
  \emph {et~al.}}]{hjetpol}%
  \BibitemOpen
  \bibfield  {author} {\bibinfo {author} {\bibfnamefont {H.}~\bibnamefont
  {Okada}}, \bibinfo {author} {\bibfnamefont {I.}~\bibnamefont {Alekseev}},
  \bibinfo {author} {\bibfnamefont {A.}~\bibnamefont {Bravar}}, \bibinfo
  {author} {\bibfnamefont {G.}~\bibnamefont {Bunce}}, \bibinfo {author}
  {\bibfnamefont {S.}~\bibnamefont {Dhawan}}, \bibinfo {author} {\bibfnamefont
  {R.}~\bibnamefont {Gill}}, \bibinfo {author} {\bibfnamefont {W.}~\bibnamefont
  {Haeberli}}, \bibinfo {author} {\bibfnamefont {O.}~\bibnamefont {Jinnouchi}},
  \bibinfo {author} {\bibfnamefont {A.}~\bibnamefont {Khodinov}}, \bibinfo
  {author} {\bibfnamefont {Y.}~\bibnamefont {Makdisi}},  \emph {et~al.},\
  }\href@noop {} {}\Eprint {http://arxiv.org/abs/0601001} {arXiv:0601001
  [hep-ex]} \BibitemShut {NoStop}%
\bibitem [{\citenamefont {O.Jinnouchi}\ \emph {et~al.}()\citenamefont
  {O.Jinnouchi}, \citenamefont {I.G.Alekseev}, \citenamefont {A.Bravar},
  \citenamefont {G.Bunce}, \citenamefont {S.Dhawan}, \citenamefont {H.Huang},
  \citenamefont {G.Igo}, \citenamefont {V.P.Kanavets}, \citenamefont
  {K.Kurita}, \citenamefont {H.Okada} \emph {et~al.}}]{pCpol}%
  \BibitemOpen
  \bibfield  {author} {\bibinfo {author} {\bibnamefont {O.Jinnouchi}}, \bibinfo
  {author} {\bibnamefont {I.G.Alekseev}}, \bibinfo {author} {\bibnamefont
  {A.Bravar}}, \bibinfo {author} {\bibnamefont {G.Bunce}}, \bibinfo {author}
  {\bibnamefont {S.Dhawan}}, \bibinfo {author} {\bibnamefont {H.Huang}},
  \bibinfo {author} {\bibnamefont {G.Igo}}, \bibinfo {author} {\bibnamefont
  {V.P.Kanavets}}, \bibinfo {author} {\bibnamefont {K.Kurita}}, \bibinfo
  {author} {\bibnamefont {H.Okada}},  \emph {et~al.},\ }\href@noop {} {}\Eprint
  {http://arxiv.org/abs/0412053} {arXiv:0412053 [nucl-ex]} \BibitemShut
  {NoStop}%
\bibitem [{\citenamefont {Ackermann}\ \emph {et~al.}(2003)\citenamefont
  {Ackermann}, \citenamefont {Adams}, \citenamefont {Adler}, \citenamefont
  {Ahammed}, \citenamefont {Ahmad}, \citenamefont {Allgower}, \citenamefont
  {Amonett}, \citenamefont {Amsbaugh}, \citenamefont {Anderson}, \citenamefont
  {Anderson} \emph {et~al.}}]{rhicstar}%
  \BibitemOpen
  \bibfield  {author} {\bibinfo {author} {\bibfnamefont {K.}~\bibnamefont
  {Ackermann}}, \bibinfo {author} {\bibfnamefont {N.}~\bibnamefont {Adams}},
  \bibinfo {author} {\bibfnamefont {C.}~\bibnamefont {Adler}}, \bibinfo
  {author} {\bibfnamefont {Z.}~\bibnamefont {Ahammed}}, \bibinfo {author}
  {\bibfnamefont {S.}~\bibnamefont {Ahmad}}, \bibinfo {author} {\bibfnamefont
  {C.}~\bibnamefont {Allgower}}, \bibinfo {author} {\bibfnamefont
  {J.}~\bibnamefont {Amonett}}, \bibinfo {author} {\bibfnamefont
  {J.}~\bibnamefont {Amsbaugh}}, \bibinfo {author} {\bibfnamefont
  {B.}~\bibnamefont {Anderson}}, \bibinfo {author} {\bibfnamefont
  {M.}~\bibnamefont {Anderson}},  \emph {et~al.},\ }\href@noop {} {\bibfield
  {journal} {\bibinfo  {journal} {Nucl. Instr. and Meth. A}\ }\textbf {\bibinfo
  {volume} {499}},\ \bibinfo {pages} {624} (\bibinfo {year}
  {2003})}\BibitemShut {NoStop}%
\bibitem [{\citenamefont {Anderson}\ \emph {et~al.}(2003)\citenamefont
  {Anderson}, \citenamefont {Berkovitz}, \citenamefont {Betts}, \citenamefont
  {Bossingham}, \citenamefont {Bieser}, \citenamefont {Brown}, \citenamefont
  {Burks}, \citenamefont {de~la Barca~Sánchez}, \citenamefont {Cebra},
  \citenamefont {Cherney} \emph {et~al.}}]{startpc}%
  \BibitemOpen
  \bibfield  {author} {\bibinfo {author} {\bibfnamefont {M.}~\bibnamefont
  {Anderson}}, \bibinfo {author} {\bibfnamefont {J.}~\bibnamefont {Berkovitz}},
  \bibinfo {author} {\bibfnamefont {W.}~\bibnamefont {Betts}}, \bibinfo
  {author} {\bibfnamefont {R.}~\bibnamefont {Bossingham}}, \bibinfo {author}
  {\bibfnamefont {F.}~\bibnamefont {Bieser}}, \bibinfo {author} {\bibfnamefont
  {R.}~\bibnamefont {Brown}}, \bibinfo {author} {\bibfnamefont
  {M.}~\bibnamefont {Burks}}, \bibinfo {author} {\bibfnamefont {M.~C.}\
  \bibnamefont {de~la Barca~Sánchez}}, \bibinfo {author} {\bibfnamefont
  {D.}~\bibnamefont {Cebra}}, \bibinfo {author} {\bibfnamefont
  {M.}~\bibnamefont {Cherney}},  \emph {et~al.},\ }\href@noop {} {\bibfield
  {journal} {\bibinfo  {journal} {Nucl. Instr. and Meth. A}\ }\textbf {\bibinfo
  {volume} {499}},\ \bibinfo {pages} {659} (\bibinfo {year}
  {2003})}\BibitemShut {NoStop}%
\bibitem [{\citenamefont {Beddo}\ \emph {et~al.}(2003)\citenamefont {Beddo},
  \citenamefont {Bielick}, \citenamefont {Fornek}, \citenamefont {Guarino},
  \citenamefont {Hill}, \citenamefont {Krueger}, \citenamefont {LeCompte},
  \citenamefont {Lopiano}, \citenamefont {Spinka}, \citenamefont {Underwood}
  \emph {et~al.}}]{starbemc}%
  \BibitemOpen
  \bibfield  {author} {\bibinfo {author} {\bibfnamefont {M.}~\bibnamefont
  {Beddo}}, \bibinfo {author} {\bibfnamefont {E.}~\bibnamefont {Bielick}},
  \bibinfo {author} {\bibfnamefont {T.}~\bibnamefont {Fornek}}, \bibinfo
  {author} {\bibfnamefont {V.}~\bibnamefont {Guarino}}, \bibinfo {author}
  {\bibfnamefont {D.}~\bibnamefont {Hill}}, \bibinfo {author} {\bibfnamefont
  {K.}~\bibnamefont {Krueger}}, \bibinfo {author} {\bibfnamefont
  {T.}~\bibnamefont {LeCompte}}, \bibinfo {author} {\bibfnamefont
  {D.}~\bibnamefont {Lopiano}}, \bibinfo {author} {\bibfnamefont
  {H.}~\bibnamefont {Spinka}}, \bibinfo {author} {\bibfnamefont
  {D.}~\bibnamefont {Underwood}},  \emph {et~al.},\ }\href@noop {} {\bibfield
  {journal} {\bibinfo  {journal} {Nucl. Instr. and Meth. A}\ }\textbf {\bibinfo
  {volume} {499}},\ \bibinfo {pages} {725} (\bibinfo {year}
  {2003})}\BibitemShut {NoStop}%
\bibitem [{\citenamefont {Allgower}\ \emph {et~al.}(2003)\citenamefont
  {Allgower}, \citenamefont {Anderson}, \citenamefont {Baldwin}, \citenamefont
  {Balewski}, \citenamefont {Belt-Tonjes}, \citenamefont {Bland}, \citenamefont
  {Brown}, \citenamefont {Cadman}, \citenamefont {Christie}, \citenamefont
  {Cyliax} \emph {et~al.}}]{stareemc}%
  \BibitemOpen
  \bibfield  {author} {\bibinfo {author} {\bibfnamefont {C.}~\bibnamefont
  {Allgower}}, \bibinfo {author} {\bibfnamefont {B.}~\bibnamefont {Anderson}},
  \bibinfo {author} {\bibfnamefont {A.}~\bibnamefont {Baldwin}}, \bibinfo
  {author} {\bibfnamefont {J.}~\bibnamefont {Balewski}}, \bibinfo {author}
  {\bibfnamefont {M.}~\bibnamefont {Belt-Tonjes}}, \bibinfo {author}
  {\bibfnamefont {L.}~\bibnamefont {Bland}}, \bibinfo {author} {\bibfnamefont
  {R.}~\bibnamefont {Brown}}, \bibinfo {author} {\bibfnamefont
  {R.}~\bibnamefont {Cadman}}, \bibinfo {author} {\bibfnamefont
  {W.}~\bibnamefont {Christie}}, \bibinfo {author} {\bibfnamefont
  {I.}~\bibnamefont {Cyliax}},  \emph {et~al.},\ }\href@noop {} {\bibfield
  {journal} {\bibinfo  {journal} {Nucl. Instr. and Meth. A}\ }\textbf {\bibinfo
  {volume} {499}},\ \bibinfo {pages} {740} (\bibinfo {year}
  {2003})}\BibitemShut {NoStop}%
\bibitem [{\citenamefont {Kiryluk}()}]{starbbc}%
  \BibitemOpen
  \bibfield  {author} {\bibinfo {author} {\bibfnamefont {J.}~\bibnamefont
  {Kiryluk}},\ }\href@noop {} {}\Eprint {http://arxiv.org/abs/0501072}
  {arXiv:0501072 [hep-ex]} \BibitemShut {NoStop}%
\bibitem [{\citenamefont {Adler}\ \emph {et~al.}(2001)\citenamefont {Adler},
  \citenamefont {Denisov}, \citenamefont {Garcia}, \citenamefont {Murray},
  \citenamefont {Strobele},\ and\ \citenamefont {White}}]{starzdc}%
  \BibitemOpen
  \bibfield  {author} {\bibinfo {author} {\bibfnamefont {C.}~\bibnamefont
  {Adler}}, \bibinfo {author} {\bibfnamefont {A.}~\bibnamefont {Denisov}},
  \bibinfo {author} {\bibfnamefont {E.}~\bibnamefont {Garcia}}, \bibinfo
  {author} {\bibfnamefont {M.}~\bibnamefont {Murray}}, \bibinfo {author}
  {\bibfnamefont {H.}~\bibnamefont {Strobele}}, \ and\ \bibinfo {author}
  {\bibfnamefont {S.}~\bibnamefont {White}},\ }\href@noop {} {\bibfield
  {journal} {\bibinfo  {journal} {Nucl. Instr. and Meth. A}\ }\textbf {\bibinfo
  {volume} {470}},\ \bibinfo {pages} {433} (\bibinfo {year}
  {2001})}\BibitemShut {NoStop}%
\bibitem [{\citenamefont {Llope}\ \emph {et~al.}()\citenamefont {Llope},
  \citenamefont {Zhou}, \citenamefont {Nussbaum}, \citenamefont {Hoffmann},
  \citenamefont {Asselta}, \citenamefont {Brandenburg}, \citenamefont
  {Butterworth}, \citenamefont {Camarda}, \citenamefont {Christie},
  \citenamefont {Crawford} \emph {et~al.}}]{starvpd}%
  \BibitemOpen
  \bibfield  {author} {\bibinfo {author} {\bibfnamefont {W.}~\bibnamefont
  {Llope}}, \bibinfo {author} {\bibfnamefont {J.}~\bibnamefont {Zhou}},
  \bibinfo {author} {\bibfnamefont {T.}~\bibnamefont {Nussbaum}}, \bibinfo
  {author} {\bibfnamefont {G.}~\bibnamefont {Hoffmann}}, \bibinfo {author}
  {\bibfnamefont {K.}~\bibnamefont {Asselta}}, \bibinfo {author} {\bibfnamefont
  {J.}~\bibnamefont {Brandenburg}}, \bibinfo {author} {\bibfnamefont
  {J.}~\bibnamefont {Butterworth}}, \bibinfo {author} {\bibfnamefont
  {T.}~\bibnamefont {Camarda}}, \bibinfo {author} {\bibfnamefont
  {W.}~\bibnamefont {Christie}}, \bibinfo {author} {\bibfnamefont
  {H.}~\bibnamefont {Crawford}},  \emph {et~al.},\ }\href@noop {} {}\Eprint
  {http://arxiv.org/abs/1403.6855} {arXiv:1403.6855 [physics.ins-det]}
  \BibitemShut {NoStop}%
\bibitem [{run(2013)}]{run12polnote}%
  \BibitemOpen
  \href@noop {} {\bibfield  {journal} {\bibinfo  {journal} {The RHIC
  Polarimetry Group, http://public.bnl.gov/docs/cad/Documents/RHIC polarization
  for Runs 9-12.pdf}\ } (\bibinfo {year} {2013})}\BibitemShut {NoStop}%
\bibitem [{run(2012)}]{run12polresult}%
  \BibitemOpen
  \href@noop {} {\bibfield  {journal} {\bibinfo  {journal} {The RHIC
  Polarimetry Group, https://wiki.bnl.gov/rhicspin/Run\_12\_polarization}\ }
  (\bibinfo {year} {2012})}\BibitemShut {NoStop}%
\bibitem [{\citenamefont {Cronin-Hennessy}\ \emph {et~al.}(2000)\citenamefont
  {Cronin-Hennessy}, \citenamefont {Beretvas},\ and\ \citenamefont
  {Derwent}}]{rellumcdf2000}%
  \BibitemOpen
  \bibfield  {author} {\bibinfo {author} {\bibfnamefont {D.}~\bibnamefont
  {Cronin-Hennessy}}, \bibinfo {author} {\bibfnamefont {A.}~\bibnamefont
  {Beretvas}}, \ and\ \bibinfo {author} {\bibfnamefont {P.~F.}\ \bibnamefont
  {Derwent}},\ }\href@noop {} {\bibfield  {journal} {\bibinfo  {journal} {Nucl.
  Instr. and Meth. A}\ }\textbf {\bibinfo {volume} {A}},\ \bibinfo {pages}
  {443} (\bibinfo {year} {2000})}\BibitemShut {NoStop}%
\bibitem [{rel(2012)}]{rellum2009}%
  \BibitemOpen
  \href@noop {} {\bibfield  {journal} {\bibinfo  {journal} {J. Hayes-Wehle, J.
  Seele and B. Surrow, STAR relative luminosity analysis note
  https://drupal.star.bnl.gov/STAR/starnotes/private/psn0570}\ } (\bibinfo
  {year} {2012})}\BibitemShut {NoStop}%
\bibitem [{\citenamefont {Cacciari}\ \emph {et~al.}(2012)\citenamefont
  {Cacciari}, \citenamefont {Salam},\ and\ \citenamefont {Soyez}}]{fastjet}%
  \BibitemOpen
  \bibfield  {author} {\bibinfo {author} {\bibfnamefont {M.}~\bibnamefont
  {Cacciari}}, \bibinfo {author} {\bibfnamefont {G.~P.}\ \bibnamefont {Salam}},
  \ and\ \bibinfo {author} {\bibfnamefont {G.}~\bibnamefont {Soyez}},\
  }\href@noop {} {\bibfield  {journal} {\bibinfo  {journal} {Euro. Phys. J.}\
  }\textbf {\bibinfo {volume} {C72}},\ \bibinfo {pages} {1896} (\bibinfo {year}
  {2012})}\BibitemShut {NoStop}%
\bibitem [{\citenamefont {Abelev}\ \emph {et~al.}(2015)\citenamefont {Abelev}
  \emph {et~al.}}]{alicecone}%
  \BibitemOpen
  \bibfield  {author} {\bibinfo {author} {\bibfnamefont {B.}~\bibnamefont
  {Abelev}} \emph {et~al.} (\bibinfo {collaboration} {ALICE Collaboration}),\
  }\href@noop {} {\bibfield  {journal} {\bibinfo  {journal} {Phys. Rev. D}\
  }\textbf {\bibinfo {volume} {91}},\ \bibinfo {pages} {112012} (\bibinfo
  {year} {2015})}\BibitemShut {NoStop}%
\bibitem [{\citenamefont {Sjostrand}\ \emph {et~al.}(2006)\citenamefont
  {Sjostrand}, \citenamefont {Mrenna},\ and\ \citenamefont
  {Skands}}]{pythia2006}%
  \BibitemOpen
  \bibfield  {author} {\bibinfo {author} {\bibfnamefont {T.}~\bibnamefont
  {Sjostrand}}, \bibinfo {author} {\bibfnamefont {S.}~\bibnamefont {Mrenna}}, \
  and\ \bibinfo {author} {\bibfnamefont {P.}~\bibnamefont {Skands}},\
  }\href@noop {} {\bibfield  {journal} {\bibinfo  {journal} {JHEP}\ }\textbf
  {\bibinfo {volume} {0605}},\ \bibinfo {pages} {026} (\bibinfo {year}
  {2006})}\BibitemShut {NoStop}%
\bibitem [{\citenamefont {Agostinelli}\ \emph {et~al.}(2003)\citenamefont
  {Agostinelli} \emph {et~al.}}]{geant4}%
  \BibitemOpen
  \bibfield  {author} {\bibinfo {author} {\bibfnamefont {S.}~\bibnamefont
  {Agostinelli}} \emph {et~al.} (\bibinfo {collaboration} {Geant4
  Collaboration}),\ }\href@noop {} {\bibfield  {journal} {\bibinfo  {journal}
  {Nucl. Instr. and Meth. A}\ }\textbf {\bibinfo {volume} {506}},\ \bibinfo
  {pages} {250} (\bibinfo {year} {2003})}\BibitemShut {NoStop}%
\bibitem [{\citenamefont {Skands}(2010)}]{perugia2010}%
  \BibitemOpen
  \bibfield  {author} {\bibinfo {author} {\bibfnamefont {P.}~\bibnamefont
  {Skands}},\ }\href@noop {} {\bibfield  {journal} {\bibinfo  {journal} {Phys.
  Rev. D}\ }\textbf {\bibinfo {volume} {82}},\ \bibinfo {pages} {074018}
  (\bibinfo {year} {2010})}\BibitemShut {NoStop}%
\bibitem [{\citenamefont {Agakishiev}\ \emph {et~al.}(2006)\citenamefont
  {Agakishiev} \emph {et~al.}}]{starpipm2006}%
  \BibitemOpen
  \bibfield  {author} {\bibinfo {author} {\bibfnamefont {G.}~\bibnamefont
  {Agakishiev}} \emph {et~al.} (\bibinfo {collaboration} {STAR
  Collaboration}),\ }\href@noop {} {\bibfield  {journal} {\bibinfo  {journal}
  {Phys. Lett. B}\ }\textbf {\bibinfo {volume} {637}},\ \bibinfo {pages} {161}
  (\bibinfo {year} {2006})}\BibitemShut {NoStop}%
\bibitem [{\citenamefont {Agakishiev}\ \emph {et~al.}(2012)\citenamefont
  {Agakishiev} \emph {et~al.}}]{starpipm2012}%
  \BibitemOpen
  \bibfield  {author} {\bibinfo {author} {\bibfnamefont {G.}~\bibnamefont
  {Agakishiev}} \emph {et~al.} (\bibinfo {collaboration} {STAR
  Collaboration}),\ }\href@noop {} {\bibfield  {journal} {\bibinfo  {journal}
  {Phys. Rev. Lett.}\ }\textbf {\bibinfo {volume} {108}},\ \bibinfo {pages}
  {072302} (\bibinfo {year} {2012})}\BibitemShut {NoStop}%
\bibitem [{\citenamefont {Combridge}\ \emph {et~al.}(1977)\citenamefont
  {Combridge}, \citenamefont {Kripfganz},\ and\ \citenamefont
  {Ranft}}]{locrssection1977}%
  \BibitemOpen
  \bibfield  {author} {\bibinfo {author} {\bibfnamefont {B.~L.}\ \bibnamefont
  {Combridge}}, \bibinfo {author} {\bibfnamefont {J.}~\bibnamefont
  {Kripfganz}}, \ and\ \bibinfo {author} {\bibfnamefont {J.}~\bibnamefont
  {Ranft}},\ }\href@noop {} {\bibfield  {journal} {\bibinfo  {journal} {Phys.
  Lett. B}\ }\textbf {\bibinfo {volume} {70}},\ \bibinfo {pages} {234}
  (\bibinfo {year} {1977})}\BibitemShut {NoStop}%
\bibitem [{\citenamefont {Adams}\ \emph {et~al.}(2004)\citenamefont {Adams}
  \emph {et~al.}}]{MarciaET}%
  \BibitemOpen
  \bibfield  {author} {\bibinfo {author} {\bibfnamefont {J.}~\bibnamefont
  {Adams}} \emph {et~al.} (\bibinfo {collaboration} {STAR Collaboration}),\
  }\href@noop {} {\bibfield  {journal} {\bibinfo  {journal} {Phys. Rev. C}\
  }\textbf {\bibinfo {volume} {70}},\ \bibinfo {pages} {054907} (\bibinfo
  {year} {2004})}\BibitemShut {NoStop}%
\end{thebibliography}%
